\documentclass[fleqn,usenatbib]{mnras}
\usepackage{supertabular}
\usepackage{caption}
\usepackage{newtxtext,newtxmath}

\usepackage[T1]{fontenc}

\usepackage{ragged2e}

\usepackage{comment}

\DeclareRobustCommand{\VAN}[3]{#2}
\let\VANthebibliography\thebibliography
\def\thebibliography{\DeclareRobustCommand{\VAN}[3]{##3}\VANthebibliography}

\newcommand{\rmax}{\ensuremath{r_\mathrm{max}}}
\newcommand{\tpeak}{\ensuremath{\tau_\mathrm{peak}}}
\newcommand{\tcent}{\ensuremath{\tau_\mathrm{cent}}}
\newcommand{\tjav}{\ensuremath{\tau_\mathrm{JAV}}}
\newcommand{\javelin}{{\sc javelin}\,}
\newcommand{\pyccf}{{\sc PyCCF}\,}

\newcommand{\sw}[1]{\textit{Swift}}

\newcommand{\orcid}[1]{\textsuperscript{\href{http://orcid.org/#1}{
\hskip2pt\includegraphics[width=8pt]{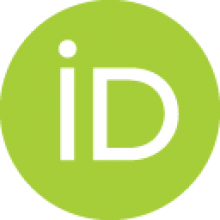}}}}

\usepackage{graphicx}	
\usepackage{amsmath}	
\usepackage{lineno}  

\title[Standard Disk Predictions]{Departures from Standard Disk Predictions in Intensive Ground-Based Monitoring of Three AGN}

\author[Gonzalez-Buitrago et al.]{
D. Gonzalez-Buitrago\orcid{0000-0002-9280-1184},$^{1,2}$\thanks{E-mail: dgonzalez@astro.unam.mx}
A. J. Barth\orcid{0000-0002-3026-0562},$^{2}$
R. Edelson\orcid{0000-0001-8598-1482},$^{3}$
J.~V. Hern{\'a}ndez Santisteban\orcid{0000-0002-6733-5556},$^{4}$
Keith Horne\orcid{0000-0003-1728-0304},$^{4}$
\newauthor
T. Schmidt\orcid{0000-0002-2772-8160},$^{2,3}$
Yan-Rong Li\orcid{0000-0001-5841-9179},$^{5}$
Hengxiao Guo\orcid{0000-0001-8416-7059},$^{6}$
M. D. Joner\orcid{0000-0003-0634-8449},$^{7}$
E. Cackett\orcid{0000-0002-8294-9281},$^{8}$
J. Gelbord\orcid{0000-0001-9092-8619},$^{9}$ 
M.C. Bentz\orcid{0000-0002-2816-5398},$^{10}$
\newauthor
W.N. Brandt\orcid{0000-0002-0167-2453},$^{11,12,13}$
M. Goad,$^{14}$
K. Korista\orcid{0000-0003-0944-1008},$^{15}$
M. Vestergaard\orcid{0000-0001-9191-9837},$^{16,17}$
C. Villforth\orcid{0000-0002-8956-6654},$^{18}$ 
A. Breeveld\orcid{0000-0002-0001-7270},$^{19}$ 
\newauthor
T.~G.~Brink\orcid{0000-0001-5955-2502},$^{20}$
E. M. Corsini\orcid{0000-0003-3460-5633},$^{21,22}$ 
E. Dalla Bont\`a\orcid{0000-0001-9931-8681},$^{21,22,23}$
Gary J Ferland\orcid{0000-0003-4503-6333},$^{24}$
A.~V.~Filippenko\orcid{0000-0003-3460-0103},$^{20}$
\newauthor
Ma. T. Garc\'ia-D\'iaz\orcid{0000-0002-9772-5555},$^{1}$
M. Hallum\orcid{0000-0002-6634-9673},$^{7,25}$
J. C.  Horst$^{26}$
M. Kim\orcid{0000-0002-3560-0781},$^{27}$
Y. Krongold,$^{28}$
J. Kruger,$^{7}$
B. Kuhn$^{29}$
\newauthor
S. Kumar\orcid{0000-0001-8367-7591},$^{30}$
M. Mehdipour\orcid{0000-0002-4992-4664},$^{31}$
L. Morelli\orcid{0000-0001-6890-3503},$^{32}$
S. Mathur,$^{33,34}$
H. Netzer,$^{35}$
P. Ochner\orcid{0000-0001-5578-8614},$^{21,22}$
I. Pagotto\orcid{0000-0002-3701-1892},$^{21}$
\newauthor
A. Pizzella\orcid{0000-0001-9585-417X},$^{21,22}$ 
D.~J. Sand\orcid{0000-0003-4102-380X},$^{22}$
A. Siviero,$^{21,22}$
M. Spencer,$^{7}$
H. Sung\orcid{0000-0001-9515-3584},$^{36}$
S. Vaughan,$^{37}$
H. Winkler\orcid{0000-0003-2662-0526},$^{38}$
\newauthor
W. Zheng\orcid{0000-0002-2636-6508},$^{20,39}$\\
\textit{Affiliations listed at the end of the manuscript.}
\\
}

\date{Accepted XXX. Received YYY; in original form ZZZ}

\pubyear{\the\year{}}

\begin{document}
\label{firstpage}
\pagerange{\pageref{firstpage}--\pageref{lastpage}}
\maketitle

\begin{abstract}
We present ground-based, multi-band light curves of the AGN Mrk~509, NGC\,4151, and NGC\,4593 obtained contemporaneously with \sw\, monitoring.
We measure cross-correlation lags relative to \sw\, UVW2 (1928~\AA) and test the standard prediction for disk reprocessing, which assumes a geometrically thin, optically thick accretion disk where continuum interband delays follow the relation \( \tau(\lambda) \propto \lambda^{4/3} \).
For Mrk~509 the 273-d \sw\, campaign gives well-defined lags that increase with wavelength as $\tau(\lambda)\propto\lambda^{2.17\pm0.2}$, steeper than the thin-disk prediction, and the optical lags are a factor of $\sim5$ longer than expected for a simple disk-reprocessing model.
This ``disk-size discrepancy'' as well as excess lags in the $u$ and $r$ bands (which include the Balmer continuum and H$\alpha$, respectively) suggest a mix of short lags from the disk and longer lags from nebular continuum originating in the broad-line region.
The shorter \sw\, campaigns, 69~d on NGC\,4151 and 22~d on NGC\,4593, yield less well-defined, shorter lags $<2$~d.
The NGC\,4593 lags are consistent with $\tau(\lambda) \propto \lambda^{4/3}$ but with uncertainties too large for a strong test.
For NGC\,4151 the \sw\, lags match $\tau(\lambda) \propto \lambda^{4/3}$, with a small $U$-band excess, but the ground-based lags in the $r$, $i$, and $z$ bands are significantly shorter than the $B$ and $g$ lags, and also shorter than expected from the thin-disk prediction.
The interpretation of this unusual lag spectrum is unclear.
Overall these results indicate significant diversity in the $\tau-\lambda$ relation across the optical/UV/NIR, which differs from the more homogeneous behavior seen in the \sw\, bands.
\end{abstract}



\section{Introduction}
\label{sec:intro}

Although it is now  accepted that nearly   all galaxies harbor a supermassive black hole (SMBH) at their centers, in the local universe only a small fraction display an active galactic nucleus (AGN), with core luminosities exceeding  $10^{42}$ erg s$^{-1}$, sometimes outshining the rest of their host galaxy. These large luminosities are thought to arise from gravitational potential energy released in a “central engine” composed of a SMBH, a compact X-ray-emitting corona, and a larger, flatter UV/optical-emitting accretion disk. This core structure is further surrounded by the broad-line region (BLR), a dusty torus, and the narrow-line region (NLR) \citep{Urry1995}.

In recent decades, our knowledge about AGNs has increased considerably, yet many open questions remain regarding their structure, geometry, and behavior \citep[e.g.,][]{Lawrence:2018, Davis2020}. Determining the architecture and conditions of the central engine is critical for understanding the processes that contribute to their extraordinary luminosity and dominate their accretion mechanisms, as well as potentially testing general relativity in the strong gravity regime. Unfortunately, the SMBH/corona/disk system in highly accreting AGN is far too small to be spatially resolved directly \citep[though near-infrared interferometry has resolved the BLR in a few cases;][]{GravityColl:2018,GravityColl:2020}, so indirect methods have been employed for its study. One such approach is gravitational microlensing \citep{Morgan10,Blackburne:2011,Chartas:2016,Kochanek:2020}, which requires a rare intervening lens. Another widely used technique is Reverberation Mapping (RM), first proposed by \citet{Blandford82}, which has long served as a powerful tool to probe the BLR surrounding the central engine.
This involves using UV or optical spectroscopy to monitor AGN over long durations, and then measuring ``light echoes,'' or time delays, in the continuum light curve (thought to arise in the accretion disk) as it is reprocessed into broad emission lines and diffuse continuum produced in the BLR. These time lags represent the delay between variations in the continuum and the corresponding response in the emission lines, allowing BLR size measurements under the assumption that the dominant timescale is the light-travel time between the regions \citep[see][for a recent review]{Cackett:2021}.

More recently this same basic technique has been used to probe the accretion disk through measurement of continuum reverberation lags, following early detections of wavelength-dependent continuum reverberation in the UV and optical \citep{Collier98, Sergeev05, Fausnaugh17, Jiang2024}. AGN intensive broadband reverberation mapping (IBRM) monitoring began with the 2014 ground-based/\sw/Hubble Space Telescope (\textit{HST}) ``AGN-STORM'' campaign  NGC\,5548, which monitored the X-ray/UV/optical continuum at high cadence (1-3 times per day) and long duration (120-180 days).
This groundbreaking campaign yielded a number of important results \citep{Edelson15,Fausnaugh16,deRosa:2015}. These include:
1) Strong UV/optical correlations with time lags increasing with wavelength, consistent with the expected $\tau(\lambda) \propto \lambda^{4/3}$ slope from the standard thin disk model \citep{ShakuraSunyaev73};
2) However, the observed lags were $\sim$3 times longer than predicted, suggesting that the accretion disk is larger than the standard model would imply;
3) The $U$ band  (which includes a contribution from the Balmer recombination continuum) shows an excess lag most likely associated with diffuse BLR continuum emission \citep{Korista00}; 
4) The correlation between X-ray and UV variations is much weaker than the correlations between UV and optical bands \citep{McHardy2016}, casting doubt on the standard reprocessing model that links the X-ray corona and UV-emitting accretion disk.
\sw\, monitoring studies of other AGN broadly confirm these findings \citep[e.g.,][]{Edelson17, McHardy18, Edelson19, Cackett:2020, Hernandez20, Vincentelli:2021, Gonzalez2022, Kara2023, Gonzalez-Buitrago2023}. In all, over a dozen AGN have now been subjected to \sw\, IBRM, giving results generally consistent with the trends found for NGC 5548. Analogous results have been obtained over more limited wavelength ranges for individual AGN targeted in ground-based monitoring programs \citep[e.g.,][]{Fausnaugh2018, PozoNunez2019, Fian2022}, and for large samples of quasars observed across a broad range of redshifts in time-domain surveys, where results show significant object-to-object variation in lag behavior, including cases with strong diffuse continuum contributions from the BLR \citep[e.g.,][]{Jiang17,Mudd18,Homayouni19,Yu2020:DES,  Jha2022,WGuo2022,HGuo2022, Sharp:2024, Mandal:2025}. A more extensive discussion and further references can be found in \citet{Cackett:2021}.
Many targets from intensive \sw, programs are also monitored with simultaneous ground-based optical observations. These ground-based campaigns complement \sw, data by extending lag measurements to longer wavelengths and providing higher S/N photometry in the \sw, optical filters (U, B, and V), allowing for more robust testing of previous results. 

In this paper, we report optical photometry of three \sw\ ~targets from \citet{Edelson19}, using coordinated ground-based observations to broaden the wavelength coverage for continuum lag measurements, with the aim of evaluating the standard disk reprocessing model and exploring possible additional contributions, such as diffuse emission from the BLR continuum or intrinsic disk fluctuations.
The paper is organized as follows: The observations are presented in Section~\ref{sec:Observation}, the data reduction in Section~\ref{sec:reduction}, measurement of reverberation lags in Section~\ref{sec:analysis}, and examination of wavelength-dependent variability behavior in Sections~\ref{sec:fluxflux} and \ref{sec:lagfits}. A summary and brief concluding remarks are given in Section~\ref{sec:summary}.

\section{Sample and Observations}
\label{sec:Observation}

\begin{table*}
    \centering
    \caption{AGN Properties and Campaign Dates}
    \label{tab:AGNproperties}
    \begin{tabular}{lcccccccc}
        \hline\hline
        Object & Redshift  & Distance & $L_{5100}$  & $M_{\mathrm{BH}}$ & $\dot{m}_{\mathrm{Edd}}$ & $E(B-V)$ & Date Range  & Date Range  \\
         &  &  &  &  &  &  &  \sw\, & Ground \\
         &  ($z$) &  (Mpc) & (erg s$^{-1}$) &  ($M_{\odot}$) & $\dot{m}_{\mathrm{Edd}}$ & (mag) & (MJD) &  (MJD) \\
        (1) & (2) & (3) & (4) & (5) & (6) & (7) &  (8) & (9) \\    
        \hline
        Mrk~509    & 0.0344  & 147   & $1.2 \times 10^{44}$ & $1.1 \times 10^{8}$  & 0.095 & 0.051 & 57829.8 - 58102.5 & 57827.8 - 58092.5 \\
        NGC\,4151  & 0.0033  & 15.8  & $2.7 \times 10^{42}$ & $2.3 \times 10^{7}$  & 0.014 & 0.024 & 57438.0 - 57507.3 & 57416.9 - 57608.3 \\
        NGC\,4593  & 0.0090  & 37.7  & $2.3 \times 10^{42}$ & $8.2 \times 10^{6}$  & 0.081 & 0.022 & 57582.7 - 57605.3 & 57540.2 - 57613.4 \\
        \hline
    \end{tabular}
    \vspace{0.2cm}
    \begin{minipage}{\textwidth}
        {\small
        \textbf{Notes:} Luminosity distances for Mrk~509 and NGC\,4593 are from the AGN Black Hole Mass Database \citep{bentz2015}. For NGC\,4151, the Cepheid distance is from \citet{Yuan2020}. 
        Luminosities at 5100 Å for Mrk~509 and NGC\,4593 are measured from HST STIS spectra obtained during the \sw\, campaigns; $L_{5100}$ refers to $\lambda L_\lambda$ measured at $\lambda_{\mathrm{rest}} = 5100$ Å. 
        For Mrk~509, this is measured from our STIS spectrum taken at MJD 58048.86, and for NGC\,4593, the luminosity is determined from the mean spectrum of the STIS monitoring campaign described by \citet{Cackett18}. 
        The $L_{5100}$ value for NGC\,4151 is from \citet{bentz13}, rescaled to the updated Cepheid distance from \citet{Yuan2020}. 
        Sources for black hole masses and Eddington ratio estimates ($\dot{m}_{\mathrm{Edd}}$) are described in Section~\ref{sec:sample}. 
        Galactic foreground reddening $E(B-V)$ values are based on extinction data from NED \citep{Schlafly11} with the assumption of $R_V=3.1$.
        }
    \end{minipage}
\end{table*}

\subsection{Sample and AGN Properties}
\label{sec:sample} 
This study presents data from the ground-based photometric campaigns coordinated with the \sw\, programs for three of the four AGN from the IBRM sample of \citet{Edelson19}: Mrk~509, NGC\,4151, and NGC\,4593 \citep{McHardy18}. For the fourth object in the \citet{Edelson19} sample, NGC\,5548, the ground-based light curves and lag measurements were presented by \citet{Fausnaugh16}. Table~\ref{tab:AGNproperties} lists properties of the three AGN included here. Measurements of black-hole mass were obtained from H$\beta$ reverberation mapping, as compiled by \citet{bentz2015}. Masses were calculated from H$\beta$ lags and velocity widths $\sigma_{\rm line}$ assuming a virial $f$-factor of 4.3 \citep{Grier13}, using reverberation measurements from \citet{Peterson04} for Mrk~509, \citet{Bentz06} and \citet{derosa18} for NGC\,4151, and \citet{Denney06} and \citet{Barth15} for NGC\,4593. We also include in Table~\ref{tab:AGNproperties} estimates of the Eddington ratio $\dot{m}_\mathrm{Edd} = L_\mathrm{bol}/L_\mathrm{Edd}$ for each object, adopting values directly from previous studies based on multiwavelength SED modeling. Specifically, we use $\dot{m} = 0.095$ for Mrk~509 \citep{Vasudevan:2009}, 0.014 for NGC\,4151 \citep{Mahmoud2020}, and 0.081 for NGC\,4593 \citep{McHardy18}. For Mrk~509, the estimate is based on contemporaneous SED observations, which help reduce the impact of AGN variability. In contrast, the SEDs used for NGC\,4151 and NGC\,4593 are not contemporaneous with the Swift monitoring campaigns, so additional uncertainty due to intrinsic variability should be considered. An uncertainty of at least a factor of $\sim$3 should be assumed for these Eddington-ratio estimates \citep{Onken04}, stemming from uncertainty in the black-hole masses, bolometric luminosities, and variability.

\subsection{\sw\, Campaigns}

Results from the \sw\, programs on these three targets have been published previously, and we summarize the campaign properties here.  
\sw\, carried out  intensive monitoring of Mrk509 (in 2017), NGC4151 (in 2016) and NGC4593 (in 2016) for 272.6, 69.3, and 22.6 days, respectively, with mean sampling intervals of 1.07, 0.22, and 0.12 days.
Observations for each object were obtained with the \sw\, X-ray Telescope \citep[XRT;][]{Burrows2005} and the UV-Optical Telescope \citep[UVOT;][]{Roming05} in the UVW2, UVM2, UVW1, U, B, and V filter bands. Table~\ref{tab:AGNproperties} lists properties of the targets and the dates of the observing campaigns.   The \sw\, campaigns and reverberation lags have previously been described by \citet{Edelson17} for NGC\,4151, \citet{McHardy18} for NGC\,4593, and \citet{Edelson19} for all three of these objects. Independent analyses of the \sw\, campaign data for NGC\,4593 and Mrk~509 were also carried out by \citet{Pal2018} and \citet{Kumari2021}, respectively.

\begin{table*}
    \centering
    \caption{Telescopes, Filters, and Locations}
    \label{tab:AGNphotometry}
    \begin{tabular}{lllc}
        \hline\hline
        Object & Telescope (Location) & Filters & \# Nights \\
        (1) & (2) & (3) & (4) \\
        \hline
        Mrk\,509  & \sw  & & X-ray, UVW2, UVM2, UVW1, U, B, V  & 273 \\
                  & LCOGT (Global Network)  & $V$, $u$, $g$, $r$, $i$, $z_s$  & 264 \\
                  & LT (La Palma, Spain)  & $u$, $g$, $r$, $i$, $z$    & 266  \\
                  & RATIR (San Pedro Mártir, Mexico)  & $Z$, $Y$, $J$          & 120 \\
                  & WMO (Wyoming, USA)  & $V$               & 216    \\
                  & HST  & G140L, G230L, G430L, G750L     & 1    \\
        \hline
        NGC\,4151 & \sw  &  & X-ray, UVW2, UVM2, UVW1, U, B, V   & 69  \\
                 & Asiago (Asiago, Italy)  & $B$, $V$, $R$, $I$   & 159  \\
                  & FTN (Hawaii, USA)  & $B$, $V$, $R$, $I$, $g$, $r$, $i$, $z_s$  & 204  \\
                  & KAIT (Lick Observatory, USA)  & $B$, $V$, $R$, $I$    & 168  \\
                  & LCOGT (Global Network)  & $B$, $V$, $R$, $I$, $u$, $g$, $r$, $i$, $z_s$ & 182 \\
                  & LT (La Palma, Spain)  & $u$, $g$, $r$, $i$, $z$     & 173 \\
                  & MLO (Montana, USA)  & $V$            & 64  \\
                  & Nickel (Lick Observatory, USA)  & $B$, $V$, $R$, $I$ & 166   \\
                  & RDS (Rozhen, Bulgaria)  & $V$, $R$  & 77   \\
                  & SW (Sonneberg, Germany)  & $g$, $r$, $i$  & 46   \\
                  & WMO (Wyoming, USA)  & $B$, $V$, $R$, $I$    & 133  \\
        \hline
        NGC\,4593 & \sw  &  &  X-ray, UVW2, UVM2, UVW1, U, B, V   & 22    \\
                  & LCOGT (Global Network)  & $u$, $g$, $r$, $i$, $z_s$   & 73    \\
                  & FTN (Hawaii, USA)  & $g$, $r$, $i$, $z_s$       & 2   \\
                  & FTS (Australia)  & $u$, $g$, $r$, $i$, $z_s$   & 38    \\
        \hline
    \end{tabular}
    \vspace{0.2cm}
    \begin{minipage}{\linewidth}
    \justifying
        {\small
        \textbf{Notes:} Telescope names are as listed in Section~\ref{sec:Telescopes}, and Table~\ref{tab:Telescopes1} lists the properties of the telescopes and cameras. Column~4 lists the number of unique observing epochs, where each epoch corresponds to a distinct calendar date with one or more exposures. The locations correspond to the primary sites of each telescope.
        }
    \end{minipage}
\end{table*}

\subsection{Ground-based Monitoring}
\label{sec:Telescopes}
In concurrent with the \sw\, observation campaigns for these three targets, intensive photometric monitoring was carried out with multiple ground-based telescopes to broaden the wavelength coverage of the dataset and to provide light curves of higher S/N in the optical bands. Each ground-based campaign was organized independently, resulting in heterogeneous coverage (in terms of telescopes and filter sets) for the three targets. The filter sets used for each monitoring campaign are listed in Table~\ref{tab:AGNphotometry}.  
Table~\ref{tab:Telescopes1} lists other properties of the telescopes and CCD cameras used in the campaigns. The ground-based campaign for Mrk~509 spanned approximately the same range of dates as the \sw\, program, while for NGC\,4151 and NGC\,4593 the ground-based monitoring programs covered a longer duration extending both before and after the \sw\, campaigns (Table~\ref{tab:AGNproperties}).

\begin{table*}
    \centering
    \caption{Telescope and Detector Properties}
    \label{tab:Telescopes1}
    \begin{tabular}{lccccccc}
        \hline\hline
         & Primary  & Pixel &  &  Readout & Pixel &  & Field  \\
        Telescope &  Mirror & Scale & Gain  &  Noise &  Array &  Binning  &  of View  \\
         &  Diameter (m) &   ($^{\prime\prime}$/pix) &  (e$^-$/ADU) &  (e$^-$) & (pix $\times$ pix) &  & ($^{\prime} \times ^{\prime}$) \\
        \hline
        \sw{} UVOT  & 0.3  & 0.5  & --  & --  & 1024 $\times$ 1024  & 1 $\times$ 1 & 17 $\times$ 17 \\
        LCOGT       & 1.0  & 0.38  & 1.0  & 12.0  & 4096 $\times$ 4096  & 1 $\times$ 1 & 26.5 $\times$ 26.5 \\
        FTN/FTS     & 2.0  & 0.27  & 2.0  & 4.0   & 2048 $\times$ 2048  & 2 $\times$ 2 & 4.7 $\times$ 4.7  \\
        LT          & 2.0  & 0.15  & 1.6  & 8.0   & 4096 $\times$ 4112  & 2 $\times$ 2 & 10 $\times$ 10  \\
        WMO         & 0.9  & 0.61  & 1.4  & 12.0  & 2000 $\times$ 2000  & 1 $\times$ 1 & 20.5 $\times$ 20.5   \\
        Nickel      & 1.0  & 0.18  & 1.7  & 8.3   & 2048 $\times$ 2048  & 2 $\times$ 2 & 6.3 $\times$ 6.3  \\
        KAIT        & 0.76 & 0.80  & 3.8  & 12.0  & 500  $\times$ 500   & 1 $\times$ 1 & 6.8 $\times$ 6.8 \\
        MLO         & 1.0  & 0.39  & 2.5  & 7.8   & 2048 $\times$ 2048  & 1 $\times$ 1 & 13.3 $\times$ 13.3 \\
        Asiago      & 0.92 & 0.86  & 0.9  & 12.0  & 4008 $\times$ 2672  & 1 $\times$ 1 & 57 $\times$ 38 \\
        SW          & 0.4  & 0.72  & 0.4  & 13.2  & 1676 $\times$ 1266  & 2 $\times$ 2 & 19 $\times$ 15 \\
        RDS         & 0.5  & 0.43  & 1.3  & 12.0  & 4096 $\times$ 4096  & 1 $\times$ 1 & 29.4 $\times$ 29.4 \\
        RATIR (C2)  & 1.5  & 0.30  & 2.2  & 14.7  & 2048 $\times$ 2048  & 2 $\times$ 2 & 10 $\times$ 10 \\
        RATIR (C3)  & 1.5  & 0.30  & 2.4  & 11.2  & 2048 $\times$ 2048  & 2 $\times$ 2 & 10 $\times$ 10 \\
        \hline
    \end{tabular}
    \vspace{0.2cm}
    \begin{minipage}{\linewidth}
    \justifying
        {\small
        \textbf{Notes:} Telescope names are as listed in Section~\ref{sec:Telescopes}. For RATIR, the properties listed refer to the HAWAII-2RG detectors in the C2 and C3 near-infrared imaging channels.
        }
    \end{minipage}
\end{table*}

\emph{Las Cumbres Observatory Global Telescope network (LCOGT):} The largest contributions of data to these ground-based monitoring programs are from the robotic 1\,m telescopes of Las Cumbres Observatory Global Telescope network \citep{Brown13, Boroson14}, obtained as part of the LCOGT AGN Key Project \citep{Valenti15, hlabathe2020, Hernandez20}. These telescopes are located at  Siding Spring Observatory, the South African Astronomical Observatory, Cerro Tololo Inter-American Observatory, and McDonald Observatory. 

\emph{Faulkes Telescope North/South (FTN/FTS):}  The LCOGT AGN Key Project also obtained observations with the robotic 2\,m Faulkes Telescope North (FTN) located  Haleakala Observatory on Maui, and the Faulkes Telescope South (FTS) located at Siding Spring Observatory, which are part of the LCOGT network.

\emph{Liverpool Telescope (LT):} The robotic 2\,m Liverpool Telescope at the Observatorio del Roque de los Muchachos on La Palma in the Canary Islands was used for observations with the IO:O camera as part of the monitoring campaign.

\emph{West Mountain Observatory (WMO):} The 0.9\,m telescope at West Mountain Observatory (WMO) in Utah.

\emph{Lick Observatory:} The 1\,m Nickel Telescope (Nickel) and the robotic 0.76\,m Katzman Automatic Imaging Telescope \citep[KAIT;][]{Filippenko01} at Lick Observatory.

\emph{Mount Laguna Observatory (MLO):} The 1\,m telescope at Mount Laguna Observatory participated in the monitoring campaign.

\emph{Asiago Astrophysical Observatory (Asiago):} The 0.92\,m telescope at the Asiago Astrophysical Observatory was used for photometric observations during the campaign.

\emph{Southwestern University Observatory (SW):} The 0.4\,m Fountainwood telescope, operated by Southwestern University in Georgetown, Texas, contributed to the observational campaign.

\emph{Rancho Del Sol Observatory (RDS):} The 0.5\,m telescope at Rancho Del Sol Observatory in Camino, California, was used for observations during the campaign.

\emph{Reionization and Transients InfraRed Telescope (RATIR):} The 1.5\,m RATIR telescope at the Observatorio Astron\'omico Nacional de San Pedro M\'artir, M\'exico \citep{Watson12}, provided simultaneous optical and near-infrared imaging. Observations were obtained in the $Z$, $Y$ (C2 channel) and $J$ bands (C3 channel), though the $H$-band data had insufficient S/N. While RATIR covered only part of the campaign, it extended the wavelength range beyond the optical bands used at other facilities.

\subsection{Filter Sets}
 
Filters used to monitor each AGN at each telescope are listed in Table~\ref{tab:AGNphotometry}. For the \emph{BVRI} (italicized text) filter bands, data included both \citet{Bessell90} filters (LCOGT, FTN, Liverpool, Asiago, KAIT, Nickel) and  Johnson-Cousins filters (WMO). For brevity, we will refer to all of these (Bessell and Johnson-Cousins) as Johnson bands hereafter.  Data were obtained at several telescopes using Sloan Digital Sky Survey (SDSS) \emph{u$^\prime$g$^\prime$r$^\prime$i$^\prime$} filters \citep{Fukugita1996}. (We omit the primes on the SDSS filter names hereafter in this paper.)

For the $z$ band, three different filter variants were used. These include LCOGT and FTN/FTS observations with a Pan-STARRS $z_s$ filter ($\lambda_c = 8700$ \AA), LT observations using an SDSS $z^\prime$ filter ($\lambda_c = 9134$ \AA), and RATIR observations in a $Z$ filter ($\lambda_c = 9200$ \AA). 

\subsection{Hubble Space Telescope Spectroscopy}\label{sec:hst}

\begin{figure*}
    \centering
    \includegraphics[width=\textwidth]{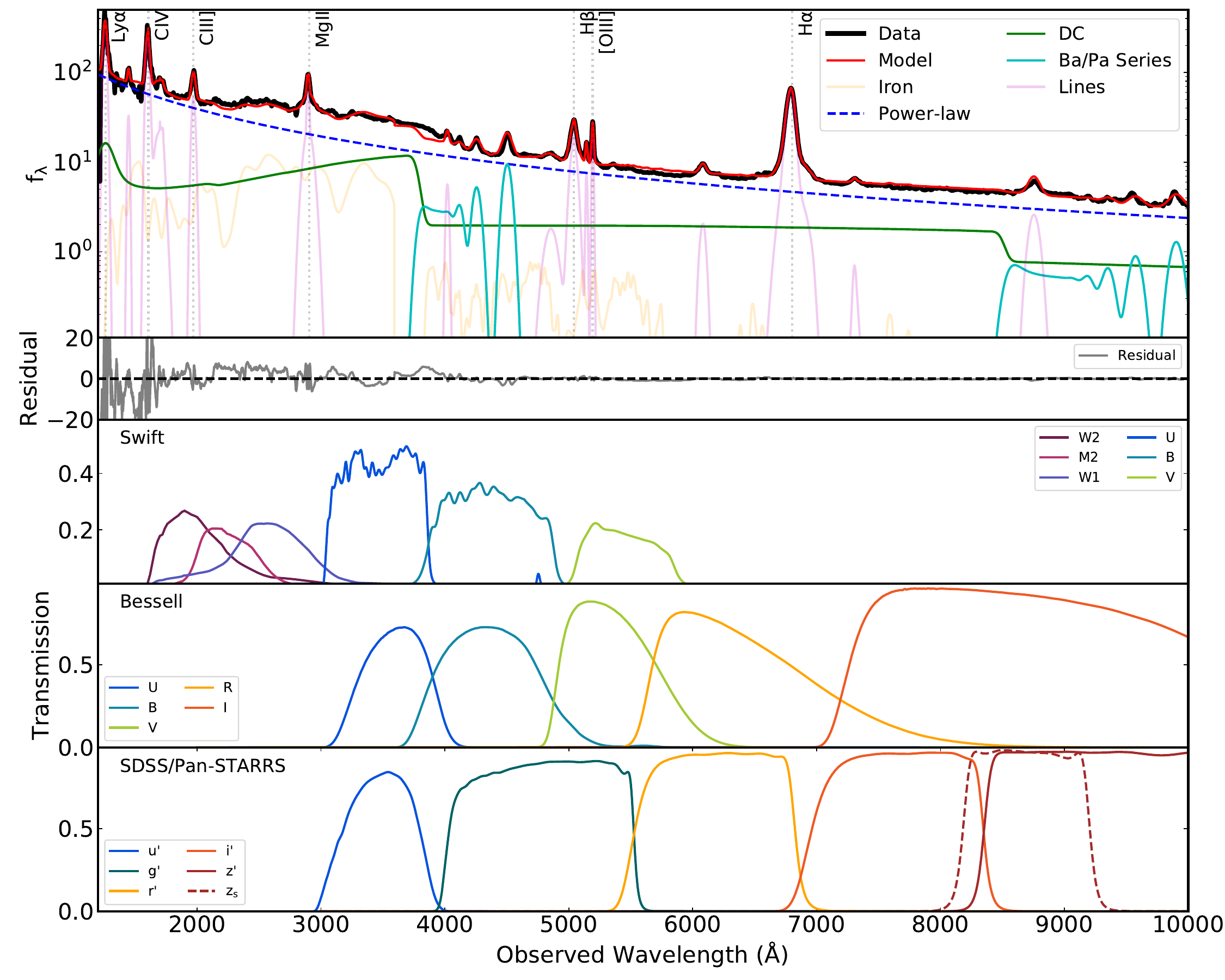}
    \caption{UV and optical spectrum of Mrk~509 and filter passbands used in the monitoring campaign, illustrating the range of spectral features that contribute to each band. \emph{Upper panel:} HST STIS spectrum and multi-component model fit (described in Section \ref{sec:Mrk509specfit}), including power-law continuum, emission lines, \ion{Fe}{II} blends, and diffuse continuum (DC). Units are $10^{-15}$ erg cm$^{-2}$ s$^{-1}$ \AA$^{-1}$. Residuals of the fit are displayed below the spectrum. \emph{Lower panels:} Filter transmission curves for \sw\, and ground-based passbands. Ground-based transmission curves represent the $u^\prime g^\prime r^\prime i^\prime z_s$ filter bands used at LCOGT telescopes, and the $z^\prime$ band used at some of the other sites. The displayed transmission curves do not include the effects of atmospheric absorption or detector quantum efficiency.}
    \label{fig:stis_with_filters}
\end{figure*}

We additionally observed Mrk~509 with the Space Telescope Imaging Spectrograph (STIS) on the Hubble Space Telescope (HST) during the \sw\, campaign, on 2017 October 22 (program GO-15124), to determine the relative contributions of the emission-line and continuum components to the broad-band filters used in the \sw\, and ground-based monitoring programs. Observations were taken with the gratings  G140L (with the FUV-MAMA detector), G230L (with the NUV-MAMA), and the optical G430L and G750L gratings (with the STIS CCD detector). Together, these settings provide continuous coverage over $\sim1150$--10000 \AA\ at spectral resolving power of $R\sim 500$. Total exposure times were 2208, 1150, 400, and 200 s with G140L, G230L, G430L, and G750L, respectively. For each grating the observation was split into 2, 3, or 4 dithered exposures to facilitate removal of cosmic-ray hits and bad pixels.  For the optical CCD exposures, the AGN was centered at the E1 aperture position to minimize charge-transfer losses. Additional CCD fringe flats were taken in the G750L setting at the end of the observing sequence. 

The STIS data were processed using the STIS calibration pipeline tools. For the UV exposures, we used the 1-d extracted spectra produced by the task {\sc{x1d}}. The optical exposures (G430L and G750L) required additional cleaning due to the prevalence of bad pixels and charge-transfer trails on the CCD.  For these exposures, we followed the HST pipeline procedure in {\sc{PyRAF}}\footnote{The standard HST calibration pipeline is available at \url{https://github.com/spacetelescope/hstcal}.}, performing overscan subtraction, flat fielding, bias corrections, and dark corrections with the {\sc{basic2d}} task without any cosmic-ray rejection. We then used {\sc{l.a.cosmic}} \citep{vanDokkum01} to eliminate the remaining bad pixels and cosmic-ray hits, as this performed a better cleaning of the CCD images than the native pipeline cosmic ray rejection for our data.  For the G750L data, we also corrected for CCD fringing by applying the {\sc{normspflat}}, {\sc{mkfringeflat}}, and {\sc{defringe}} tasks and using fringe flat exposures taken at the end of the HST visit. After combining the individual cleaned exposures with the {\sc{imshift}} and {\sc{imcombine}} tasks to account for the slight dithers between the images, we applied the standard pipeline calibration steps to extract the combined 1-d spectrum using the {\sc{x1d}} task. 

The complete extracted STIS spectrum of Mrk\,509 is shown in Fig.~\ref{fig:stis_with_filters}, along with the \sw\, and ground-based filter passbands in the UV and optical ranges, to illustrate the spectral features and continuum regions that contribute to each band. The continuum S/N per pixel in the final STIS spectrum is $\sim68$, 88, 136, and 82 at wavelengths 1400, 2600, 5300, and 7500 \AA, respectively. The model spectrum in Figure~\ref{fig:stis_with_filters} is discussed in Section~6.3.
 
For NGC\,4593, 26 epochs of STIS UV and optical monitoring observations coordinated with its \sw\, campaign  were carried out in HST program GO-14121, and a continuum reverberation analysis of these spectra was presented by \citet{Cackett18}. No HST spectroscopic observations of NGC\,4151 were obtained during its \sw\, monitoring campaign.

\section{Light Curve Measurement}
\label{sec:reduction}

\subsection{Ground-Based Data: Initial Processing}

Standard data-processing steps including overscan subtraction and flat-fielding were applied to each image either by an automated pipeline (where provided by the observatory facility) or otherwise by the observers. For telescopes that did not include world coordinate system (WCS) solutions as part of their standard processing pipelines, we  used \texttt{astrometry.net} \citep{Lang10} to derive WCS solutions and add this information to the FITS headers. 

The RATIR near-infrared data were reduced through standard bias and dark subtraction, flat fielding, and cosmic ray corrections using the PhotoPipe pipeline.\footnote{\url{https://github.com/maxperry/photometrypipeline}}

\subsection{Photometry}

Light curves for each AGN were measured using the automated aperture photometry procedure described by \citet{Pei14}. RATIR light curves were measured separately using different procedures, described below. This photometry pipeline, written in IDL, is implemented as a set of wrapper routines for the aperture photometry routines in the IDL Astronomy User's Library \citep{Landsman1993}, and is designed to operate efficiently and consistently on large and heterogeneous data sets consisting of images from multiple telescopes and cameras having different plate scales and CCD properties. In each image, the AGN and several comparison stars (at least six for each image) are identified automatically using the WCS information in the image header, and the AGN and comparison star instrumental magnitudes are measured through a circular aperture with background sky measured in a surrounding annulus. Magnitude offsets are then derived for each image in order to minimize the scatter of the comparison star light curves relative to the mean magnitude. Applying these magnitude offsets to the AGN photometry yields a light curve in instrumental magnitudes for each telescope and filter. The error estimates returned by the photometry procedure incorporate photon-counting uncertainties, background sky noise, and CCD readout noise, but do not include other (sometimes dominant) error sources such as flat-fielding variations and point-spread function variations across the field of view. For consistency with \citet{Edelson19}, dates used for light curves in this paper are given as Modified Julian Date (MJD), but we also include Heliocentric Julian Date (HJD) for each data point in Table~\ref{tab:data} for completeness (The heliocentric correction to the observation times is much smaller than the uncertainties in the measured reverberation lags.)

\begin{table*}
    \centering
    \caption{Photometric Data}
    \label{tab:data}
    \begin{tabular}{lccccc c}
        \hline\hline
        Name & Filter & MJD & THJD & $f_\lambda$ & $\sigma$ & Telescope \\
             &        &     &  &  &  &  \\
         (1)  & (2)& (3) & (4) & (5) & (6) & (7) \\
        \hline
        Mrk~509 & $u$ & 57847.6201 & 7848.1176 & 2.336 & 0.010 &  LCOGT \\
        Mrk~509 & $u$ & 57847.6240 & 7848.1215 & 2.358 & 0.011 &  LCOGT \\
        Mrk~509 & $u$ & 57848.6394 & 7849.1370 & 2.340 & 0.007 &  LCOGT \\
        \hline
    \end{tabular}
    \vspace{0.2cm}
    \begin{minipage}{\linewidth}
    \justifying
        {\small
        \textbf{Notes:} This table includes both ground-based and \sw\, photometric data used in this paper. For each data point, Column~3 lists the Modified Julian Date (MJD), used in the time-series analysis in this paper. Column~4 lists the truncated Heliocentric Julian Date (THJD = Heliocentric Julian Date$-$2450000) for completeness. Columns 5 and 6 list the flux density ($f_\lambda$) and uncertainty in units of $10^{-14}$ erg cm$^{-2}$ s$^{-1}$ \AA$^{-1}$ for the UV, optical, and near-IR bands, while for the \sw\, X-ray bands (HX, SX) the data are in units of counts s$^{-1}$. This table is available online in its entirety in machine-readable form.
        }
    \end{minipage}
\end{table*}

A photometric aperture radius of 4\arcsec\ was used for Mrk~509 and NGC\,4593, and a 3\arcsec\ aperture was used for NGC\,4151, chosen to optimize the signal-to-noise ratio while minimizing host-galaxy contamination. In all cases, the background sky annulus spanned a radial range of 10\arcsec\ to 20\arcsec\ around each object.  The resulting light curves include a residual contribution of host-galaxy light within the photometric aperture that should be constant across the light curve, except for small changes resulting from differences in seeing.  For data points that were large outliers in the light curves, the images were inspected visually. In most cases, the discrepant photometric measurements could be traced back to defects in the data such as poor focus, guiding errors, poor flatfielding, or high background noise levels resulting from bright moonlight and clouds, and those faulty images were then discarded from the final light curve measurements. Such outliers accounted for fewer than 2\% of the total data points.

To place the light curves on a calibrated magnitude scale, we obtained magnitudes from SDSS and from the APASS catalog \citep{henden2019} for the comparison stars, and used a weighted average over the comparison stars to calculate the magnitude offset between instrumental and calibrated magnitudes. For filters that are not included in the APASS survey ($R$ and $I$), approximate calibrations were obtained from the $r$ and $i$-band comparison star magnitudes using color-dependent transformations \citep{Fukugita1996}. The comparison stars were selected to span a similar magnitude range as the target AGN (typically $14 < V < 17$). We visually inspected their light curves and excluded any stars showing signs of variability during the monitoring period. The AGN light curves were then converted from magnitudes to flux densities (erg~cm$^{-2}$~s$^{-1}$~\AA$^{-1}$) using standard photometric zeropoints \citep{Fukugita1996, bessell1998}. 

Photometry was performed separately on the RATIR data using the IRAF {\sc qphot} and {\sc apphot} tasks with an aperture radius of 5\arcsec\ and a sky annulus between 7\arcsec\ and 10\arcsec. The magnitude scale for the RATIR $J$-band was calibrated using 2MASS magnitudes of comparison stars. Calibration for the $Z$ and $Y$ bands was carried out using the color equations from \cite{Hodgkin09}, based on the 2MASS $J$ and $H$ magnitudes of the same comparison stars used for the $J$-band calibration.  

\subsection{Light Curve Intercalibration}

Light curves measured in a given filter at different telescopes often show systematic offsets from one another that result from differences in filter transmission curves or wavelength-dependent detector quantum efficiency. Some of our light curves exhibited clear differences between data points from different telescopes that exceeded the level of the photometric uncertainties. These differences are particularly noticeable in light curves of very high S/N, such as the Mrk~509 data, where systematic offsets of $\sim0.02$ mag between LCOGT and LT data were present in some filter bands. Either additive or multiplicative rescaling of the light curves is insufficient to remove fully these discrepancies, but a combination of additive and multiplicative rescaling can usually provide a satisfactory intercalibration. 
Minor differences in filter transmission and effective wavelengths may cause small systematic offsets, but any required color corrections are likely negligible compared to other sources of error.

To combine the light curves from multiple telescopes, we used the Bayesian code PyCALI\footnote{\url{https://github.com/LiyrAstroph/PyCALI}} \citep{Li2014}.  PyCALI models light curves using a damped random walk (DRW) process \citep{Kelly:2009,Kozlowski:2016,Kozlowski:2017}, and applies an additive and multiplicative rescaling to each telescope's data in order to carry out the intercalibration.
Additionally, PyCALI expands the photometric error bars by adding a systematic term in quadrature to the photometric flux uncertainties for each telescope's data. These final error estimates also incorporate uncertainties
and covariances of the additive and multiplicative factors for each telescope's data. For our data, PyCALI almost entirely removes the systematic differences between light curves from different telescopes, implying that the intercalibrations are viable, although small residual differences can still be discerned in the highest S/N light curves, for example between the LCOGT and LT observations of Mrk~509. These residuals are far below the overall variability amplitude in the light curves, and their impact on the lag measurements is negligible. For input to PyCALI, each 1\,m telescope of the LCOGT network was treated as an independent telescope. Although these telescopes and cameras have identical design, there can be noticeable systematic offsets in light curves between them, particularly in bluer filters, as discussed by \citet{Hernandez20}.

For Mrk~509 and NGC\,4151, we combined the SDSS $z^\prime$ and Pan-STARRS $z_s$ data into a single $z$-band light curve, and PyCALI achieved a successful intercalibration between these two bands. The RATIR $Z$-band data for Mrk~509, having lower S/N and a shorter monitoring duration, were analyzed separately and not included in the $z$-band intercalibration.

\begin{figure*}
    \centering
    \includegraphics[trim=1.0cm 5.5cm 1.0cm 8.5cm, clip, width=\textwidth]{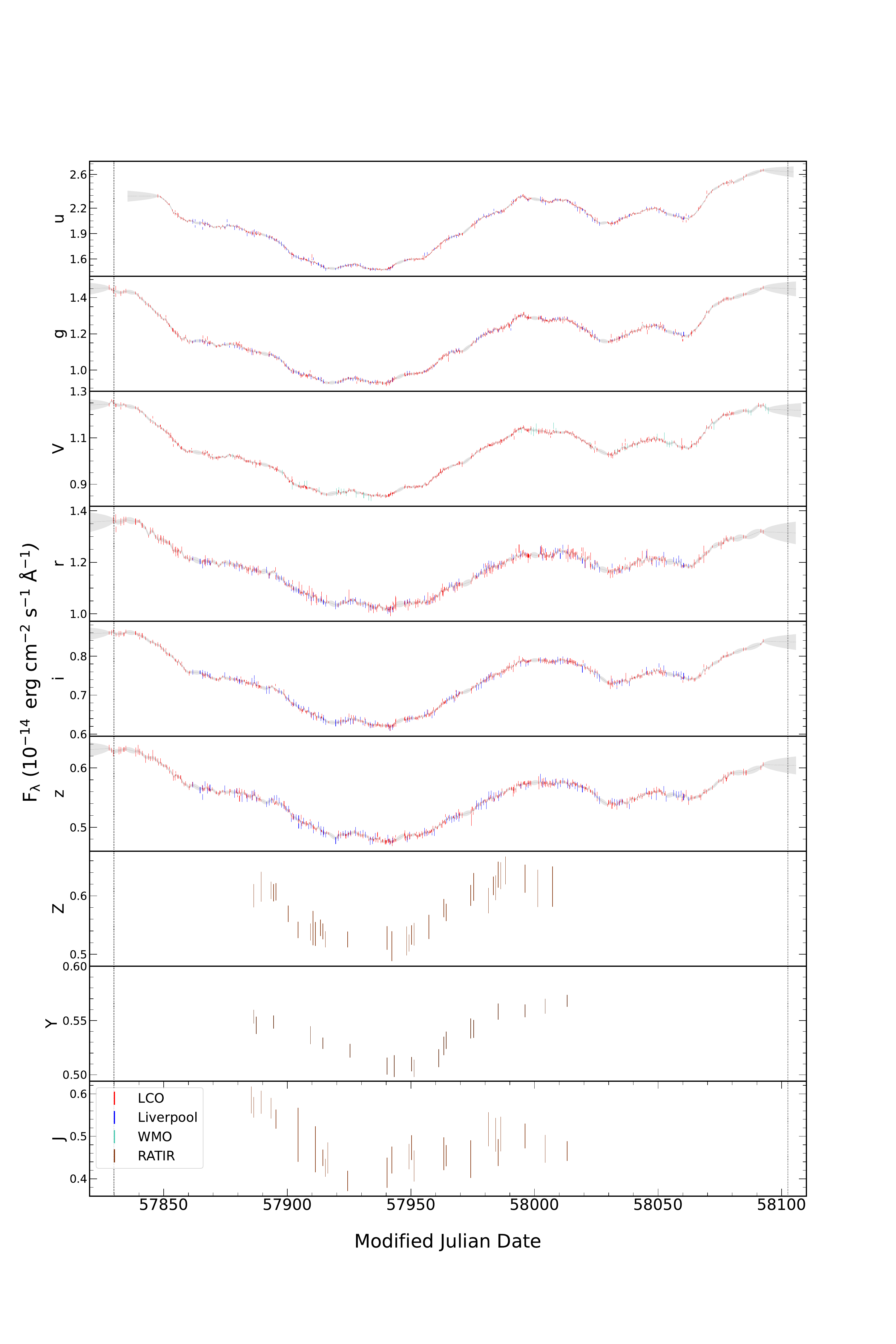}
    \caption{Light curves of Mrk~509 in all ground-based filter bands. The legend in the lower panel lists the colors used to denote each telescope.  Vertical dotted lines correspond to the starting and ending dates of the \sw\, campaign. The gray curves and surrounding shaded regions illustrate the mean and standard deviation of the DRW models from the PyCALI intercalibration. No intercalibration was performed for the \emph{Z}, \emph{Y}, and \emph{J} bands since these data were taken from a single telescope (RATIR).}
    \label{fig:lcMrk509}
\end{figure*}

\begin{figure*}
\begin{center}
    \includegraphics[trim=1.0cm 6.2cm 1.0cm 8.5cm, clip, width=16cm]{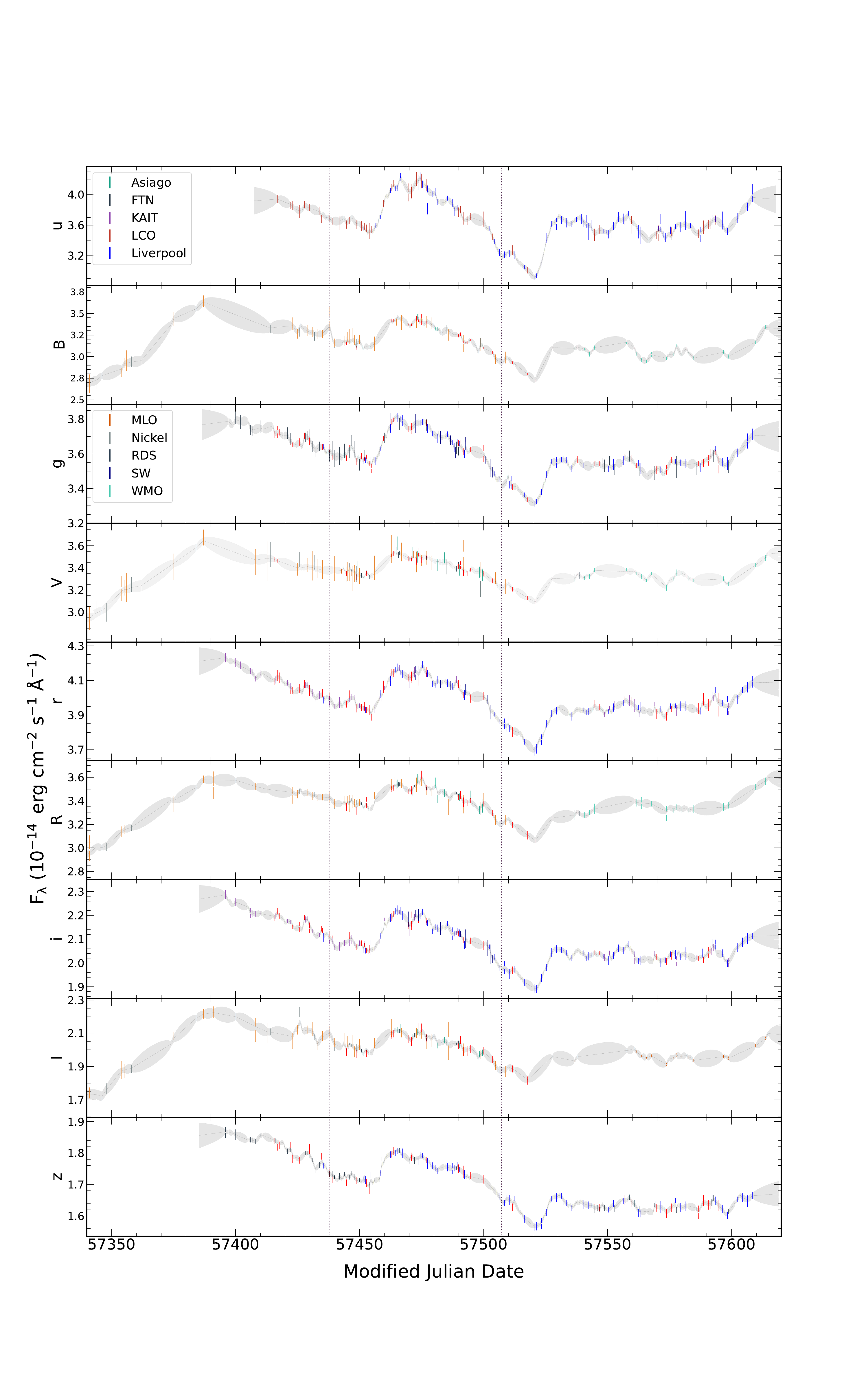}
\end{center}
    \caption{Light curves of NGC\,4151 in all ground-based filter bands. The legends in the first and third panels list the colors used to denote each telescope.  Vertical dotted lines correspond to the starting and ending dates of the \sw\, campaign. The gray curves and surrounding shaded regions illustrate the mean and standard deviation of the DRW models from the PyCALI intercalibration.}
    \label{fig:lcNGC4151}
\end{figure*}

\begin{figure*}
\begin{center}
    \includegraphics[trim=1.0cm 3.cm 1.0cm 6.0cm, clip, width=17cm]{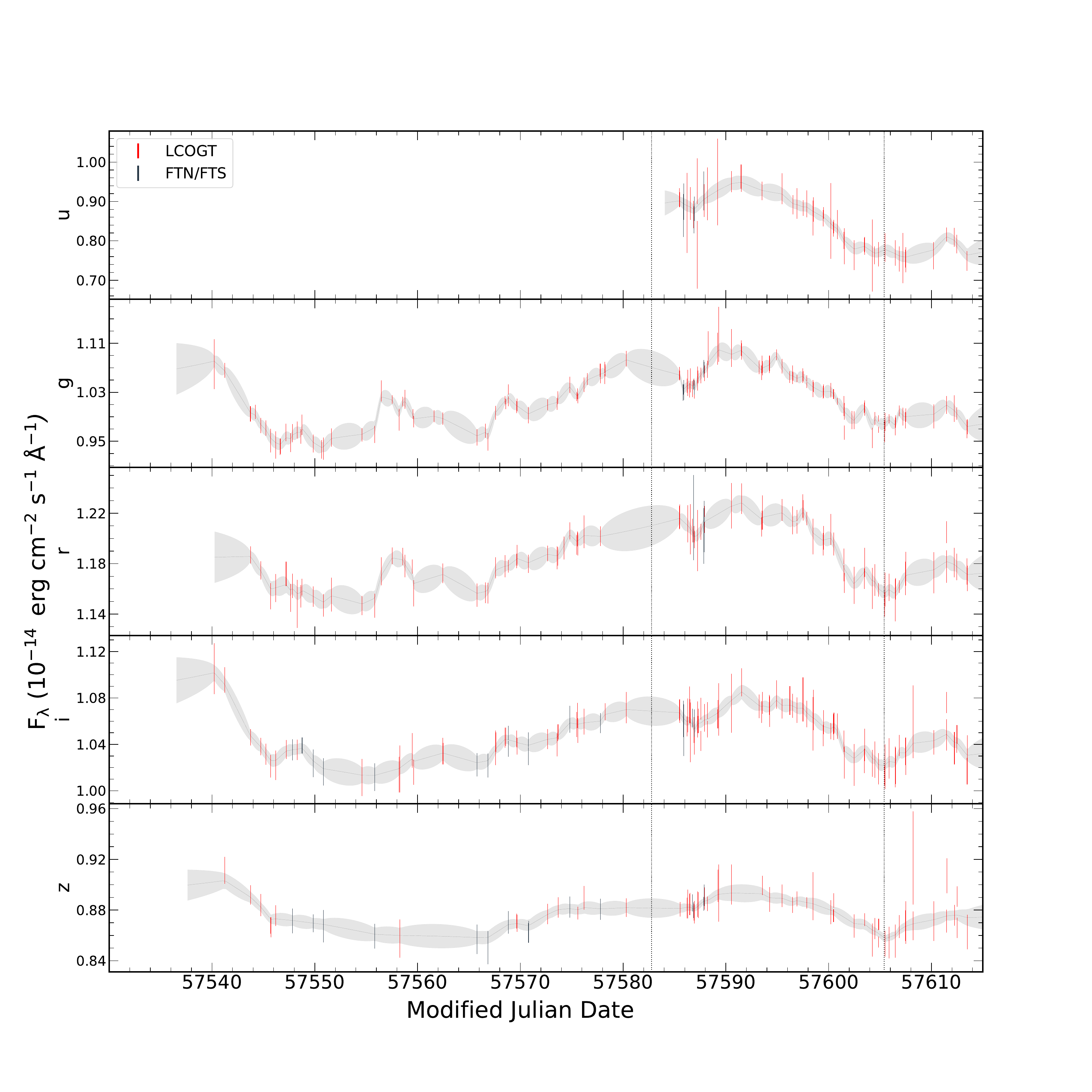}
\end{center}
    \caption{Light curves of NGC\,4593 in all ground-based filter bands. The legend in the upper panel lists the colors used to denote each telescope.  Vertical dotted lines correspond to the starting and ending dates of the \sw\, campaign. The gray curves and surrounding shaded regions illustrate the mean and standard deviation of the DRW models from the PyCALI intercalibration.}
    \label{fig:lcNGC4593}
\end{figure*}

The final ground-based light curves and PyCALI DRW models are displayed in Figures~\ref{fig:lcMrk509},  \ref{fig:lcNGC4151}, and \ref{fig:lcNGC4593} for Mrk~509, NGC\,4151, and NGC\,4593, respectively. The optical light curves of Mrk~509 have the best quality among the three AGN, thanks to a combination of high observational cadence, very high variability amplitude, and the relatively small contribution of the AGN host galaxy. For NGC\,4151, the \emph{ugriz} light curves have substantially higher S/N and better temporal sampling than the \emph{BVRI} light curves. For NGC\,4593, ground-based observations were obtained on most nights during the \sw\, campaign, but the combination of the short duration of the \sw\, campaign (23 d) and a relatively low variability amplitude for the longer wavelength optical light curves limit the utility of the data for measurement of accurate optical lags relative to the UV.

\subsection{\sw\, data}

As mentioned earlier, \sw\, data were obtained using the XRT (0.3--10 keV) and the UVOT (in six broad UV/optical photometric band). For the UVOT data, we have carried out new measurements of the \sw\, data in order to use  improved instrumental calibrations and dropout filtering, resulting in very small changes to the light curves compared with those presented by \citet{Edelson19}. The UVOT reduction follows the same general procedure described by \citet{Hernandez20}, so the reader is referred to that paper for details. As part of this reduction, the standard UVOT pipeline corrections for coincidence loss were applied. We also verified that the count rates for all targets remained well below the known saturation limits.
We applied this method to the UVOT data on the three targets in this paper and present the results alongside the ground-based data in Table~\ref{tab:data}.
Finally the XRT data were reduced using the technique of \cite{Evans09}  as discussed in \citet{Hernandez20}.


\section{Time-Series Analysis} \label{sec:analysis}
We carried out reverberation lag measurements using two techniques:  1) the ``Interpolated Cross-Correlation Function'' (ICCF) method of \cite{Gaskell87}, as detailed in \cite{Peterson04}, and
2) the ``Just Another Vehicle for Estimating Lags In Nuclei'' (JAVELIN) method developed by \cite{Zu11}.  These analyses are detailed in the following subsections. We use the \sw\, UVW2 band light curve as the driving band for lag measurements, as it is the shortest wavelength UV band available.

\subsection{Interpolated Cross-Correlation Function}\label{sec:iccf}

The ICCF measurements were carried out using the \pyccf code \citep{pyccf}.  
This method uses linear interpolation between adjacent points in unevenly sampled time series to determine the CCF between the driving and responding light curves. 
The peak amplitude of the CCF is denoted $r_\mathrm{max}$, and \tpeak\ is the lag corresponding to this peak. 
The CCF centroid, measured over points in the CCF exceeding  $0.8\,r_\mathrm{max}$, is denoted \tcent. We use \tcent\ as the best estimate of the overall time delay since it is more closely related to the centroid of the transfer function than \tpeak\ \citep{Peterson04}. We make no detrending corrections, apart from subtracting the mean, before measuring the ICCF lags.

The uncertainty in lag is estimated through a Monte Carlo resampling procedure, the ``flux randomization/random subset selection'' (FR/RSS) method \citep{Peterson98, White:1994}. From an ensemble of FR/RSS realizations of the light curves, the distribution of \tcent\ is determined (the cross-correlation centroid distribution, or CCCD). 
The median of the CCCD is used as the final value of \tcent, and the uncertainties in \tcent\ are determined by confidence intervals equivalent to 68\% of the FR/RSS realisations of the CCCD respectively.

\begin{figure*}
    \centering
    \includegraphics[trim=0.1cm 0.1cm 0.1cm 0.2cm, clip, scale=0.28]{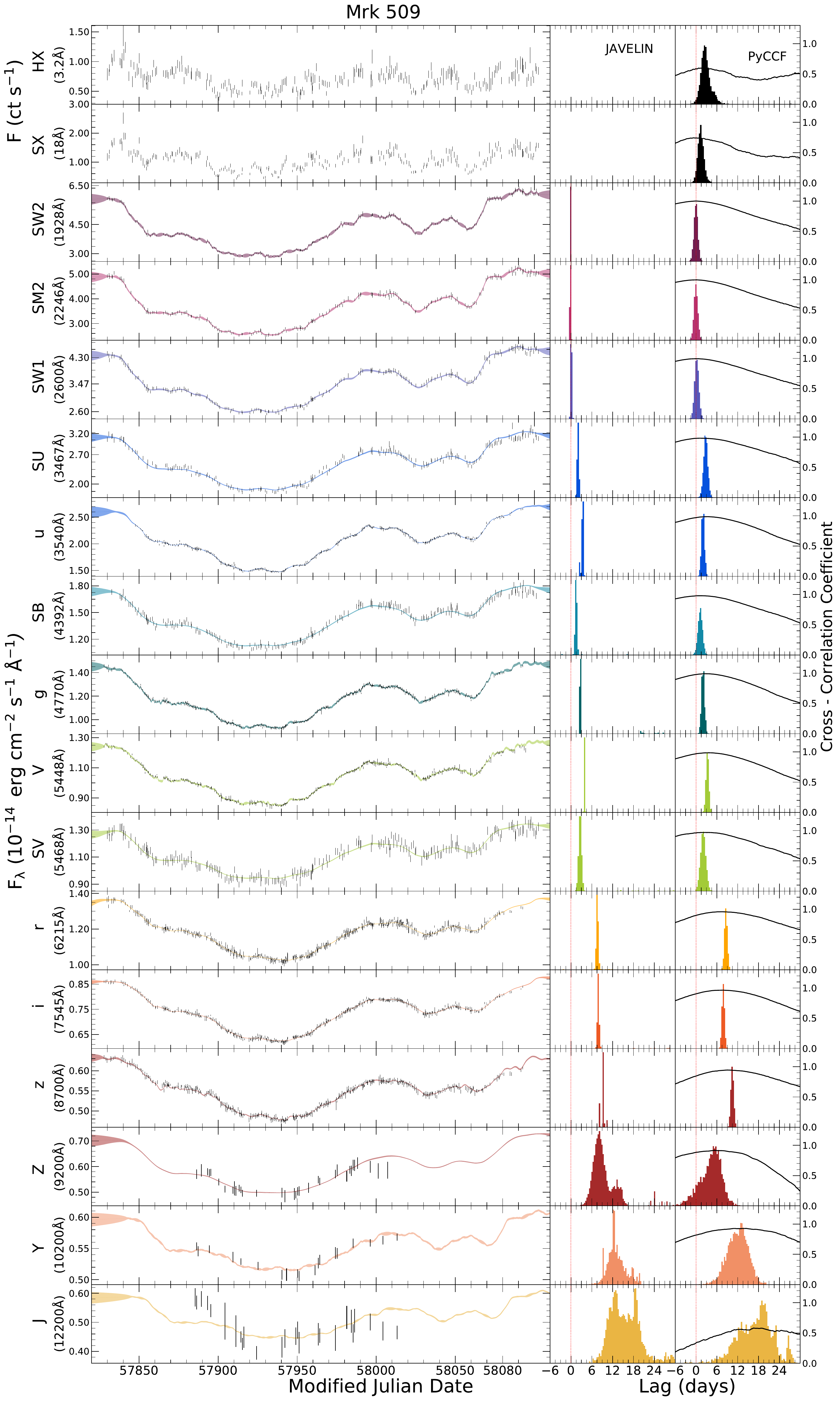}
    \caption{Lag measurements for Mrk~509. Left panels show the light curve for each band (black points with error bars) and the \javelin uncertainty band. For each wavelength band, the two panels at right show the \javelin posterior lag distribution and the cross-correlation results. In the \pyccf panels, the black curve is the CCF, and the histogram shows the cross-correlation centroid distribution. The CCF $y$-axis values range from 0 to 1. All lags are measured in the observed frame relative to the \sw\, UVW2-band light curve. The red vertical line in the \javelin and \pyccf panels denotes zero lag. \sw\, UVOT bands are denoted with an S, as in SV, SB, and SU. \javelin fits are omitted for the X-ray bands.}
    \label{fig:lag_Mrk509}
\end{figure*}

\begin{figure*}
\begin{center}
    \includegraphics[trim=0.2cm 7.5cm 1.5cm 7.2cm, clip, width=17cm]{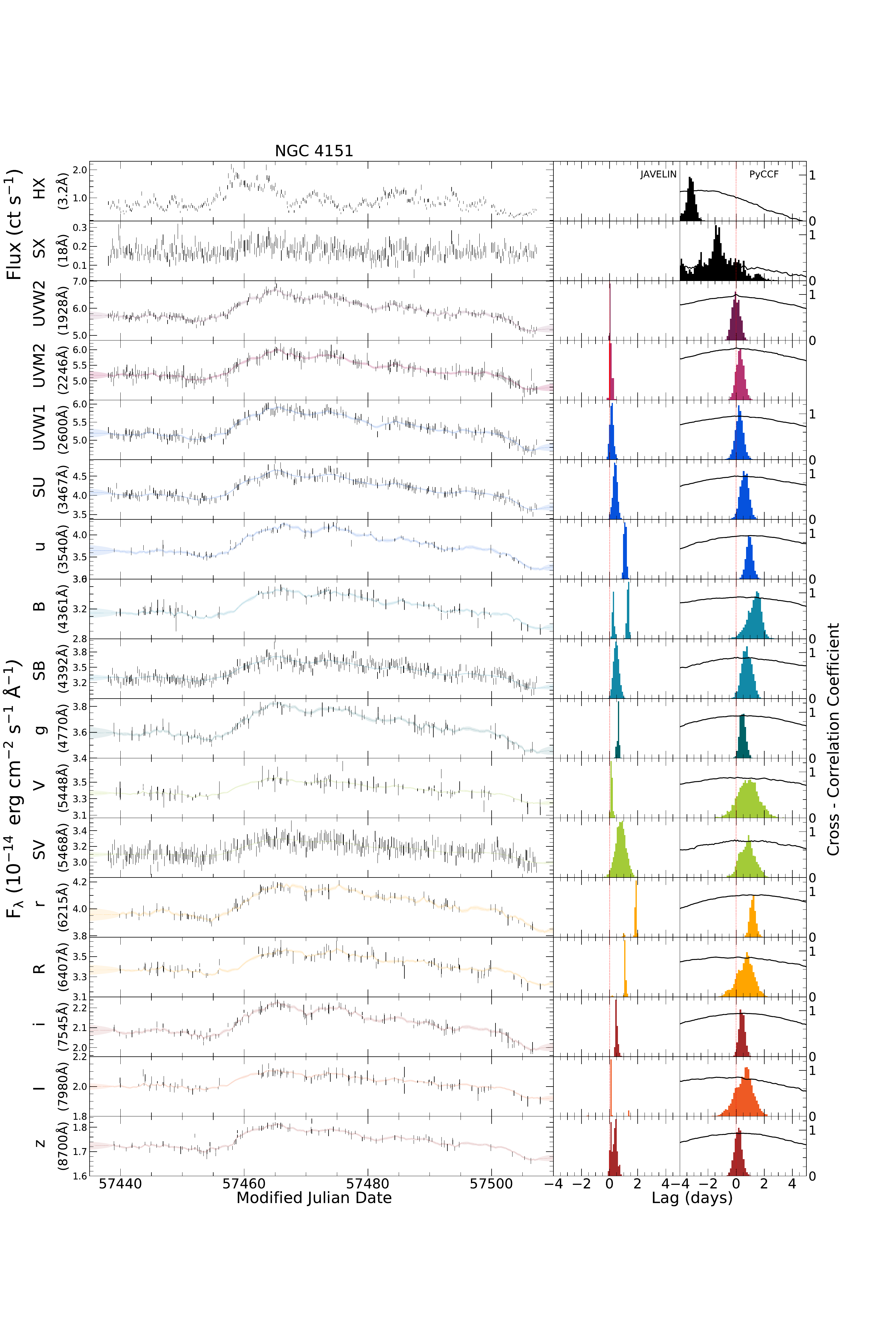}
\end{center}
    \caption{Lightcurve fits (left) and lag measurements (right) as in as in Fig.~\ref{fig:lag_Mrk509} but for NGC\,4151. 
    }
    \label{fig:lag_NGC4151}
\end{figure*}

\begin{figure*}
\begin{center}
    \includegraphics[trim=0.3cm 6.8cm 1.5cm 7.2cm, clip, width=17cm]{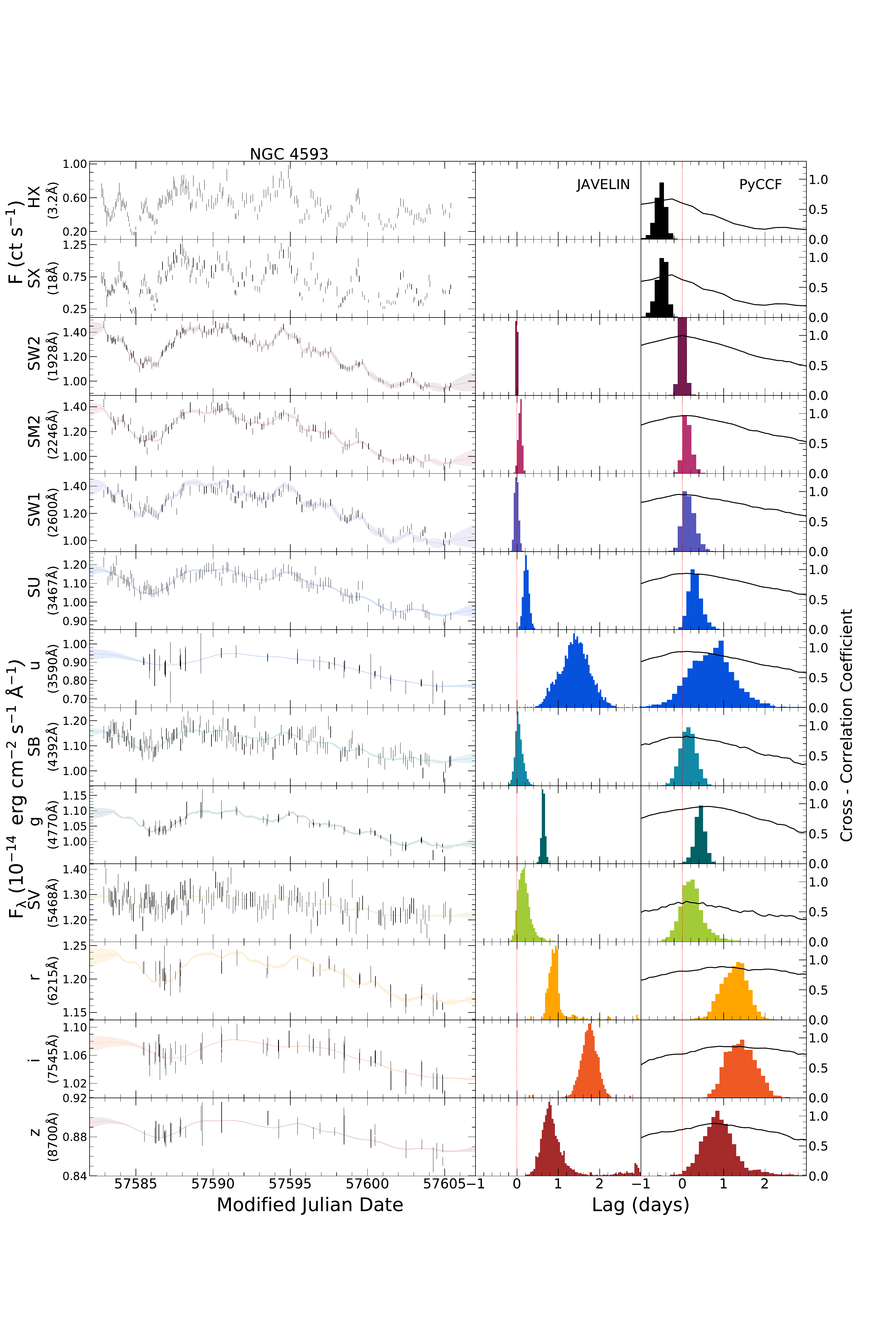}
\end{center}
    \caption{Lightcurve fits (left) and lag measurements (right) as in Fig.~\ref{fig:lag_Mrk509} but for NGC\,4593. 
    }
    \label{fig:lag_NGC4593}
\end{figure*}

CCFs were calculated over a search range of $\pm200$ days for Mrk\,509, $\pm70$ days for NGC\,4151 and $\pm15$ days for NGC\,4593,  with a lag increment of 0.25 days for Mrk~509 and an increment of 0.1 days for NGC\,4151 and NGC\,4593.  For NGC\,4151 and NGC\,4593, the ground-based monitoring duration was longer than the duration of the \sw\, campaign. In these cases the portion of the ground-based light curves used for lag measurement was restricted to the range of dates spanned by the \sw\, UVW2 light curves (as listed in Table~\ref{tab:AGNproperties} and shown as dotted vertical lines in Figures \ref{fig:lcNGC4151} and \ref{fig:lcNGC4593}) to ensure consistency of lag measurements between the \sw\, and ground-based bands.  Tables~\ref{tab:lagMrk509}, \ref{tab:lag4151} and \ref{tab:lagNGC4593} show the resulting ICCF lags and uncertainties.
The CCFs and CCCDs are displayed in Figs.~\ref{fig:lag_Mrk509}, \ref{fig:lag_NGC4151}, and \ref{fig:lag_NGC4593}.

For NGC\,4151, the lower quality of the \emph{BVRI} light curves (relative to the \emph{ugriz} bands) led to poor cross-correlation results and unphysical negative lags for three of the bands, when using only the ground-based data from the same range of dates as the \sw\, campaign. We found that more reasonable (although still highly uncertain) cross-correlation results could be obtained by using the full-duration ground-based light curves, so for the \emph{BVRI} bands the ICCF results presented in Figure\,\ref{fig:lag_NGC4151} and Table\,\ref{tab:lag4151} are based on the full-duration ground-based light curves as shown in Figure\,\ref{fig:lcNGC4151}. (The anomalous lags measured from the restricted-duration light curves are displayed in Figure\,\ref{fig:comparison_JavPyCC} for comparison with the values measured from the full-duration data.) Overall, the low-quality \emph{BVRI} light curves for NGC\,4151 unfortunately add little value to our analysis and we caution against drawing any strong conclusions based on the \emph{BVRI} data, but we include them for completeness.

\begin{table}
    \centering
    \caption{Mrk~509 Reverberation Lags  (Observed Frame)}
    \label{tab:lagMrk509}
    \renewcommand{\arraystretch}{1.4}
    \begin{tabular}{lccccc}
        \hline
        Band & $\lambda_\mathrm{cen}$  & $r_\mathrm{max}$ & \tcent  & \tjav  & JAV$_{\rm width}$ \\
            &\AA & &days& days& days \\
        (1) & (2) & (3) & (4) & (5) & (6) \\
        \hline
        HX & 3.2 & 0.59 & $2.69^{+1.14}_{-1.65}$  & --- & --- \\
        SX & 18.0 & 0.74 & $1.36^{+0.82}_{-0.89}$ & --- & --- \\
        UVW2 & 1928 & 1.00 & $0.00^{+0.56}_{-0.57}$ & $0.00^{+0.02}_{-0.02}$ & $0.13^{+0.18}_{-0.08}$ \\
        UVM2 & 2246 & 0.99 & $-0.00^{+0.56}_{-0.58}$ & $-0.17^{+0.12}_{-0.13}$ & $2.68^{+1.10}_{-0.10}$ \\
        UVW1 & 2600 & 0.99 & $0.22^{+0.59}_{-0.61}$ & $0.11^{+0.15}_{-0.06}$ & $6.00^{+5.39}_{-1.15}$ \\
        \sw\, U & 3467 & 0.98 & $2.79^{+0.59}_{-0.60}$ & $2.03^{+0.22}_{-0.22}$ & $12.29^{+1.08}_{-1.19}$ \\
        $u$ & 3540 & 0.98 & $1.99^{+0.44}_{-0.46}$ & $3.41^{+0.21}_{-0.28}$ & $6.10^{+1.34}_{-0.11}$ \\
        \sw\, B & 4392 & 0.97 & $1.23^{+0.69}_{-0.62}$ & $1.48^{+0.25}_{-0.26}$ & $15.27^{+1.25}_{-1.39}$ \\
        $g$ & 4770 & 0.99 & $2.76^{+0.46}_{-0.36}$ & $2.66^{+0.05}_{-0.04}$ & $0.008^{+0.006}_{-0.010}$ \\
        $V$ & 5448 & 0.98 & $3.28^{+0.37}_{-0.36}$ & $3.83^{+0.02}_{-0.02}$ & $0.010^{+0.030}_{-0.007}$ \\
        \sw\, V & 5468 & 0.96 & $2.06^{+0.74}_{-0.77}$ & $2.65^{+0.44}_{-0.41}$ & $13.21^{+2.25}_{-2.84}$ \\
        $r$ & 6215 & 0.95 & $8.58^{+0.37}_{-0.44}$ & $7.59^{+0.28}_{-0.26}$ & $20.34^{+0.64}_{-0.68}$ \\
        $i$ & 7545 & 0.96 & $7.86^{+0.35}_{-0.36}$ & $7.86^{+0.25}_{-0.22}$ & $19.87^{+0.52}_{-0.57}$ \\
        $z$ & 8700 & 0.94 & $10.44^{+0.38}_{-0.38}$ & $9.36^{+0.07}_{-1.09}$ & $3.51^{+0.10}_{-0.10}$ \\
        $Z$ & 9200 & 0.91 & $4.64^{+3.73}_{-2.44}$ & $8.57^{+4.11}_{-1.88}$ & $20.07^{+65.54}_{-11.06}$ \\
        $Y$ & 10200 & 0.92 & $12.15^{+3.12}_{-2.72}$ & $12.56^{+3.18}_{-1.80}$ & $0.75^{+1.84}_{-0.58}$ \\
        $J$ & 12200 & 0.58 & $17.57^{+5.68}_{-4.35}$ & $15.18^{+4.31}_{-3.78}$ & $1.76^{+4.83}_{-1.31}$  \\
        \hline
    \end{tabular}
    \vspace{0.2cm}
    \begin{minipage}{\linewidth}
        \justifying
        {\small
        \textbf{Notes:} Lags are measured relative to the UVW2 reference band. Columns are (1) X-ray band or filter band; (2) central wavelength; (3) CCF peak height $r_\mathrm{max}$; (4) \pyccf\ lag \tcent; (5) \javelin lag \tjav; (6) \javelin width ($w$), representing the top-hat response function width.  Lags are listed in the observed frame.}
    \end{minipage}
\end{table}

\begin{table}
    \centering
    \caption{NGC\,4151 Reverberation Lags  (Observed Frame)}
    \label{tab:lag4151}
    \renewcommand{\arraystretch}{1.4}
    \begin{tabular}{lccccc}
        \hline
        Band & $\lambda_\mathrm{cen}$  & $r_\mathrm{max}$ & \tcent & \tjav  & JAV$_{\rm widths}$ \\
    &\AA & &days& days& days \\
        \hline
        
        HX & 3.2 & 0.66 & $-3.23^{+0.31}_{-0.27}$ & --- & --- \\
        SX & 18.0 & 0.34 & $-1.56^{+3.09}_{-1.46}$ & --- & --- \\
        UVW2 & 1928 & 1.00 & $0.00^{+0.28}_{-0.29}$ & $0.00^{+0.00}_{-0.01}$ & $0.03^{+0.05}_{-0.02}$ \\
        UVM2 & 2246 & 0.95 & $0.28^{+0.28}_{-0.28}$ & $0.08^{+0.09}_{-0.07}$ & $0.46^{+0.33}_{-0.33}$ \\
        UVW1 & 2600 & 0.95 & $0.23^{+0.28}_{-0.26}$ & $0.16^{+0.27}_{-0.14}$ & $0.90^{+0.88}_{-0.43}$ \\
        \sw\, U & 3467 & 0.94 & $0.59^{+0.30}_{-0.31}$ & $0.39^{+0.16}_{-0.16}$ & $2.14^{+0.81}_{-0.38}$ \\
        $u$ & 3540 & 0.94 & $0.94^{+0.23}_{-0.23}$ & $1.10^{+0.10}_{-0.10}$ & $0.07^{+0.13}_{-0.06}$ \\
        $B$ & 4361 & 0.89 & $1.35^{+0.56*}_{-0.38}$ & $1.22^{+0.11}_{-0.98}$ & $0.03^{+0.04}_{-0.02}$ \\
        \sw\, B & 4392 & 0.91 & $0.76^{+0.36}_{-0.39}$ & $0.47^{+0.20}_{-0.19}$ & $2.20^{+1.39}_{-1.02}$ \\
        $g$ & 4770 & 0.93 & $0.46^{+0.20}_{-0.22}$ & $0.63^{+0.07}_{-0.10}$ & $0.02^{+0.03}_{-0.01}$ \\
        $V$ & 5448 & 0.88 & $0.87^{+0.75*}_{-0.70}$ & $0.13^{+0.06}_{-0.04}$ & $0.01^{+0.02}_{-0.00}$ \\
        \sw\, V & 5468 & 0.81 & $0.82^{+0.58}_{-0.52}$ & $0.80^{+0.38}_{-0.36}$ & $4.77^{+2.07}_{-1.81}$ \\
        $r$ & 6215 & 0.92 & $1.18^{+0.20}_{-0.20}$ & $1.86^{+0.04}_{-0.06}$ & $0.02^{+0.05}_{-0.01}$ \\
        $R$ & 6407 & 0.86 & $0.70^{+0.65*}_{-0.53}$ & $1.11^{+0.07}_{-0.04}$ & $0.01^{+0.03}_{-0.007}$ \\
        $i$ & 7545 & 0.94 & $0.40^{+0.20}_{-0.22}$ & $0.48^{+0.09}_{-0.05}$ & $0.02^{+0.03}_{-0.01}$ \\
        $I$ & 7980 & 0.85 & $0.63^{+0.71*}_{-0.55}$ & $0.11^{+0.01}_{-0.01}$ & $0.007^{+0.004}_{-0.007}$ \\
        $z$ & 8700 & 0.93 & $0.15^{+0.28}_{-0.26}$ & $0.32^{+0.17}_{-0.26}$ & $0.02^{+0.05}_{-0.02}$ \\
        \hline
    \end{tabular}
    \vspace{0.2cm}
    \begin{minipage}{\linewidth}
        \justifying
        {\small
        \textbf{Notes:} Columns are as described for Table~\ref{tab:lagMrk509}. The  $^*$ symbol for the \emph{BVRI} bands denotes lags that were measured using the full duration of the ground-based light curves rather than the duration corresponding to the dates of the \sw\, campaign.
        }
    \end{minipage}
\end{table}

\begin{table}
    \centering
    \caption{NGC\,4593 Reverberation Lags  (Observed Frame)}
    \label{tab:lagNGC4593}
    \renewcommand{\arraystretch}{1.3}
    \begin{tabular}{lccccc}
        \hline
        Band & $\lambda_\mathrm{cen}$& $r_\mathrm{max}$ & \tcent & \tjav  & JAV$_{\rm widths}$ \\
            &\AA & &days& days& days \\
        \hline
        HX & 3.2 & 0.67 & $-0.49^{+0.12}_{-0.12}$ & --- & --- \\
        SX & 18.0 & 0.70 & $-0.49^{+0.12}_{-0.12}$ & --- & --- \\
        UVW2 & 1928 & 1.00 & $0.00^{+0.10}_{-0.12}$ & $0.00^{+0.00}_{-0.00}$ & $0.03^{+0.03}_{-0.01}$ \\
        UVM2 & 2246 & 0.96 & $0.11^{+0.10}_{-0.12}$ & $0.07^{+0.06}_{-0.10}$ & $0.44^{+0.27}_{-0.23}$ \\
        UVW1 & 2600 & 0.95 & $0.14^{+0.14}_{-0.14}$ & $-0.00^{+0.04}_{-0.04}$ & $0.45^{+0.33}_{-0.39}$ \\
        \sw\, U & 3467 & 0.93 & $0.29^{+0.14}_{-0.18}$  & $0.22^{+0.06}_{-0.06}$ & $1.31^{+0.36}_{-0.29}$ \\
        $u$ & 3540 & 0.88 & $0.72^{+0.55}_{-0.52}$ & $1.40^{+0.34}_{-0.38}$ & $4.72^{+0.79}_{-1.15}$ \\
        \sw\, B & 4392 & 0.82 & $0.14^{+0.19}_{-0.19}$ & $0.04^{+0.10}_{-0.07}$ & $0.54^{+0.78}_{-0.42}$ \\
        $g$ & 4770 & 0.95 & $0.41^{+0.10}_{-0.13}$ & $0.64^{+0.04}_{-0.04}$ & $0.12^{+0.29}_{-0.09}$ \\
        \sw\, V & 5468 & 0.67 & $0.21^{+0.25}_{-0.31}$ & $0.15^{+0.15}_{-0.13}$ & $0.49^{+0.89}_{-0.36}$ \\
        $r$ & 6215 & 0.88 & $1.27^{+0.34}_{-0.29}$ & $0.90^{+0.28}_{-0.14}$ & $0.34^{+0.85}_{-0.29}$ \\
        $i$ & 7545 & 0.86 & $1.41^{+0.35}_{-0.37}$ & $1.75^{+0.23}_{-0.22}$ & $3.27^{+0.57}_{-0.87}$ \\
        $z$ & 8700 & 0.87 & $0.90^{+0.39}_{-0.64}$ & $0.82^{+0.37}_{-0.20}$ & $1.69^{+1.88}_{-1.36}$ \\
        \hline
    \end{tabular}
    \vspace{0.2cm}
    \begin{minipage}{\linewidth}
        \justifying
        {\small
        \textbf{Notes:} Columns are as described for Table~\ref{tab:lagMrk509}.
        }
    \end{minipage}
\end{table}

\subsection{JAVELIN Analysis}\label{sec:javelin}

The \javelin method fits a model to the light curve data in the driving and responding bands to determine the time delay between them. The driving light curve  is modelled by a DRW process and the responding light curve, typically at a lower frequency, is related to it via convolution with a top-hat transfer function parameterized by its scale factor, width and central delay time.
The lag (\tjav) is the centroid of this transfer function.
\javelin implements a Bayesian framework using Markov Chain Monte Carlo (MCMC) sampling of the posterior parameter distributions.
We used 20\,000 MCMC burn-in and sampling iterations for Mrk~509 and NGC\,4151, and 10\,000 for NGC\,4593.  Figs.~\ref{fig:lag_Mrk509}, \ref{fig:lag_NGC4151}, and \ref{fig:lag_NGC4593} show the resulting \javelin fits to the lightcurve data and the posterior \tjav\ distributions for comparison with the \tcent\ distributions from \pyccf.
Our estimates of \tjav\ and corresponding uncertainties, derived as the median and inner 68\% confidence interval
based on the MCMC samples,
are listed in Tables~\ref{tab:lagMrk509}, \ref{tab:lag4151} and \ref{tab:lagNGC4593}.

The validity of \javelin's DRW assumption as an accurate model for AGN UV/optical variability has been a topic of considerable investigation \citep[e.g.,][]{Kasliwal15, Kozlowski:2016, Kozlowski:2017, Guo17, Yu2020:sim}. 
Kepler data indicate that AGN show a broad range of  power spectral density (PSD) slopes at high temporal frequencies, which are generally steeper than the $P\propto f^{-2}$ high-frequency behavior of the DRW model \citep{Mushotzky11,Edelson14,Aranzana18}. 
However, simulations by \citet{Yu2020:sim} carried out for the objects from \citet{Edelson19} conclude that the DRW assumption does not lead to biases in continuum lag measurements even if the underlying variability process departs from the DRW assumptions.

The derived \tjav\ error bars are often smaller and sometimes much smaller than those for \tcent\ obtained from \pyccf. Even for some of the poorest quality light curves, such as the \emph{BVRI} light curves of NGC\,4151, most of the \javelin\ lag uncertainties are below 0.1 d, while the PyCCF lag uncertainties for these bands are several tenths of a day, and the \javelin\ uncertainties appear unrealistically small given the data quality. Furthermore, some of the \javelin\ results exhibit other worrisome behavior.
For example, for NGC\,4151 the posterior distribution of $\tau_{\rm JAV}$ between the ground-based \emph{B} and \sw\, UVW2 bands (Fig.~\ref{fig:lag_NGC4151})
splits into two narrow and isolated peaks separated by $\sim1$~day
($\tjav\approx0.25\pm0.15$ or  $1.25\pm0.15$~d). 
In contrast, PyCCF yields a single-peaked CCCD for each of the optical bands.
This suggests that the \pyccf\ lag measurements may be more reliable than those we derived with \javelin.

We carried out trial \javelin\ measurements of the lags between the X-ray bands and UVW2 (using either UVW2 or the X-rays as the driving band), but obtained poor fits to the light curves for all three AGN. While there is clearly some degree of correlated structure between the X-ray and UV light curves, the correlation is relatively weak, as indicated by the low CCF \rmax\ values for the X-ray bands. The relationship between the X-ray and UVW2 light curves is evidently more complex than a linear transfer convolution, as described previously by \citet{Edelson19} for these sources. Consequently, we omit \javelin\ lag measurements for the X-ray bands in Tables \ref{tab:lagMrk509}-\ref{tab:lagNGC4593}. 
All lag measurements presented in these tables correspond to observed-frame values of both lag time and wavelengths.
For the case of the \emph{BVRI} light curves of NGC\,4151, as discussed above, we obtained very similar \javelin\ results when using either the full-duration \emph{BVRI} light curves or the light curves with duration matching that of the \sw\, campaign, and the \javelin\ results presented in Table \ref{tab:lag4151} are based on the shorter duration matching the dates of the \sw\, campaign.

\begin{figure*}
\begin{center}
\includegraphics[trim=4cm 0.0cm 2cm 1cm,clip,width=18cm]{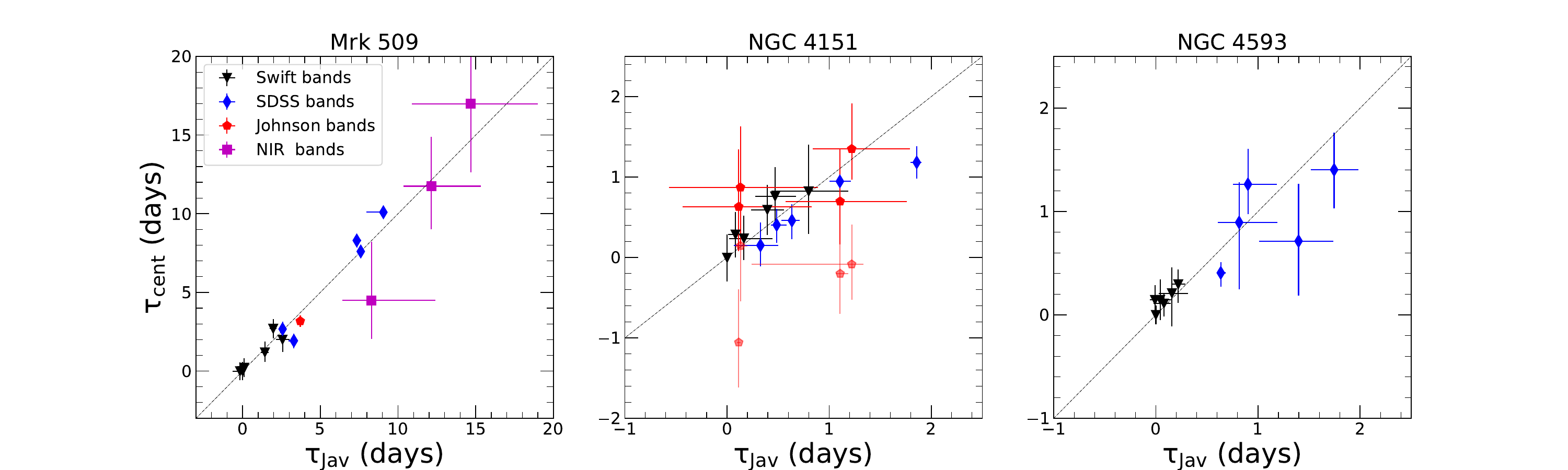}
\end{center}
\caption{Comparison between lags measured with \pyccf (\tcent) and JAVELIN (\tjav). \sw\, points are denoted in black, Johnson and Bessell filter bands in red, \emph{ugriz} bands in blue, and RATIR bands in magenta. For NGC\,4151, the light red points denote the PyCCF \emph{BVRI} lags measured using just the duration of the ground-based light curves corresponding to the dates of the \sw\, campaign, while the dark red points correspond to the \emph{BVRI} PyCCF lags measured using the full duration of the ground-based data.
}
\label{fig:comparison_JavPyCC}
\end{figure*}

Figure \ref{fig:comparison_JavPyCC} compares the \pyccf\ and \javelin\ lags for the three AGN. 
The lags obtained from the two methods are mostly similar, and indeed, a large majority ($\sim$ 85\%) agree within $1\sigma$, indicating that the reported uncertainties are likely somewhat overestimated..   \pyccf\ generally yields larger measurement uncertainties than \javelin, as expected \citep{Yu2020:sim}. For NGC\,4151, the anomalous \emph{BVRI} ICCF lags obtained using the restricted-duration light curves are shown as light red points, while the ICCF measurements obtained using the full-duration \emph{BVRI} light curves are shown in red; these are much more consistent with the \javelin\ lags although the uncertainties on both the ICCF and \javelin\ results are very large for these bands.

\section{Flux-Flux Analysis} 
\label{sec:fluxflux}

A further diagnostic of disk emission, in addition to the delay spectrum, is the shape of the spectral energy distribution (SED). The combination of AGN and host-galaxy light in the observed-frame direct flux SED complicates the isolation of the AGN component, especially when using ground-based data with large photometric apertures and limited spatial resolution. However, since the variability originates primarily in the AGN, the use of variable flux (flux difference between bright and faint states) enables a cleaner separation of the AGN spectrum from the contaminating host-galaxy light.
In this section, we describe a flux-flux analysis that follows the methodology used by \cite{Hernandez20} and \citet{Vincentelli:2021} for the analysis of multiband data from the \sw\, and ground-based campaigns.

\begin{figure*}
\centering
	\includegraphics[width=0.8\textwidth]{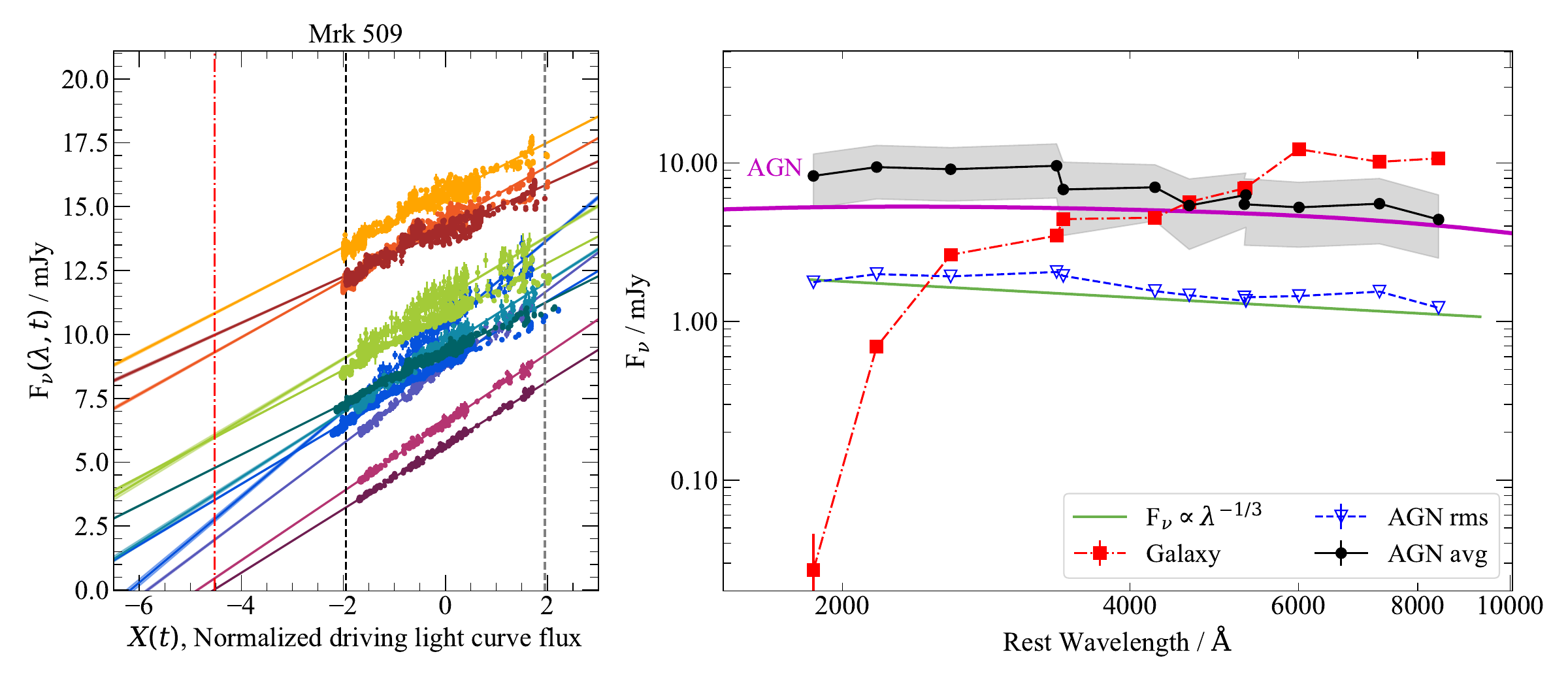}\\
	\includegraphics[width=0.8\textwidth]{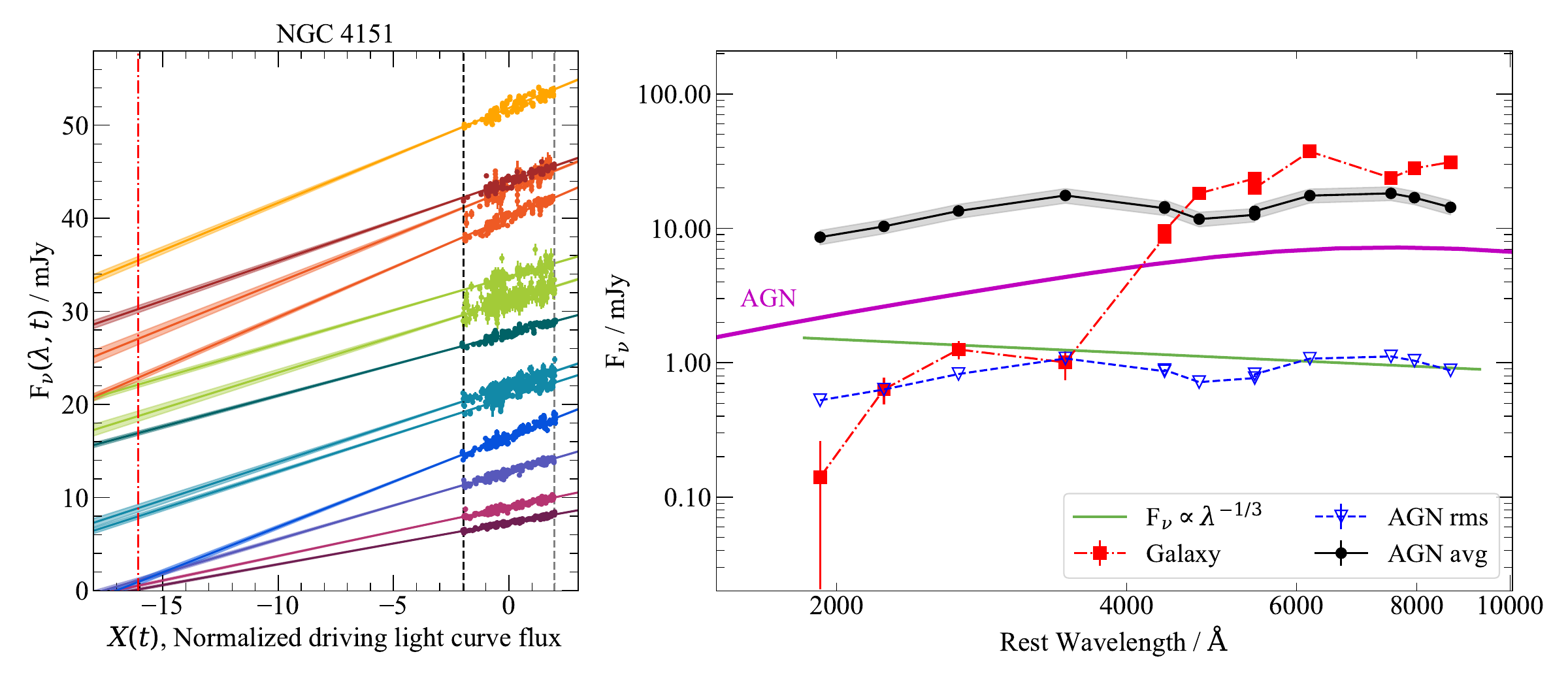}\\
 \includegraphics[width=0.8\textwidth]{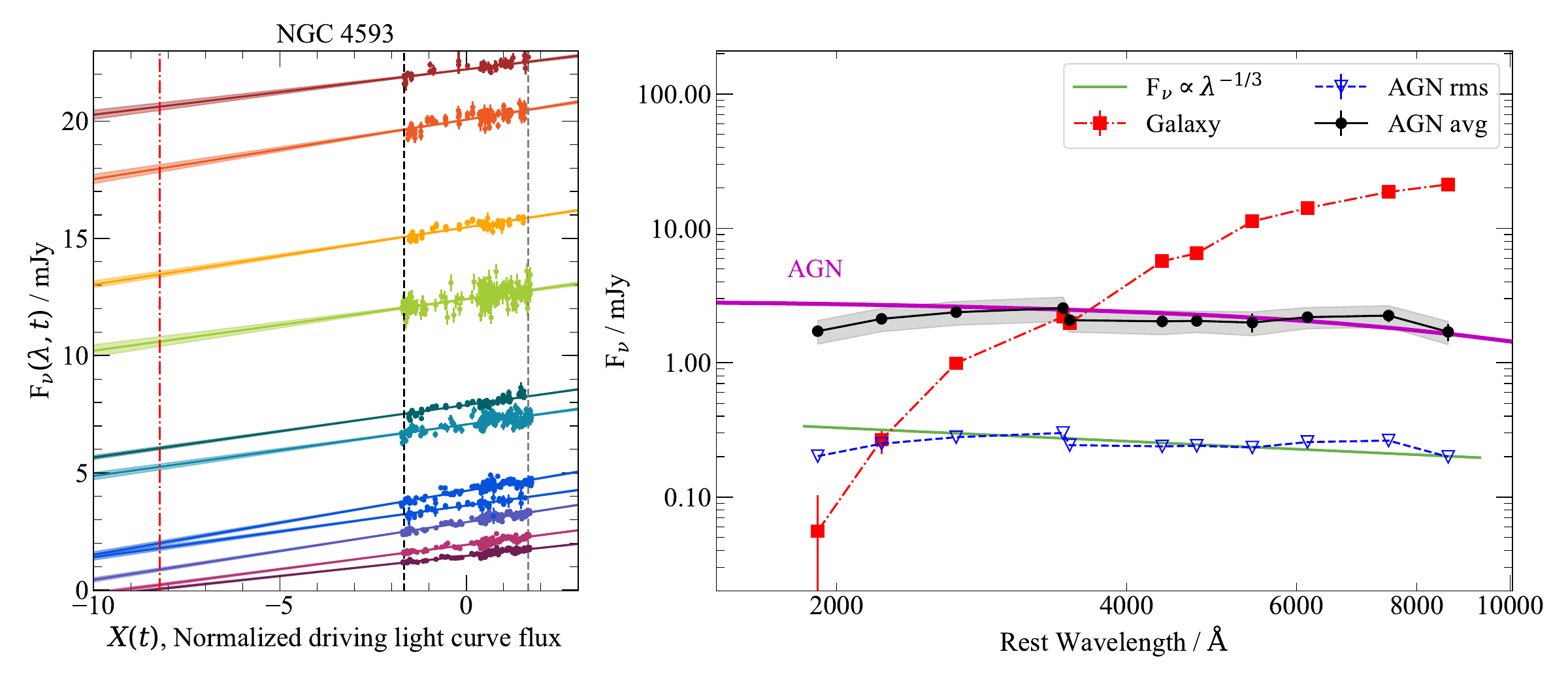}
    \caption{Flux-flux analysis for Mrk~509 (top panel), NGC\,4151 (middle panel) and NGC\,4593 (bottom panel).
    Left panels show for each AGN the observed fluxes fitted with the linear model $F(\lambda,t)=\bar{F}(\lambda)+ \Delta F(\lambda)\,X(t)$, where the dimensionless lightcurve shape $X(t)$ is normalized to $\left<X\right>=0$ and $\left<X^2\right>=1$ so that the intercept $\bar{F}(\lambda)$ at $X=0$ is the mean (galaxy+AGN) spectrum and the slope $\Delta F(\lambda)$ is the rms spectrum of the AGN variations. Vertical dotted lines delimit the maximum ($X_{\rm max}$, black) and minimum ($X_{\rm min}$, grey) of the AGN variations. A vertical red line marks $X_0$, where the UVW2 model extrapolates to $1\,\sigma$ above zero flux, defining the adopted galaxy spectrum. The colors of the lightcurves represent each filter as used in Fig.~\ref{fig:lag_Mrk509}--\ref{fig:lag_NGC4593}.
    Right panels show the resulting spectral decomposition for each object. The black line denotes the average SED of the AGN component, the grey band surrounding the black line shows the variability range (from minimum to maximum flux), and the blue dashed line shows the rms spectrum.
    The constant host-galaxy component is shown in red. An AGN accretion disk model \citep{KubotaDone18} for the parameters of each AGN is shown as the purple line for comparison with the AGN spectrum inferred from the flux-flux analysis. An arbitrarily scaled power law, $F_{\nu} \propto\lambda^{-1/3}$, is also shown as a reference in green. The SEDs have been corrected for Milky Way dust extinction using the values in Table~\ref{tab:AGNproperties}.
    }
    \label{fig:fluxflux}
\end{figure*}

\begin{table*}
    \centering
    \caption{Flux-Flux decomposition results.}
    \label{tab:fluxflux_results}
    \begin{tabular}{lc|ccc|ccc|ccc}
        \hline
        Filter & $\lambda$ (\AA) & \multicolumn{3}{c|}{Mrk~509} & \multicolumn{3}{c|}{NGC\,4151} & \multicolumn{3}{c}{NGC\,4593} \\
        \hline
        &  & $\bar{F}_{\nu}$(AGN) & $\Delta{F}_{\nu}$(AGN) & $F_{\nu}$(Host) 
           & $\bar{F}_{\nu}$(AGN) & $\Delta{F}_{\nu}$(AGN) & $F_{\nu}$(Host) 
           & $\bar{F}_{\nu}$(AGN) & $\Delta{F}_{\nu}$(AGN) & $F_{\nu}$(Host) \\
           &&mJy&mJy&mJy&mJy&mJy&mJy&mJy&mJy&mJy\\
        \hline
UVW2    & 1928 & $  8.28 \pm  0.01$ & $  3.10 \pm  0.01$ &  $  0.03 \pm  0.03$ & $  8.60 \pm  0.25$ & $  1.05 \pm  0.19$ &  $  0.14 \pm  0.14$ & $  1.72 \pm  0.02$ & $  0.34 \pm  0.01$ &  $  0.05 \pm  0.05$  \\ 
UVM2    & 2246 & $  9.42 \pm  0.05$ & $  3.49 \pm  0.04$ &  $  0.69 \pm  0.02$ & $ 10.35 \pm  0.27$ & $  1.23 \pm  0.20$ &  $  0.63 \pm  0.14$ & $  2.13 \pm  0.12$ & $  0.42 \pm  0.09$ &  $  0.26 \pm  0.06$  \\
UVW1    & 2600 & $  9.13 \pm  0.07$ & $  3.37 \pm  0.06$ &  $  2.64 \pm  0.02$ & $ 13.49 \pm  0.42$ & $  1.61 \pm  0.32$ &  $  1.25 \pm  0.20$ & $  2.39 \pm  0.14$ & $  0.48 \pm  0.11$ &  $  0.99 \pm  0.07$  \\
\sw\, U & 3467 & $  9.60 \pm  0.15$ & $  3.59 \pm  0.14$ &  $  3.48 \pm  0.02$ & $ 17.61 \pm  0.61$ & $  2.18 \pm  0.46$ &  $  1.00 \pm  0.26$ & $  2.56 \pm  0.17$ & $  0.53 \pm  0.13$ &  $  2.20 \pm  0.08$  \\
$u$     & 3540 & $  6.80 \pm  0.05$ & $  3.32 \pm  0.04$ &  $  4.43 \pm  0.02$ &      								-- &        -- &        -- & $  2.08 \pm  0.16$ & $  0.38 \pm  0.13$ &  $  1.97 \pm  0.06$  \\
$B$     & 4361 &      								-- &        -- &        -- & $ 14.14 \pm  0.48$ & $  1.65 \pm  0.36$ &  $  8.64 \pm  0.21$ &      								-- &        -- &        --  \\
\sw\, B & 4392 & $  7.02 \pm  0.12$ & $  2.72 \pm  0.11$ &  $  4.52 \pm  0.02$ & $ 14.49 \pm  0.67$ & $  1.76 \pm  0.52$ &  $  9.60 \pm  0.22$ & $  2.03 \pm  0.20$ & $  0.41 \pm  0.16$ &  $  5.70 \pm  0.06$  \\
$g$     & 4770 & $  5.39 \pm  0.03$ & $  2.54 \pm  0.03$ &  $  5.68 \pm  0.01$ & $ 11.75 \pm  0.40$ & $  1.45 \pm  0.30$ &  $ 18.22 \pm  0.18$ & $  2.06 \pm  0.14$ & $  0.37 \pm  0.11$ &  $  6.52 \pm  0.06$  \\
$V$     & 5448 & $  6.28 \pm  0.14$ & $  2.36 \pm  0.13$ &  $  6.95 \pm  0.02$ & $ 12.57 \pm  0.54$ & $  1.42 \pm  0.41$ &  $ 23.53 \pm  0.19$ & $  2.00 \pm  0.32$ & $  0.40 \pm  0.26$ &  $ 11.27 \pm  0.06$  \\
\sw\, V & 5468 & $  5.49 \pm  0.04$ & $  2.47 \pm  0.04$ &  $  6.89 \pm  0.02$ & $ 13.41 \pm  0.84$ & $  1.62 \pm  0.66$ &  $ 19.93 \pm  0.21$ &      								-- &        -- &        --  \\ 
$r$     & 6215 & $  5.24 \pm  0.07$ & $  2.30 \pm  0.06$ &  $ 12.20 \pm  0.02$ & $ 17.57 \pm  0.76$ & $  2.08 \pm  0.58$ &  $ 37.33 \pm  0.27$ & $  2.19 \pm  0.21$ & $  0.39 \pm  0.17$ &  $ 14.17 \pm  0.07$  \\ 
$i$     & 7545 & $  5.53 \pm  0.07$ & $  2.44 \pm  0.06$ &  $ 10.16 \pm  0.02$ & $ 18.22 \pm  0.64$ & $  2.12 \pm  0.48$ &  $ 23.74 \pm  0.29$ & $  2.25 \pm  0.24$ & $  0.41 \pm  0.20$ &  $ 18.67 \pm  0.07$  \\
$I$     & 7980 &      								-- &        -- &        -- & $ 16.93 \pm  1.01$ & $  1.90 \pm  0.78$ &  $ 27.99 \pm  0.27$ &      								-- &        -- &        --  \\
$z$     & 8700 & $  4.40 \pm  0.07$ & $  1.89 \pm  0.06$ &  $ 10.70 \pm  0.01$ & $ 14.35 \pm  0.64$ & $  1.72 \pm  0.49$ &  $ 31.14 \pm  0.23$ & $  1.71 \pm  0.25$ & $  0.33 \pm  0.20$ &  $ 21.24 \pm  0.06$  \\
        \hline
    \end{tabular}
    \vspace{0.2cm}
    \begin{minipage}{\textwidth}
        {\small
        \textbf{Notes:} For each AGN, the three columns list the mean AGN flux ($\bar{F}_{\nu}$), the range of variability ($\Delta{F}_{\nu}$) and host-galaxy flux density, as inferred from the flux-flux analysis. All fluxes have been corrected for Milky Way dust extinction stated in Table~\ref{tab:AGNproperties}.
        }
    \end{minipage}
\end{table*}

This method decomposes the spectral variations $F(\lambda,t)$ into a constant spectrum $\bar{F}(\lambda)$ plus a variable spectrum $\Delta F(\lambda)$ modulated in time by a dimensionless lightcurve $X(t)$.
Allowing also for wavelength-dependent time delays $\tau(\lambda)$, the resulting model of the spectral variations is
\begin{equation}\label{eq:fluxflux}
    F(\lambda,t) 
    = \bar{F}( \lambda )
    + \Delta F(\lambda)
    \, X\left( t - \tau\left(\lambda\right)\right) \ .
\end{equation}
To construct $X(t)$, we measure a running optimal average\footnote{\url{https://github.com/FergusDonnan/Running-Optimal-Average}} (using a Gaussian window function with a width of 1.5 days) of the UVW2 light curve, which is then normalized such that $\left<X\right>_t=0$
and $\left<X^2\right>_t=1$. 
We adopt the median interband lags (as measured from the ICCF method as described above) to shift the driving light curve by $\tau(\lambda)$, final fit of the light curve in each filter. This time shift is needed in particular for the larger delays in Mrk~509. The best fit for the linear relationship at each wavelength, shown in the left panels of Fig.~\ref{fig:fluxflux}, determines the mean spectrum $\bar{F}_\nu(\lambda)$,
and the rms spectrum $\Delta F_\nu(\lambda)$
in Eqn.~(\ref{eq:fluxflux}).
Note that the linear model of Eqn.~(\ref{eq:fluxflux}) is a good fit in all cases, with no strong evidence for curvature that would imply a change in the SED of the disk component as it brightens and fades. There is additional scatter in the linear fits towards longer wavelengths due to the differences in blurring from the UVW2-based driving light curve to the rest of the light curves. This additional blurring arises from the finite width of the asymmetric transfer function expected from an accretion disk \citep[e.g.,][]{Cackett07,Starkey16}.

To find the constant SED of the host galaxy, we extrapolate the best fit linear model to lower fluxes,
equivalent to turning off the AGN, until at $X_0$ it reaches zero flux within its 1$\sigma$ uncertainty. 
This provides a lower limit to the contribution of the galaxy at UVW2, since any additional non-variable emission (e.g., from the narrow-line region or extended stellar light) would be included in this estimate. 
This provides a lower limit to the contribution of the galaxy at UVW2. Thus,
evaluating Eqn.~(\ref{eq:fluxflux}) at this crossing,
as indicated by the red vertical dot-dashed on the plot~\ref{fig:fluxflux},
provides the contribution of the host galaxy to the overall light:

\begin{equation}
    F_\nu^{\rm gal}(\lambda)= \bar{F}_\nu(\lambda) + \Delta F_\nu(\lambda)  X_0 .
\end{equation}
We adopt this as the SED of the host galaxy\footnote{Strictly speaking this is a lower limit to the host galaxy SED.}.
The slope $\Delta F(\lambda)$ then gives the SED of the AGN component:
\begin{equation}
    F_\nu^{\rm AGN}(\lambda,t) = \Delta F_\nu( \lambda ) \, \left[ X(t-\tau(\lambda)) - X_0 \right]
\end{equation}
This method assumes that the shape of the AGN's SED remains constant as it becomes dimmer than the host galaxy.
For very large extrapolations (e.g., NGC~4151 due to the small variability amplitude observed during the campaign) it is possible that the AGN slope changes  and curvature appears in this analysis, invalidating the linear model. However, this linear amplitude seems to hold for AGN with large fractional amplitude such as Mrk\,509 or Fairall 9  \citep{Hernandez20}.

The right panel of Fig.~\ref{fig:fluxflux} 
shows the resulting SEDs for the AGN and host-galaxy components (in the galaxy's rest frame), where the fluxes have been dereddened by their foreground Galactic extinction as listed in Table~\ref{tab:AGNproperties}. For the AGN component, we plot the average spectrum (black line) and the range of variability around the mean (grey bands), which can be directly compared to accretion disk models. In blue, we show the rms spectrum.

The three host-galaxy SEDs, shown by the red curves in Fig.\ref{fig:fluxflux} and Table\ref{tab:fluxflux_results}, resemble the SEDs of early-type galaxies dominated by old red stars, but with additional UV flux from a younger stellar population. These host-galaxy SEDs correspond to light originating from the region close to the nucleus, within the radii of the photometric apertures. In comparison with these red host-galaxy SEDs, the AGN have relatively blue SEDs. The linearity of the flux–flux relations in Fig.\ref{fig:fluxflux} confirms that the observed color trends arise from the varying contribution of the constant red host-galaxy component, rather than from intrinsic changes in the AGN SED as it brightens and fades\citep{Sakata2010}.

The three AGN SEDs exhibit different shapes, as expected due to differences in their black-hole masses and Eddington ratios, leading to different maximum temperatures in the inner accretion disk. In Fig.\ref{fig:fluxflux} (purple line) we overlay for comparison the accretion disk model SEDs from \citet{KubotaDone18}, using values listed in Table\ref{tab:AGNproperties} for each object, assuming a black-hole spin of $a=0.9$ and an inclination angle of $i=30^{\circ}$. In all three cases, the expected SED describes the shape of the mean AGN spectrum reasonably well, albeit with some normalization offset. The AGN UV/optical SEDs may alternatively be interpreted as a power law modified by extinction and reddening by SMC-like dust in the host galaxy along the line of sight to the AGN, as illustrated by the green lines in Fig.~\ref{fig:fluxflux} \citep{Weaver22}.

It is notable that for NGC\,4151 the SED of the variable AGN component  is effectively described by an accretion disk at its low accretion rate of $<2$\% of $L_{\rm Edd}$ (albeit with a normalization discrepancy given the uncertainties on the true mass transfer rate of the disk and the inclination angle),
while its delay spectrum deviates from the expected relationship at longer wavelengths, as discussed further in Section~\ref{sec:lags}. 

\section{Lag vs. wavelength relationships}
\label{sec:lagfits}

\subsection{Power-Law fits to the lag spectra} 

The standard thin accretion-disk model \citep{ShakuraSunyaev73} predicts a relationship between disk radius and temperature of the form $T(r) \propto r^{-3/4}$. 
If the UV and optical variations are powered by disk reprocessing of emission originating from near the disk center, then this model predicts that the time delays of the reprocessed emission should follow a relationship of the form $\tau(\lambda) \propto \lambda^{4/3}$ as a result of the light-travel time between the X-ray emitting corona and the UV/optical emitting regions of the disk surface. 
Using the measured reverberation lags, we can test if the data follow this simple model prediction as well as search for deviations from a monotonic lag-wavelength trend that might arise from other line or emission components originating from the BLR. Previous fits to the lag-wavelength relations for these three AGN spanning the wavelength range of the \sw\, UV and optical bands (UVW2 through V) showed that the data were generally consistent with a $\tau \propto \lambda^{4/3}$ model, but with an excess lag in the U band and with a normalization larger than predicted by the simple disk reprocessing model given the black-hole mass and Eddington ratio of each object \citep{Edelson17, McHardy18, Edelson19}. 

\begin{figure*}[]
\begin{center}
\includegraphics[trim=4cm 0.1cm 4cm 1cm,clip,width=18cm]{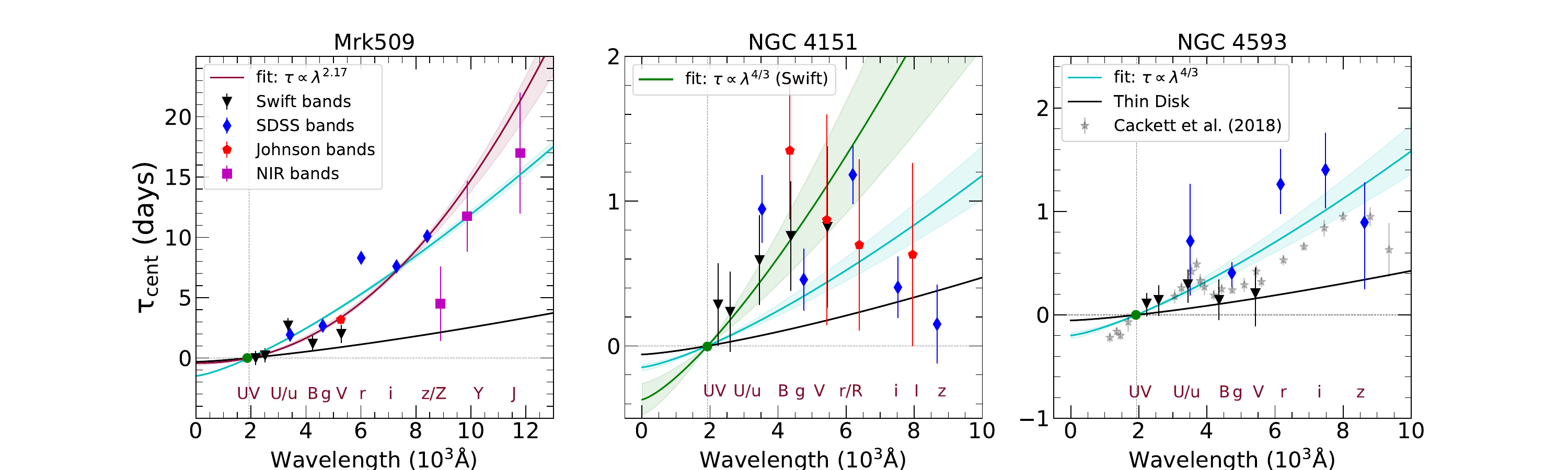}
\end{center}
\caption{Rest-frame lag-wavelength relationships for each AGN, measured relative to the UVW2 band ($\lambda_0=1928$~\AA). The three panels show lag data (points with error bars) measured with \pyccf\ (\tcent) for \sw\, UVOT bands (black) and ground-based \emph{ugriz} (blue), \emph{BVRI} (red), and the RATIR \emph{ZYJ} bands (magenta).
Curves with $\pm$$1\sigma$ uncertainty envelopes are fits of a power-law model, $\tau =\tau_0\,[\left(\lambda/\lambda_0\right)^\beta-1]$, as in  Eqn.~(\ref{eqn:cont_lag}). The constraints on $\tau_0$ and $\beta$ are given in Table~\ref{tab:infolag}. Cyan curves are fits to all data with the power-law slope $\beta$ fixed at $4/3$
(Fits~\#1, 4 and 6). For Mrk\,509, the dark red curve (Fit\,3) shows the fit omitting the lags of the U/\emph{u/r} bands and leaving $\beta$ free. For NGC\,4151, the green curve (Fit\,5) shows the fit to only the \sw\, lags, with $\beta$ fixed at 4/3. In the right panel, the light gray stars in NGC\,4593 correspond to the lag-time measurements from \citet{Cackett18} based on HST STIS data. In each panel, the black curves give predictions for a thin-disk model with the $M_\mathrm{BH}$ and $\dot{m}_\mathrm{Edd}\equiv L/L_\mathrm{Edd}$ values given in Table~\ref{tab:AGNproperties}.  
}

\label{fig:lagfit}
\end{figure*}

Here, we extend these results by combining the \sw\, and ground-based data to examine the lag-wavelength relation over a broader wavelength range, following methods applied in past work \citep[e.g.,][]{Edelson19}. We divide the observed lags and wavelengths by $(1+z)$ to correct for time dilation and to obtain the rest-frame wavelengths observed in each filter. Fig.~\ref{fig:lagfit} illustrates the resulting lags for the three AGN as a function of rest-frame wavelength.  Since \javelin\ yielded anomalous results for some filter bands, such as the double-peaked distribution for the \emph{B}-band lag of NGC\,4151, and the \javelin measurement uncertainties appear to be unrealistically small in some cases, we use the \pyccf lags to examine the lag-wavelength behavior. X-ray measurements were excluded from the fits, as they typically deviate from the trends seen in UV and optical bands \citep[e.g.,][]{Edelson19}.

\begin{table*}
    \centering
    \caption{Lag-Wavelength Fitting Results}
    \label{tab:infolag}
    \begin{tabular}{ccccccccc}
        \hline
        (1)  & (2)  & (3)  & (4)  & (5)  & (6) & (7)  & (8) \\
         &  &   &  $\tau_{0}$  &  &  &  &  \\
        Fit (Number) & AGN & (3) Lag data & (days) & $\beta$ &  dof & $\chi^2$ & $\chi^2_\nu$ \\
        \hline
        1 & Mrk~509 & all data  & $1.59\pm0.10$ & 4/3  & 13 & 85.02 & 6.54 \\
        2 & & no U/\emph{u}/\emph{r}  & $1.49\pm0.07$ & 4/3  & 10 & 34.98 & 3.49 \\
        3 & & no U/\emph{u}/\emph{r} & $0.42\pm0.13$ & $2.17\pm0.21$ & 9 & 12.84 & 1.42 \\
        \hline
        4 & NGC\,4151 & all data        & $0.14\pm0.04$ & 4/3 & 13 & 45.73 & 3.51 \\
        5 & & only \sw\, & $0.37\pm0.06$ & 4/3 & 4 & 1.10 & 0.27 \\
        \hline
        6 & NGC\,4593 & all data & $0.19\pm0.03$ & 4/3  & 9 & 9.50 & 1.05 \\
        \hline
    \end{tabular}
    \vspace{0.2cm}
    \begin{minipage}{\textwidth}
        {\small
        \textbf{Notes:} Results of fitting the power-law lag spectrum model, Eqn.~(\ref{eqn:cont_lag}), to the \pyccf\ lags, as shown in Fig.~\ref{fig:lagfit}.
        Column~(1) designates each fit with a number. Columns~(2) and (3) detail the AGN and lag data included or excluded in each fit.Column~(4) reports the lag normalization parameter $\tau_0$ and its uncertainty.Column~(5) is the power-law index $\beta$, usually fixed at 4/3 but free for Fit\,3.
        In counting the residual degrees of freedom, Column~(6), we omit the UVW2 data point since the model fit was constrained to pass through $\tau=0$ at the wavelength of UVW2. Columns~(7) and (8) give $\chi^2$ and reduced $\chi^2_\nu=\chi^2/{\rm dof}$ for each model fit. 
        }
    \end{minipage}
\end{table*}

We fit the lag spectra by minizing $\chi^2$ for a model of the form

\begin{equation}
\label{eqn:cont_lag}
\tau(\lambda) = \tau_{0} \, \left[ \left(\frac{\lambda}{\lambda_{0}}\right)^\beta - 1
\right].
\end{equation}
For the reference wavelength we use $\lambda_ {0}=1928/(1+z)$~\AA, 
corresponding to the UVW2 band used as the driving band for the CCF lag measurements.
In all but one of the fits described below, $\beta$ is fixed at $4/3$ and the single fitting parameter is $\tau_{0}$, which sets the normalization of the lag spectrum. Given the higher data quality for Mrk\,509, we also carried out a fit in which both $\tau_0$ and $\beta$ were allowed to vary. 
Figure~\ref{fig:lagfit} shows the best fits found for each AGN. 
These fitting results are summarized in Table~\ref{tab:infolag}, and the details and results of the fitting process are described for each object below.

\emph{Mrk~509} (Fig.~\ref{fig:lagfit} left panel): 
The longer lags and higher quality of the lag measurements for Mrk~509, plus the extended wavelength coverage provided by the RATIR data, provide a better test of the model and stronger constraints on the lag-wavelength fit parameters. 
The initial fit for this target (Fit\,1 as listed in Table \ref{tab:infolag}) with $\beta$ fixed to $4/3$ gives $\tau_0=1.59\pm0.10$ days, but with $\chi^2_\nu=6.5$ ($p\text{-value} = 0.92$) indicating an unacceptable fit. The large $\chi^2$ value is primarily due to a small number of strong outlier points, most notably the \emph{r}-band lag which lies well above the best-fit relation. 
There is good evidence in several other AGN that emission from the BLR can contribute significantly to lags in the \emph{u}-band region (from Balmer continuum emission) and in the \emph{r} band from the H$\alpha$ line and Paschen continuum  \citep[e.g.,][]{Korista01, Cackett18, Korista19, Edelson19}, and Fig.~\ref{fig:stis_with_filters} 
illustrates the contribution of the Balmer continuum and H$\alpha$ to the broad-band filters used in our photometry (for further details, see Section~\ref{sec:Mrk509specfit}). 
We therefore performed a second fit (Fit\,2) omitting the lag data for the  U, \emph{u}, and \emph{r} bands, similar to the approach taken by \citet{Fausnaugh16} for NGC 5548.
This fit yielded $\tau_0=1.49\pm0.07$ days with $\chi^2_\nu=3.49$ ($p\text{-value} = 0.96$), a substantial improvement over Fit\,1 but still a formally poor fit.
Finally, we performed a third fit (Fit\,3 - dark red line in Fig.~\ref{fig:lagfit}) in which $\beta$ was allowed to vary, again excluding the U, \emph{u}, and \emph{r} bands that may be substantially contaminated by BLR emission.
That fit yields $\tau_0=0.42\pm0.13$ days and $\beta=2.17\pm0.21$, achieving
a more acceptable $\chi^2_\nu=1.4$ ($p\text{-value} = 0.99$) for 9 degrees of freedom.
From this series of fits, we conclude that for Mrk\,509 the lag spectrum is not adequately fitted by the $\tau\propto\lambda^{4/3}$ prediction of the thin disk model, and a somewhat steeper power-law slope ($\beta \approx 2.17\pm 0.21$) is preferred, even after omitting the \emph{u}, \emph{U} and \emph{r} band lags thought to be most contaminated by BLR emission. This best-fit slope deviates significantly from the thin-disk expectation of $\beta = 4/3$, with a difference of $\sim$4$\sigma$ relative to the formal uncertainty, indicating a statistically significant departure from the standard model.

A lag-wavelength slope consistent with $\beta=2$ was recently found in an optical monitoring campaign for the AGN PG 2130+099 by \cite{Fian2022}, who noted that this slope is consistent with expectations for the inner regions of a slim (rather than thin) accretion disk. However, for Mrk\,509, given the scatter of the measured lags about the best-fit curves for either the $\beta=4/3$ or the $\beta=2.17$ models, our data do not provide a strong test to distinguish between these cases despite the formally better fit of the $\beta=2.17$ (free-slope) model.

For comparison, \citet{PozoNunez2019} presented results from a 2-year ground-based photometric monitoring program on Mrk\,509 spanning 2016-2017 (overlapping with our 2017 campaign), using four narrow-band filters spanning 4300-7000 \AA. They found lags of $\sim2$ days for the 6200 and 7000 \AA\ filter bands measured relative to the 4300 \AA\ band, and a lag-wavelength relation consistent with a $\beta=4/3$ slope over this limited wavelength range.

\emph{NGC\,4151} (Fig.\ \ref{fig:lagfit} middle panel): From the \sw\, data, \citet{Edelson17} and \citet{Edelson19} found a trend of increasing lag with wavelength across the \sw\, optical bands, up to a V-band lag of $\sim1$ day. In the ground-based data, however, we find surprising behavior at the longest wavelengths. The \emph{u} and \emph{g}-band lags are consistent (within the large measurement uncertainties) with the trend seen in the \sw\, data and the \emph{r}-band lag of $1.18\pm0.20$ days is slightly larger than the \sw\, V-band lag, continuing the expected trend. However, the lags for the \emph{i} and \emph{z} bands are substantially shorter than the \emph{r}-band lag. Such a downward trend in the long-wavelength lags is unexpected. While the ground-based \emph{BVRI} lags are of much poorer quality, as described previously, they also exhibit a decreasing lag-wavelength trend. The  unusual behavior seen in the \emph{i} and \emph{z} bands appears to be a genuine feature of the data over the time span of the \sw\, and ground-based campaign. The cross-correlation peaks are primarily determined by the brief, steep increase in flux around MJD 57460 (see Figure \ref{fig:lag_NGC4151}). Inspection of the light curves shows that this feature in the \emph{i} and \emph{z} bands occurs nearly simultaneously with the corresponding increase in the \sw\, UVW2 band. The decreasing trend in lags across the \emph{riz} bands is seen in the \javelin\ measurements as well as in the ICCF results.

Since the lag spectrum in NGC\,4151 evidently deviates from the normal behavior seen in other objects, the usual power-law model of Eqn.\,\ref{eqn:cont_lag} cannot adequately describe the data. For completeness, we carried out an initial fit of the $\beta=4/3$ model (Fit\,4) spanning the full wavelength range of the data, but this model fit is formally unacceptable with $\chi^2_\nu=3.51$. To describe just the increasing lag spectrum at the shorter wavelengths, we then performed another fit (Fit\,5 - green line Fig.~\ref{fig:lagfit}) to only the \sw\, data, excluding the ground-based measurements, and obtained an acceptable fit with $\chi^2_\nu=1.1$ for 4 degrees of freedom. This fit (to just the \sw\, UVOT data) is essentially the same as that reported in \cite{Edelson17, Edelson19}.

\emph{NGC\,4593} (Fig.~\ref{fig:lagfit} right panel): 
The short ($\lesssim1.5$~day) lags and relatively large uncertainties for most of the filter bands limit the utility of these data for constraining the lag-wavelength relationship beyond the results previously obtained from the \sw\, data and from the HST STIS campaign presented by \citet{Cackett18}. The ground-based \emph{ugriz} lags are all greater than the lags of the \sw\, optical bands, but in the range of wavelength overlap between the two datasets the \emph{u} and \emph{g}-band lags are consistent (within their uncertainties) with the \sw\, lags in adjacent bands.
Our $\beta=4/3$ model (Fit\,6) gives $\tau_0=0.19\pm0.03$ day with $\chi^2_\nu = 1.05$ for 9 degrees of freedom. Since this outcome is consistent with an acceptable fit and the data quality does not warrant more detailed model fitting, we did not carry out any other variants of the model fit.

\subsection{Derived disk sizes} \label{sec:sizes}

In previous IBRM studies of AGNs \citep{Fausnaugh16,Edelson15,Edelson17,Edelson19}, larger accretion disk sizes have been inferred from the measured lags than are predicted by the standard thin-disk model \citep{ShakuraSunyaev73}, if the continuum reverberation lags are interpreted as indicators of light-travel time across the disk. This discrepancy is still under debate, mainly because detailed studies have been carried out on relatively few AGN, limited in their range of black-hole mass and accretion rate, and in many cases the observational monitoring has been performed only in the UV/optical bands, with just a few cases extending
to near-infrared (NIR) wavelengths \citep{Fausnaugh17,Hernandez20}. Recent results from reverberation surveys of more luminous quasars show a broader range of disk sizes, with some consistent with or smaller than thin-disk expectations \citep[e.g.,][]{Mudd18,Homayouni19}, suggesting that the discrepancy may not be universal. We can use the results obtained from the lag-wavelength fits described in the previous section to compare the model parameter $\tau_0$ to the value predicted for lamp-post reprocessing by a thin disk to test whether these objects exhibit a ``disk size discrepancy'' similar to those found in other AGN.

The theoretical relation for $\tau(\lambda)$ is obtained
as follows \citep[see, e.g.,][]{Cackett07,Fausnaugh16,Edelson17,Edelson19}:
For the standard  model of a geometrically-thin accretion disk \citep{ShakuraSunyaev73}, the effective temperature profile at large radius $R$ approaches the power-law form
\begin{equation}
\label{eqn:tprofile}
 T_\mathrm{eff} = \left( \frac{ \left( 3 + \kappa \right)  G  M  \dot{M} }
    { 8 \pi  \sigma R^3 } \right)^{1/4},
\end{equation}
where $\kappa$ is the ratio of external irradiation to local viscous heating.
The light-travel time from the disk center out to radius $R$,
where $T_\mathrm{eff}=(hc/k\lambda X)$, is
\begin{equation}
\frac{R}{c} = 
\left( \frac{ k  \lambda  X } { h  c } \right)^{4/3}
\left( \frac{ ( 3 + \kappa )  G   M  \dot{M} }
{ 8  \pi  \sigma  c^3 } \right)^{1/3} .
\end{equation}
Here, $X$ is a dimensionless weighting factor that accounts for the range of radii in the disk that emit at wavelength $\lambda$. 
Evaluating this for
a black hole mass $M_{\mathrm{BH}} = M_8 \times 10^8~M_{\odot}$,
and accretion luminosity  $L_{\mathrm{bol}} = \eta \dot{M} c^2$,
with radiative efficiency $\eta$,
the predicted time delay spectrum is

\begin{equation}
\label{eqn:rc}
\frac{R/c }{ \rm day }= 0.0898 \, \left( X \frac{ \lambda }{ \lambda_0 }
    \right)^{4/3}
M_{8}^{2/3}\left(\frac{\dot{m}_\mathrm{Edd}}{\eta}\right)^{1/3}  .
\end{equation}

We scale wavelengths to $\lambda_0=1928/(1+z)$ \AA, corresponding to the UVW2 band, so that $R/c$ is directly comparable to the $\tau_0$ parameter determined from the power-law model fits to the measured lags. 
For the theoretical curves plotted in Fig.~\ref{fig:lagfit} as a black  line, we adopt $X=2.49$, $\eta=0.1$, and
the black hole masses and Eddington ratios 
$\dot{m}_{\rm Edd}\equiv L_{\rm bol}/L_{\rm Edd}$
listed in Table~\ref{tab:AGNproperties}. For $X=2.49$, the radius $R$ corresponds to the flux-weighted mean radius at wavelength $\lambda$, for a blackbody-emitting disk with temperature profile given by Eqn.\,\ref{eqn:tprofile} \citep{Fausnaugh16}. 

Table~\ref{tab:disksize} 
compares predicted and measured disk sizes
for the 3 AGN in our sample.
The table
collects the black hole mass and Eddington-ratio
estimates from Table~\ref{tab:AGNproperties},
which are used in Eqn.~(\ref{eqn:rc}) to compute the predicted disk size $R/c$ at $\lambda_0=1928/(1+z)$ \AA\ for comparison with the $\tau_0$ values determined from the lag-wavelength fits. We note that the values of $R/c$ listed in Table 10 differ from the corresponding values listed in Table 5 (Column 3) of \citet{Edelson19} because different values of $M_\mathrm{BH}$ and $\dot{m}$ are assumed.

Mrk\,509 has the highest mass ($M_8=1.1$) and the highest Eddington ratio ($\dot{m}_{\rm Edd}=0.095$) of the three objects, giving a predicted disk radius of $R=0.32$~lt-day at $\lambda=\lambda_0$. Comparing with Fits\,1 (all data) and 2 (omitting the U/\emph{u/r} bands), the measured lags $\tau_0=1.59\pm0.10$~day  and $1.49\pm 0.07$~day are larger by factors of 5.0 and 4.7, respectively, than the prediction from Equation \ref{eqn:rc}. The values of $\tau_0$ found for Fits 1 and 2 are nearly a factor of 2 larger than the corresponding value of $R/c=0.85$ (lt-days) obtained by \citet{Edelson19} solely from the \sw\, data.

For NGC\,4151, Equation \ref{eqn:rc} with $X=2.49$ predicts $R/c=0.053$~day. The lag measured from Fit\,4 (to all the lag data) is $\tau_0=0.14\pm0.04$~day, larger by a factor of 2.6. However, Fit\,4 does not provide a good representation of the lag spectrum over any portion of the observed wavelength range. Alternatively, Fit\,5 (using only the \sw\, UVOT lags, spanning the UVW2 through V bands)
gives $\tau_0=0.37\pm0.06$, consistent with the value 0.35 reported by \citet{Edelson19} from fitting the \sw\, lag spectrum. This is larger by a factor of 7 than the predicted $R/c$ value. 

NGC\,4593 has the lowest mass and lowest luminosity of the objects in this study. For $M_8 = 0.082$ ($10^8 M_{\odot}$)  and  Eddington ratio $\dot{m}_{\rm Edd}=0.081$, the predicted lag is $R/c=0.048$ (lt-days) .
The measured lag, $\tau_0 = 0.19 \pm0.03$ (lt-days) from Fit\,6,
exceeds the predicted lag by a factor 4.0. 
For comparison, \citet{Edelson19} found a disk size discrepancy of a factor of 2.1 using the \sw, data for NGC,4593.
Meanwhile, in the spectroscopic study conducted by \citet{Cackett18} (see Figure~\ref{fig:lagfit}, right panel), the disk size inferred from the lag measurements was approximately three times larger than that predicted by the standard thin-disk model. They also identified a significant lag excess around the 3646\,\AA\ Balmer jump, suggesting that diffuse emission from the broad-line region contributes notably to the interband lags.
The larger discrepancy found in the present work is primarily due to the substantially longer lags observed in the ground-based optical bands, which increases the normalization factor $\tau_0$ of the lag-wavelength fit relative to what would be obtained for the \sw\, data only.

\begin{table}
    \centering
    \caption{Comparison of Predicted Disk Sizes with Observed Lags}
    \label{tab:disksize}
    \begin{tabular}{cccc}
        \hline\hline
        AGN & $R/c$  & $\tau_0$ & $\tau_0\,c/R$ \\
         &  (days) & (days) &  \\
        (1) &  (2) & (3) & (4) \\
        \hline
        Mrk\,509  &  $0.318\pm0.015$ & $1.49\pm0.07$ & 4.7 \\
        NGC\,4151 &  $0.053\pm0.015$ & $0.14\pm0.04$ & 2.6 \\
        NGC\,4593 &  $0.048\pm0.008$ & $0.19\pm0.03$ & 4.0 \\
        \hline
    \end{tabular}
    \vspace{0.2cm}
    \begin{minipage}{\columnwidth}
        {\small
        \textbf{Notes:} 
        Column (1): Target name. Column (2): Theoretical value obtained using Equation~\ref{eqn:rc}. 
        Column (3): Observed value ($\tau_0$) determined from the power-law fit of the lags and presented in Table~\ref{tab:infolag}. 
        Column (4): Ratio of observed to theoretical sizes.
        }
    \end{minipage}
\end{table}

The results obtained here generally follow the trend, observed primarily in low-luminosity AGNs, of lags exceeding the predictions of simple disk reprocessing models by a factor of a few \citep[e.g.,][]{Li2021}, as found in prior \sw\, and ground-based continuum reverberation studies.
Any comparison of theoretical disk sizes with measured lags is subject to a few important caveats. First, the correct value
of $X$ to use for comparison with CCF lags is uncertain.  For consistency with previous work, we have adopted $X=2.49$, corresponding to a flux-weighted radius for a blackbody disk with $T\propto R^{-3/4}$. 
Using instead a higher value such as $X = 5.0$, as proposed by \citet{Tie2018}, which represents emission from a given radius only at the blackbody peak wavelength according to Wien's law, would increase $R/c$ by a factor of $(5.0/2.49)^{4/3} \approx 2.6$. This adjustment would enlarge the predicted disk size and help reconcile model expectations with the observed lags, particularly in sources like Mrk\,509. While such a value reduces the tension, it may imply a different physical interpretation compared to the standard flux-weighted emission scenario.
Theoretical lags also depend on the radiative efficiency $\eta$, and CCF lags may depend on the disk inclination $i$, which alters the shape of the delay distribution, but the values of $\eta$ and $i$ are generally unknown for individual AGN. Additionally, the values of both $M_\mathrm{BH}$ and $\dot{m}_\mathrm{Edd}$ are substantially uncertain (at perhaps $\sim0.5$ dex), due to the limitations of reverberation-based mass estimates \citep{Onken04}, bolometric corrections \citet{Runnoe2012}, and the impact of intrinsic variability on luminosity measurements. 
Furthermore, the poor correlation of X-ray and UV/optical variability found in previous work suggests that other processes may play a key role in driving disk variability, something different than that proposed by the standard disk reprocessing model.

\subsection{Modeling the Mrk\,509 STIS spectrum}
\label{sec:Mrk509specfit}

To derive estimates of the contributions of the Balmer continuum and emission lines to the broad-band filters in Mrk\,509, we carried out a multicomponent model fit to the STIS spectrum using the code PyQSOFit \citep{guo:pyqsofit}.\footnote{For this work we used a development version of PyQSOFit that adds additional model components that are not incorporated in the current public version of the code, including the diffuse continuum and the high-order Balmer and Paschen lines. See \citet{Guo:paschen2022} for a detailed description of these model components.} The spectrum was first corrected for Galactic extinction using the extinction law of \citet{Fitzpatrick:1999} and $R_V=3.1$. The fit components include a power-law continuum (PL), a diffuse continuum (DC) model from \citet{Korista19}, broad emission lines including the Balmer and Paschen series, \ion{Mg}{II}, \ion{C}{IV}, Ly$\alpha$, and several others,  narrow emission lines including [\ion{O}{III}] $\lambda\lambda4959,5007$, and \ion{Fe}{II} emission templates for the UV and optical regions of the data \citep{Boroson92,Vestergaard01,Tsuzuki06}.  The DC spectral model is based on locally optimally emitting cloud photoionization modeling of the BLR in NGC 5548, for a hydrogen column density $\log(N_\mathrm{H}/\mathrm{cm^{-2}}) = 23$, and density $\log(n_\mathrm{H}/\mathrm{cm^{-3}})$ ranging from 8 to 12  \citep{Korista19}.   In the fit, the DC model is broadened by convolution with a Gaussian kernel to match the velocity width of the Balmer lines ($\sigma=2290$ km s$^{-1}$). A $T=1400$ K blackbody component was included to represent hot-dust emission \citep[e.g.,][]{Honig14}, but this component had negligibly small flux in the best model fit. We also added a host-galaxy starlight component, using spectra of old stellar populations from \citet{Bruzual03}, but the model fits also drove the flux of this component to negligibly small values. We note, however, that the \sw\, and ground-based photometric data should contain a substantially higher starlight fraction than the STIS data due to the much larger photometric aperture radii in comparison with the 0\farcs2-width STIS slit. 

The best-fitting model yields a power-law slope of $\alpha=-1.74\pm0.03$ (for $f_\lambda \propto \lambda^\alpha$).
Figure \ref{fig:stis_with_filters} shows the fit components and the total model fit superposed on the STIS spectrum, along with residuals of the fit. The full model gives a reasonable fit to the overall spectral shape of the AGN although there are localized regions with substantial discrepancies, most notably near the Balmer jump. The Balmer jump region in AGN spectra is particularly challenging to fit, in part due to the broad range in gas density in the BLR which shifts and broadens the jump feature, as well as the pile-up of blended high-order Balmer lines redward of the Balmer jump \citep[as discussed by][]{Vincentelli:2021}. A more detailed analysis of the STIS spectrum will be the subject of future work, including fits incorporating a more sophisticated treatment of the Balmer jump region and more physically motivated accretion-disk spectral models \citep[e.g.,][]{DavisLaor11, KubotaDone18}.
 
In the best-fit model, the DC continuum fraction, defined as the flux ratio $f_\mathrm{DC} / (f_\mathrm{DC} + f_\mathrm{PL})$, reaches a maximum of 0.45--0.47 at wavelengths straddling the Balmer jump ($\lambda_\mathrm{rest}=3600-3730$ \AA), and then drops to a local minimum of $\sim0.13-0.14$ over $\lambda \approx 3860-4000$ \AA. Near the Paschen jump (8204 \AA), the DC continuum fraction is $\approx0.34$ shortward of the jump, dropping to 0.20 at longer wavelengths. These values are very similar to those found by \citet{Korista19} from photoionization models of NGC\,5548 (see their Figure 9).  

Following similar work by \citet{Vincentelli:2021} for Mrk 110, we used the results of the spectral decomposition to estimate the fractional contributions of each component to the broad-band filters, by performing synthetic photometry on each model component. Table \ref{tab:mrk509synphot} lists the fractional contributions to each \sw\, and ground-based filter band from the PL component, the DC component, \ion{Fe}{II}, all other emission lines in the model, and the combined fractional contribution of nebular emission (DC + \ion{Fe}{II} + lines). These results demonstrate that the nebular emission fraction is highest in the U, $u$, and $r$ bands, resulting from the contribution of the DC component in the $u$ band and the H$\alpha$ line in the $r$ band. The total nebular contribution in these bands is nearly 50\%, while in all other bands it ranges from 23\% to 41\%.  While the spectral decomposition does not directly yield a prediction for the excess lag, this supports the conclusion that the excess lags seen in the U/$u$/$r$ bands result are due to the BLR contribution to the broad-band lags. As seen in Figure \ref{fig:lagfit}, the $r$-band excess lag in Mrk\,509 is particularly pronounced. Despite the fact that the H$\alpha$ line falls at the red edge of the $r$-band filter, our spectral decomposition indicates that $\sim$30\% of the $r$-band flux is contributed by lines, and this contribution is dominated by the broad H$\alpha$ line. The Balmer-line lags in Mrk\,509 are much greater than the continuum lags: \citet{Peterson98} measured a lag of $\sim$80 days for H$\beta$ in Mrk\,509, and the H$\alpha$ lag in Seyferts is typically $\sim$50\% longer than that of H$\beta$ \citep{Bentz2010}, implying a lag of about 120 days for H$\alpha$. This contribution may provide a natural explanation for the unusually large $r$-band excess lag in Mrk\,509.

\begin{table}
    \centering
    \caption{Mrk\,509 Synthetic Photometry}
    \label{tab:mrk509synphot}
    \begin{tabular}{lccccc}
        \hline
        \multicolumn{1}{c}{} & \multicolumn{5}{c}{Fractional Contribution} \\
        \cline{2-6}
        Band & PL & DC & \ion{Fe}{II} & Lines & DC+\ion{Fe}{II}+Lines \\
        \hline
        \sw\, UVW2 & 0.704 & 0.111 & 0.079 & 0.106 & 0.296 \\
        \sw\, UVM2 & 0.714 & 0.135 & 0.120 & 0.031 & 0.286 \\
        \sw\, UVW1 & 0.589 & 0.168 & 0.171 & 0.072 & 0.411 \\
        \sw\, U    & 0.515 & 0.339 & 0.132 & 0.014 & 0.485 \\
        \sw\, B    & 0.677 & 0.134 & 0.026 & 0.163 & 0.323 \\
        \sw\, V    & 0.673 & 0.190 & 0.028 & 0.109 & 0.327 \\
        SDSS \emph{u} & 0.506 & 0.342 & 0.141 & 0.011 & 0.494 \\
        SDSS \emph{g} & 0.647 & 0.147 & 0.026 & 0.180 & 0.353 \\
        Bessell \emph{V} & 0.613 & 0.147 & 0.049 & 0.183 & 0.379 \\
        SDSS \emph{r} & 0.509 & 0.177 & 0.007 & 0.307 & 0.491 \\
        SDSS \emph{i} & 0.647 & 0.304 & 0.003 & 0.046 & 0.353 \\
        SDSS \emph{z} & 0.656 & 0.192 & 0.000 & 0.038 & 0.230 \\
        \hline
    \end{tabular}
    \vspace{0.2cm}
    \begin{minipage}{\linewidth}
        \justifying
        {\small
        \textbf{Notes:}  Synthetic photometry measurements of the fractional contributions to each filter from different spectral components, based on the PyQSOFit model fit to the Mrk\,509 STIS spectrum as shown in Figure \ref{fig:stis_with_filters}. For each filter, the fractional contributions are listed for the power-law (PL), diffuse continuum (DC), \ion{Fe}{II}, and emission-line components. The final column gives the sum of the DC, \ion{Fe}{II}, and emission-line components, indicating the total nebular contribution to each filter. The nebular fraction is greatest in the U, $u$, and $r$ bands, while the other bands their nebular contribution is around 23\% to 41\%.
        }
    \end{minipage}
\end{table}

The impact of BLR emission on the broad-band photometric reverberation lags can be assessed quantitatively by carrying out simulations of the light curves of each time-variable spectral component, using the results of the spectral decomposition combined with assumptions for the delay distributions of each spectral component. Such simulations are beyond the scope of this study but will be carried out in future work, for Mrk\,509 and other objects having high-quality STIS data available to carry out similar spectral decompositions.

\subsection{Diversity in IBRM UV/optical lag spectra} \label{sec:lags}

Here we summarize the most salient aspects of the UV/optical lag spectra derived from several IBRM campaigns based on monitoring with \sw\, and ground-based telescopes spanning UV through near-IR wavelengths.
For the three AGN analyzed in this paper:
\\
 a) Mrk~509 lags increase with wavelength with a best-fitting slope of $\beta=2.2\pm0.2$, with significant excess lags in the U, \emph{u}, and \emph{r} bands.
\\
 b) NGC\,4151 lags increase with wavelength from the UV up to the region of the $r$ band, but then decrease at longer wavelengths in the optical, inconsistent with expectations for reprocessing by a standard disk.
\\
c) The lags measured for NGC\,4593 in our campaign are consistent with a $\tau\propto\lambda^{4/3}$ relation, although the short duration of the monitoring and relatively large uncertainties limit our ability to tightly constrain the slope of the lag spectrum. A more intensive and simultaneous campaign by \citet{Cackett18} using \textit{HST}  reported similar lag behavior to that observed in our data (see Figure~\ref{fig:lagfit}, right panel), but revealed additional structure in the lag spectrum, including a prominent excess in the Balmer continuum region and a downturn in the Infrared.

In addition, several other AGN have published lag spectra from prior \sw\, and ground-based IBRM campaigns extending from UV to near-IR wavelengths: 
\\ 
d) Fairall~9 has the most well-defined optical lag spectrum published to date \citep{Hernandez20}, showing a good fit to $ \tau \propto \lambda^{4/3} $ with a clear excess in \emph{U/u} bands and possibly a weaker excess in \emph{r} and \emph{i} bands,
\\ 
e) NGC~5548 has lags of similar quality \citep{Fausnaugh16} that also show good agreement with $ \tau \propto \lambda^{4/3} $ and a clear excess in \emph{U/u} bands and possibly a weaker excess in \emph{r} and \emph{i} bands,
\\ 
f) Mrk~142 shows an acceptable fit to $ \tau \propto \lambda^{4/3} $ and a clear excess in \emph{U/u} bands but no indication of excess lags at longer wavelengths \citep{Cackett:2020}, and
\\ 
g) Mrk~110 shows an acceptable fit to $ \tau \propto \lambda^{4/3} $ with no clear evidence for excesses, although the lag errors are large as that was also a relatively short campaign \citep{Vincentelli:2021}. 
\\ 

h) Mrk817 shows a very good fit following the $\tau \propto \lambda^{4/3}$ relation, and no excess is observed in the \emph{U/u} bands \citep{Kara21}, although the analysis adopts the irradiated disk model of \citet{Kammoun2021}, which differs from the standard accretion disk framework \citep{ShakuraSunyaev73}. However, \citet{Netzer2022} report that the lag normalization at 2500,\AA\ in Mrk817, as well as in several other AGN, exceeds the prediction of this model by a factor of $\sim$2–3. \citet{Netzer2024} further demonstrate that a wind–BLR–small-disk combination provides a better fit to the observed up-and-down lag behavior. Likewise, \citet{Lewin2024} show that the measured lags depend on the continuum luminosity and the range of frequencies analyzed, further emphasizing the complexity of the accretion disk structure.
\\ 

i) In Mrk~335 the lag spectrum follows a $\tau \propto \lambda^{4/3}$ relationship, with a relatively modest excess lag in the \emph{u} and \emph{r} bands~\citep{Kara2023}. 

A summary of these AGN and their lag-wavelength characteristics is provided in Table~\ref{tab:AGNsummary}.

\begin{table*}
    \centering
    \caption{Comparison of Predicted Disk Sizes with Observed Lags}
    \label{tab:AGNsummary}
    \begin{tabular}{lcccll}
        \hline
        AGN & Lag-Wavelength Relationship & $\tau_{\rm obs}/\tau_{\rm SS}$ & Excess Lags & Bands with Excess & Reference \\
        (1) & (2) & (3) & (4) & (5) & (6) \\
        \hline
        Mrk\,509  &  $\beta = 2.2\pm0.2$ &  $4.7$ & Yes & U, \emph{u}, \emph{r} & This work \\
        NGC\,4151 & $\tau\propto\lambda^{4/3}$, Increases up to \emph{r}, then decreases & $2.6$ & \textbf{?} & -- & This work \\
        NGC\,4593 & $\tau\propto\lambda^{4/3}$, Increases up to \emph{r}, then decreases & $4.0$ & Yes & U, \emph{u}, \emph{r} & This work; \citet{Cackett18} \\
        Fairall~9 &  $\tau\propto\lambda^{4/3}$ & $ 3.0 $ & Yes & U, \emph{u}, \emph{r}, \emph{i} & \citet{Hernandez20}\\ 
        NGC~5548 & $\tau\propto\lambda^{4/3}$ & $3.0$ & Yes & U, \emph{u}, \emph{r}, \emph{i} & \citet{Fausnaugh16}\\
        Mrk~142 &  $\tau\propto\lambda^{4/3}$ & $3.4$ & Yes & U, \emph{u} & \citet{Cackett:2020} \\
        Mrk~110 &  $\tau\propto\lambda^{4/3}$ & $0.1$ & No & -- & \citet{Vincentelli:2021} \\
        Mrk~817 &  $\tau\propto\lambda^{4/3}$ & $2.5$ & \textbf{Yes?} & U, \emph{u} & \citet{Kara21,Netzer2024} \\
        Mrk~335 & $\tau\propto\lambda^{4/3}$ & $9.0$ & Yes & \emph{u}, \emph{r} & \citet{Kara2023} \\
        \hline
    \end{tabular}
\end{table*}

Overall these results show a greater level of diversity than was seen in earlier analyses that were restricted to the \sw\, data only. 
Perhaps the most unexpected behavior is that of NGC\,4151,  where the standard trend of increasing lag as a function of wavelength extends from the UV only up to the $r$-band region, and the longer-wavelength lag spectrum turns over to a negative slope (see Fig.~\ref{fig:lagfit} right panel). The reason for this is unclear. Interestingly, this behavior is not accompanied by any unusual features in the broadband SED, which appears relatively normal as noted in Section~\ref{sec:fluxflux}. It is likely that this overall behavior is intrinsic to the target, although whether this represents persistent behavior or whether it is merely a short-term anomaly occuring during the monitoring period is unknown. Based on the earlier (\sw\ -only) multi-wavelength timing analysis, \citet{Mahmoud2020} proposed that NGC\,4151 lacks an optically thick inner accretion disk, being replaced by a hot and optically thin gas.  The UV/optical emission would then arise from (quasi-)co-spatial clouds in the BLR. 
The breakdown of the standard disk relation in these optical data may be further confirmation of this picture, although this does not provide a straightforward explanation of the unusually short near-IR lags.
A recent study by \citet{Zhou2024} shows that the delays in NGC\,4151 can vary significantly depending on its luminosity state. In particular, they found that in the high-flux state, the continuum delay is $3.8^{+1.8}_{-1.0}$ times larger than in the low-flux state and is $14.9 \pm 2.0$ times longer than predicted by the standard thin disk model. Based on these results \citet{Zhou2024} suggests that the accretion disk in NGC\,4151 exhibits a "breathing" behavior, expanding in high-flux states and contracting in low-flux states.
Given the brief monitoring period and the limitations in the data quality for the NGC\,4151 campaign, a longer-duration \sw\, and ground-based program on this exceptionally bright AGN would be highly valuable to obtain a clearer picture of the continuum reverberation behavior in this object.

Recent ground-based continuum reverberation mapping programs on the extreme low-mass AGN NGC\,4395 (with $\log[M_\mathrm{BH}/M_\odot] \sim  4-5.5$) provide an interesting point of comparison for the long-wavelength turnover seen in the lag spectrum of NGC\,4151. The continuum variations in NGC\,4395 are so rapid that optical continuum lags are of order several minutes, and optical lags extending out to the $z$ band were recently reported from observations with the MuSCAT3 multi-channel \emph{griz} imager at the Faulkes North Telescope \citep{Montano2022} and from the \emph{ugriz} HiPERCAM imager at the Gran Telescopio Canarias \citep{McHardy2023}. Remarkably, NGC\,4395 exhibited distinctly different long-wavelength lag behavior on three nights having high-quality monitoring data. For one night of HiPERCAM monitoring in 2018, the lag spectrum in NGC\,4395 rose between \emph{u} and \emph{r} with a break to a nearly flat slope between the \emph{r} and \emph{z} bands, and \citet{McHardy2023} interpreted this long-wavelength flattening as evidence for an outer truncation in the accretion disk in NGC\,4395. During the two nights of MuSCAT3 \emph{griz} monitoring in 2022, \citet{Montano2022} found a lag spectrum closely consistent with the standard $\tau\propto\lambda^{4/3}$ trend on the first night. However on the following night the lag spectrum was markedly different, exhibiting a turnover to negative slope between the \emph{i} and \emph{z} bands, qualitatively similar to the long-wavelength turnover seen in NGC\,4151. The reason for this rapid change is unclear, but this provides a note of caution that lag spectra measured from short-duration campaigns may not be representative of longer-term behavior \citep[e.g.,][]{Su2024}. 

Another example of an anomalous lag-wavelength spectrum was recently reported for NGC\,6814 based on intensive \sw\, monitoring. For this object, \citet{Gonzalez2023} found that the lags increased from the UV through the U band up to a U-band lag of $\sim$0.3 d, but the lag spectrum turned over to a slightly negative slope between the U and V bands. They concluded that a combination of super-Eddington accretion and a disk with an outer truncation could match the overall shape of the lag spectrum, but that no combination of disk parameters could simultaneously provide a good fit to both the lag spectrum and the AGN SED inferred from flux-flux analysis. These results for NGC\,4395 and NGC\,6814, together with our findings for NGC\,4151, underscore the need for further intensive, better quality and long-duration monitoring programs focused on Seyferts of relatively low luminosity to better understand the origin of these unusual reverberation signatures.

\section{Summary and Future work } \label{sec:summary}

This paper presents the results of intensive optical/UV/X-ray monitoring of three bright AGN: Mrk~509, NGC\,4151 and NGC\,4593. In many ways our results are consistent with those reported previously for other targets of \sw\, and ground-based continuum reverberation mapping:
all show stronger correlations within the optical/UV than between the UV and X-rays, and all show evidence that at least some of the observed UV/optical correlations and lags arise from ``diffuse continuum'' emission from the BLR. Both Mrk~509 and NGC\,4593 show UV/optical lags that increase with wavelength, as expected for disk reprocessing, although the slope of the lag spectrum in Mrk\,509 appears steeper than the $\tau \propto \lambda^{4/3}$ relation predicted by the standard thin-disk model. The inferred disk sizes (typically a factor of 2-5 larger than predicted, albeit with comparable systematic uncertainties) and the poor UV/X-ray correlations present a challenge to the reprocessing picture. However, the more fundamental thin disk model may survive this challenge for example if we assume that the driving continuum differs from the observed X-ray band and if the measured lags are boosted by a significant contribution of diffuse continuum emission from the larger BLR. Our spectral decomposition of the Mrk\,509 STIS spectrum supports the hypothesis of a substantial BLR contribution to the broad-band lags, with the BLR contributing $\gtrsim30\%$ of the flux across most of the spectrum and nearly 50\% of the light in the U, $u$, and $r$ bands.

However, NGC\,4151 shows more complex interband lags, with the \emph{i} and \emph{z}-band lags significantly below that of the \emph{r} band and below the lag-wavelength trend extrapolated from the \sw\, lags at shorter wavelengths.  This is qualitatively different than observed in any of the other AGN having \sw\, and ground-based data, and is not currently understood, although somewhat analogous behavior has been seen in optical data for the dwarf AGN NGC\,4395 \citep{Montano2022} and in \sw\, data for NGC\,6814 \citep{Gonzalez2023}. To learn more about this behavior in NGC\,4151, we have continued monitoring of this bright AGN from the ground with LCOGT. Although lacking simultaneous \sw\, monitoring, intensive monitoring in the \emph{BVugriz} bands (and ideally extending to longer-wavelength IR monitoring if possible) can test whether this peculiar near-IR lag behavior is a persistent feature in NGC\,4151. It should be interesting to see if this unusual interband lag behavior persists for NGC\,4151, or if it reverts to the type of lag structure seen in other AGN, as that could help shed light on the physical properties of the accretion disk and other spatially unresolved structures in AGN.

Finally, the results presented here for these early IBRM campaigns provide instructive lessons to inform the planning of future intensive continuum monitoring programs. While the ground-based optical campaign on Mrk\,509 obtained high-S/N light curves yielding precise continuum lag measurements, the ground-based campaigns coordinated with the shorter-duration \sw\, programs on NGC\,4151 and NGC\,4593 (69 and 23 days, respectively) were less successful. In both cases the lag measurements rested largely on the detection of a single strong feature in the \sw\, light curves, highlighting a considerable element of risk when carrying out such short-duration campaigns. Furthermore, the ground-based campaigns for these targets were underpowered for the goal of measuring high-quality reverberation lags. For NGC\,4593, both the S/N and the cadence of the ground-based data that were obtainable at the time of the campaign were less than ideal, and for NGC\,4151 the coverage in the \emph{BVRI} bands was insufficient for measurement of reliable light curves and lags. Future \sw\, campaigns should, in general, be based on longer monitoring durations that will improve the chances of detecting multiple features and higher overall variability amplitudes in the light curves. 

For targets such as NGC\,4151 and NGC\,4593 having optical continuum lags of $\lesssim1$ day, precision lag measurement will ideally require obtaining multiple observations per day at ground-based facilities spanning long durations (several months), without major temporal gaps in photometric coverage. Although these are bright targets that can be observed with small telescopes, it still remains a logistical challenge to obtain light curves of the requisite cadence and quality for such objects. Fortunately, the number of observatory sites in the LCOGT network has grown in recent years, and additional robotic facilities have been established that are capable of making valuable contributions to intensive monitoring programs.  Current goals in continuum reverberation studies include measurement of frequency-resolved lags that can potentially disentangle reverberation signals on different timescales originating from different reprocessing regions \citep{Cackett2022, Yao2023, Lewin2023}, and detection of temperature fluctuations propagating through the accretion disk \citep{Neustadt2022}. The data quality required for such investigations is far greater than what is needed just for measurement of reverberation lags alone, and a combination of long monitoring durations, high cadence, and high S/N will be needed to recover these subtle signals in the data from future campaigns. The next generation of multiwavelength IBRM studies will be planned to deliver the data quality needed  to optimally carry out these investigations.

\nolinenumbers

\section*{Acknowledgements}

This work makes use of observations from the Las Cumbres Observatory global telescope network. Support for \emph{Hubble Space Telescope} program \#15124 was provided by NASA through a grant from the Space Telescope Science Institute, which is operated by the Association of Universities for Research in Astronomy, Inc., under NASA contract NAS 5-26555. 
Research by the UC Irvine group has been supported by NSF grants AST-1412693 and AST-1907290, and by grant HST-GO-15124.001-A from the Space Telescope Science Institute. 
DHGB acknowledges CONACYT support \#319800 and of the researchers program for Mexico. 
JVHS and KDH acknowledge support from STFC grant ST/R000824/1. 
MCB gratefully acknowledges support from the NSF through grant AST-2407802. 
YRL acknowledges support from NSF through grant 11922304 and from the Youth innovation Promotion Association CAS. 
M.V. gratefully acknowledges financial support from the Independent Research Fund Denmark via grant numbers DFF 8021-00130 and DFF 3103-00146 and the Carlsberg Foundation via grants CF21-0649  and CF23-0417.
EDB, EMC, and AP acknowledge the support from MIUR grant PRIN 2017 20173ML3WW-001 and Padua University
grants DOR 2022-2024; they are also funded by INAF through grant PRIN 2022 C53D23000850006. EMC acknowledges the support by the MIUR grant PRIN 2022 2022383WFT “SUNRISE” (CUP C53D23000850006).
DCL gratefully acknowledges support from the NSF through grant
AST-1210311
M.K. acknowledges the support of the National Research Foundation of Korea (NRF) grant (No. RS-2024-00347548).
YK acknowledges support from grant PAPIIT-UNAM IN102023.
The Liverpool Telescope is operated on the island of La Palma by Liverpool John Moores University in the Spanish Observatorio del Roque de los Muchachos of the Instituto de Astrofisica de Canarias with financial support from the UK Science and Technology Facilities Council.

Bradley Peterson contributed to the planning of these monitoring programs.
We thank Timothy Ross, Carolina Gould, Minkyu Kim, Heechan Yuk, Kevin Hayakawa, Goni Halevy, Jeffrey D. Molloy, Sanyum Channa, Andrew Rikhter, and Maxime de Kouchkovsky for assistance with observations at Lick Observatory. 
We also thank Dale Mudd for his active participation in this work during his time as a postdoctoral researcher at UCI.
This research has made use of the NASA/IPAC Extragalactic Database, which is funded by the National Aeronautics and Space Administration and operated by the California Institute of Technology.

\section*{Data Availability}

Access to the data used in this article is subject to request to the author and will only be provided with the authorization of the Principal Investigators (PIs) of the IBRM project.



\bibliographystyle{mnras}
\bibliography{example.bib} 

\begin{thebibliography}{}
\makeatletter
\relax
\def\mn@urlcharsother{\let\do\@makeother \do\$\do\&\do\#\do\^\do\_\do\%\do\~}
\def\mn@doi{\begingroup\mn@urlcharsother \@ifnextchar [ {\mn@doi@}
  {\mn@doi@[]}}
\def\mn@doi@[#1]#2{\def\@tempa{#1}\ifx\@tempa\@empty \href
  {http://dx.doi.org/#2} {doi:#2}\else \href {http://dx.doi.org/#2} {#1}\fi
  \endgroup}
\def\mn@eprint#1#2{\mn@eprint@#1:#2::\@nil}
\def\mn@eprint@arXiv#1{\href {http://arxiv.org/abs/#1} {{\tt arXiv:#1}}}
\def\mn@eprint@dblp#1{\href {http://dblp.uni-trier.de/rec/bibtex/#1.xml}
  {dblp:#1}}
\def\mn@eprint@#1:#2:#3:#4\@nil{\def\@tempa {#1}\def\@tempb {#2}\def\@tempc
  {#3}\ifx \@tempc \@empty \let \@tempc \@tempb \let \@tempb \@tempa \fi \ifx
  \@tempb \@empty \def\@tempb {arXiv}\fi \@ifundefined
  {mn@eprint@\@tempb}{\@tempb:\@tempc}{\expandafter \expandafter \csname
  mn@eprint@\@tempb\endcsname \expandafter{\@tempc}}}

\bibitem[\protect\citeauthoryear{{Aranzana}, {K{\"o}rding}, {Uttley},
  {Scaringi}  \& {Bloemen}}{{Aranzana} et~al.}{2018}]{Aranzana18}
{Aranzana} E.,  {K{\"o}rding} E.,  {Uttley} P.,  {Scaringi} S.,   {Bloemen} S.,
   2018, \mn@doi [\mnras] {10.1093/mnras/sty413}, \href
  {https://ui.adsabs.harvard.edu/abs/2018MNRAS.476.2501A} {476, 2501}

\bibitem[\protect\citeauthoryear{{Astropy Collaboration} et~al.,}{{Astropy
  Collaboration} et~al.}{2018}]{Astropy2018}
{Astropy Collaboration} et~al., 2018, \mn@doi [\aj] {10.3847/1538-3881/aabc4f},
  \href {https://ui.adsabs.harvard.edu/abs/2018AJ....156..123A} {156, 123}

\bibitem[\protect\citeauthoryear{{Barth} et~al.,}{{Barth}
  et~al.}{2015}]{Barth15}
{Barth} A.~J.,  et~al., 2015, \mn@doi [\apjs] {10.1088/0067-0049/217/2/26},
  \href {https://ui.adsabs.harvard.edu/abs/2015ApJS..217...26B} {217, 26}

\bibitem[\protect\citeauthoryear{{Bentz} \& {Katz}}{{Bentz} \&
  {Katz}}{2015}]{bentz2015}
{Bentz} M.~C.,  {Katz} S.,  2015, \mn@doi [\pasp] {10.1086/679601}, \href
  {https://ui.adsabs.harvard.edu/abs/2015PASP..127...67B} {127, 67}

\bibitem[\protect\citeauthoryear{{Bentz} et~al.,}{{Bentz}
  et~al.}{2006}]{Bentz06}
{Bentz} M.~C.,  et~al., 2006, \mn@doi [\apj] {10.1086/507417}, \href
  {https://ui.adsabs.harvard.edu/abs/2006ApJ...651..775B} {651, 775}

\bibitem[\protect\citeauthoryear{{Bentz} et~al.,}{{Bentz}
  et~al.}{2010}]{Bentz2010}
{Bentz} M.~C.,  et~al., 2010, \mn@doi [\apj] {10.1088/0004-637X/716/2/993},
  \href {https://ui.adsabs.harvard.edu/abs/2010ApJ...716..993B} {716, 993}

\bibitem[\protect\citeauthoryear{{Bentz} et~al.,}{{Bentz}
  et~al.}{2013}]{bentz13}
{Bentz} M.~C.,  et~al., 2013, \mn@doi [\apj] {10.1088/0004-637X/767/2/149},
  \href {https://ui.adsabs.harvard.edu/abs/2013ApJ...767..149B} {767, 149}

\bibitem[\protect\citeauthoryear{{Bessell}}{{Bessell}}{1990}]{Bessell90}
{Bessell} M.~S.,  1990, \mn@doi [\pasp] {10.1086/132749}, \href
  {https://ui.adsabs.harvard.edu/abs/1990PASP..102.1181B} {102, 1181}

\bibitem[\protect\citeauthoryear{{Bessell}, {Castelli}  \& {Plez}}{{Bessell}
  et~al.}{1998}]{bessell1998}
{Bessell} M.~S.,  {Castelli} F.,   {Plez} B.,  1998, \aap, \href
  {https://ui.adsabs.harvard.edu/abs/1998A&A...333..231B} {333, 231}

\bibitem[\protect\citeauthoryear{{Blackburne}, {Pooley}, {Rappaport}  \&
  {Schechter}}{{Blackburne} et~al.}{2011}]{Blackburne:2011}
{Blackburne} J.~A.,  {Pooley} D.,  {Rappaport} S.,   {Schechter} P.~L.,  2011,
  \mn@doi [\apj] {10.1088/0004-637X/729/1/34}, \href
  {https://ui.adsabs.harvard.edu/abs/2011ApJ...729...34B} {729, 34}

\bibitem[\protect\citeauthoryear{{Blandford} \& {McKee}}{{Blandford} \&
  {McKee}}{1982}]{Blandford82}
{Blandford} R.~D.,  {McKee} C.~F.,  1982, \mn@doi [\apj] {10.1086/159843},
  \href {https://ui.adsabs.harvard.edu/abs/1982ApJ...255..419B} {255, 419}

\bibitem[\protect\citeauthoryear{{Boroson} \& {Green}}{{Boroson} \&
  {Green}}{1992}]{Boroson92}
{Boroson} T.~A.,  {Green} R.~F.,  1992, \mn@doi [\apjs] {10.1086/191661}, \href
  {https://ui.adsabs.harvard.edu/abs/1992ApJS...80..109B} {80, 109}

\bibitem[\protect\citeauthoryear{{Boroson} et~al.,}{{Boroson}
  et~al.}{2014}]{Boroson14}
{Boroson} T.,  et~al., 2014, in {Peck} A.~B.,  {Benn} C.~R.,   {Seaman} R.~L.,
  eds,  Society of Photo-Optical Instrumentation Engineers (SPIE) Conference
  Series Vol. 9149, Observatory Operations: Strategies, Processes, and Systems
  V. p. 91491E, \mn@doi{10.1117/12.2054776}

\bibitem[\protect\citeauthoryear{{Brown} et~al.,}{{Brown}
  et~al.}{2013}]{Brown13}
{Brown} T.~M.,  et~al., 2013, \mn@doi [\pasp] {10.1086/673168}, \href
  {https://ui.adsabs.harvard.edu/abs/2013PASP..125.1031B} {125, 1031}

\bibitem[\protect\citeauthoryear{{Bruzual} \& {Charlot}}{{Bruzual} \&
  {Charlot}}{2003}]{Bruzual03}
{Bruzual} G.,  {Charlot} S.,  2003, \mn@doi [\mnras]
  {10.1046/j.1365-8711.2003.06897.x}, \href
  {https://ui.adsabs.harvard.edu/abs/2003MNRAS.344.1000B} {344, 1000}

\bibitem[\protect\citeauthoryear{{Burrows} et~al.,}{{Burrows}
  et~al.}{2005}]{Burrows2005}
{Burrows} D.~N.,  et~al., 2005, \mn@doi [\ssr] {10.1007/s11214-005-5097-2},
  \href {https://ui.adsabs.harvard.edu/abs/2005SSRv..120..165B} {120, 165}

\bibitem[\protect\citeauthoryear{{Cackett}, {Horne}  \& {Winkler}}{{Cackett}
  et~al.}{2007}]{Cackett07}
{Cackett} E.~M.,  {Horne} K.,   {Winkler} H.,  2007, \mn@doi [\mnras]
  {10.1111/j.1365-2966.2007.12098.x}, \href
  {https://ui.adsabs.harvard.edu/abs/2007MNRAS.380..669C} {380, 669}

\bibitem[\protect\citeauthoryear{{Cackett}, {Chiang}, {McHardy}, {Edelson},
  {Goad}, {Horne}  \& {Korista}}{{Cackett} et~al.}{2018}]{Cackett18}
{Cackett} E.~M.,  {Chiang} C.-Y.,  {McHardy} I.,  {Edelson} R.,  {Goad} M.~R.,
  {Horne} K.,   {Korista} K.~T.,  2018, \mn@doi [\apj]
  {10.3847/1538-4357/aab4f7}, \href
  {https://ui.adsabs.harvard.edu/abs/2018ApJ...857...53C} {857, 53}

\bibitem[\protect\citeauthoryear{{Cackett} et~al.,}{{Cackett}
  et~al.}{2020}]{Cackett:2020}
{Cackett} E.~M.,  et~al., 2020, arXiv e-prints, \href
  {https://ui.adsabs.harvard.edu/abs/2020arXiv200503685C} {p. arXiv:2005.03685}

\bibitem[\protect\citeauthoryear{{Cackett}, {Bentz}  \& {Kara}}{{Cackett}
  et~al.}{2021}]{Cackett:2021}
{Cackett} E.~M.,  {Bentz} M.~C.,   {Kara} E.,  2021, \mn@doi [iScience]
  {10.1016/j.isci.2021.102557}, \href
  {https://ui.adsabs.harvard.edu/abs/2021iSci...24j2557C} {24, 102557}

\bibitem[\protect\citeauthoryear{{Cackett}, {Zoghbi}  \& {Ulrich}}{{Cackett}
  et~al.}{2022}]{Cackett2022}
{Cackett} E.~M.,  {Zoghbi} A.,   {Ulrich} O.,  2022, \mn@doi [\apj]
  {10.3847/1538-4357/ac3913}, \href
  {https://ui.adsabs.harvard.edu/abs/2022ApJ...925...29C} {925, 29}

\bibitem[\protect\citeauthoryear{{Chartas} et~al.,}{{Chartas}
  et~al.}{2016}]{Chartas:2016}
{Chartas} G.,  et~al., 2016, \mn@doi [Astronomische Nachrichten]
  {10.1002/asna.201612313}, \href
  {https://ui.adsabs.harvard.edu/abs/2016AN....337..356C} {337, 356}

\bibitem[\protect\citeauthoryear{{Collier} et~al.,}{{Collier}
  et~al.}{1998}]{Collier98}
{Collier} S.~J.,  et~al., 1998, \mn@doi [\apj] {10.1086/305720}, \href
  {https://ui.adsabs.harvard.edu/abs/1998ApJ...500..162C} {500, 162}

\bibitem[\protect\citeauthoryear{{Davis} \& {Laor}}{{Davis} \&
  {Laor}}{2011}]{DavisLaor11}
{Davis} S.~W.,  {Laor} A.,  2011, \mn@doi [\apj] {10.1088/0004-637X/728/2/98},
  \href {https://ui.adsabs.harvard.edu/abs/2011ApJ...728...98D} {728, 98}

\bibitem[\protect\citeauthoryear{{Davis} \& {Tchekhovskoy}}{{Davis} \&
  {Tchekhovskoy}}{2020}]{Davis2020}
{Davis} S.~W.,  {Tchekhovskoy} A.,  2020, \mn@doi [\araa]
  {10.1146/annurev-astro-081817-051905}, \href
  {https://ui.adsabs.harvard.edu/abs/2020ARA&A..58..407D} {58, 407}

\bibitem[\protect\citeauthoryear{{De Rosa} et~al.,}{{De Rosa}
  et~al.}{2015}]{deRosa:2015}
{De Rosa} G.,  et~al., 2015, \mn@doi [\apj] {10.1088/0004-637X/806/1/128},
  \href {https://ui.adsabs.harvard.edu/abs/2015ApJ...806..128D} {806, 128}

\bibitem[\protect\citeauthoryear{{De Rosa} et~al.,}{{De Rosa}
  et~al.}{2018}]{derosa18}
{De Rosa} G.,  et~al., 2018, \mn@doi [\apj] {10.3847/1538-4357/aadd11}, \href
  {https://ui.adsabs.harvard.edu/abs/2018ApJ...866..133D} {866, 133}

\bibitem[\protect\citeauthoryear{{Denney} et~al.,}{{Denney}
  et~al.}{2006}]{Denney06}
{Denney} K.~D.,  et~al., 2006, \mn@doi [\apj] {10.1086/508533}, \href
  {https://ui.adsabs.harvard.edu/abs/2006ApJ...653..152D} {653, 152}

\bibitem[\protect\citeauthoryear{{Edelson}, {Vaughan}, {Malkan}, {Kelly},
  {Smith}, {Boyd}  \& {Mushotzky}}{{Edelson} et~al.}{2014}]{Edelson14}
{Edelson} R.,  {Vaughan} S.,  {Malkan} M.,  {Kelly} B.~C.,  {Smith} K.~L.,
  {Boyd} P.~T.,   {Mushotzky} R.,  2014, \mn@doi [\apj]
  {10.1088/0004-637X/795/1/2}, \href
  {https://ui.adsabs.harvard.edu/abs/2014ApJ...795....2E} {795, 2}

\bibitem[\protect\citeauthoryear{{Edelson} et~al.,}{{Edelson}
  et~al.}{2015}]{Edelson15}
{Edelson} R.,  et~al., 2015, \mn@doi [\apj] {10.1088/0004-637X/806/1/129},
  \href {https://ui.adsabs.harvard.edu/abs/2015ApJ...806..129E} {806, 129}

\bibitem[\protect\citeauthoryear{{Edelson} et~al.,}{{Edelson}
  et~al.}{2017}]{Edelson17}
{Edelson} R.,  et~al., 2017, \mn@doi [\apj] {10.3847/1538-4357/aa6890}, \href
  {https://ui.adsabs.harvard.edu/abs/2017ApJ...840...41E} {840, 41}

\bibitem[\protect\citeauthoryear{{Edelson} et~al.,}{{Edelson}
  et~al.}{2019}]{Edelson19}
{Edelson} R.,  et~al., 2019, \mn@doi [\apj] {10.3847/1538-4357/aaf3b4}, \href
  {https://ui.adsabs.harvard.edu/abs/2019ApJ...870..123E} {870, 123}

\bibitem[\protect\citeauthoryear{{Evans} et~al.,}{{Evans}
  et~al.}{2009}]{Evans09}
{Evans} P.~A.,  et~al., 2009, \mn@doi [\mnras]
  {10.1111/j.1365-2966.2009.14913.x}, \href
  {https://ui.adsabs.harvard.edu/abs/2009MNRAS.397.1177E} {397, 1177}

\bibitem[\protect\citeauthoryear{{Fausnaugh} et~al.,}{{Fausnaugh}
  et~al.}{2016}]{Fausnaugh16}
{Fausnaugh} M.~M.,  et~al., 2016, \mn@doi [\apj] {10.3847/0004-637X/821/1/56},
  \href {https://ui.adsabs.harvard.edu/abs/2016ApJ...821...56F} {821, 56}

\bibitem[\protect\citeauthoryear{{Fausnaugh} et~al.,}{{Fausnaugh}
  et~al.}{2017}]{Fausnaugh17}
{Fausnaugh} M.~M.,  et~al., 2017, \mn@doi [\apj] {10.3847/1538-4357/aa6d52},
  \href {https://ui.adsabs.harvard.edu/abs/2017ApJ...840...97F} {840, 97}

\bibitem[\protect\citeauthoryear{{Fausnaugh} et~al.,}{{Fausnaugh}
  et~al.}{2018}]{Fausnaugh2018}
{Fausnaugh} M.~M.,  et~al., 2018, \mn@doi [\apj] {10.3847/1538-4357/aaaa2b},
  \href {https://ui.adsabs.harvard.edu/abs/2018ApJ...854..107F} {854, 107}

\bibitem[\protect\citeauthoryear{{Fian}, {Chelouche}, {Kaspi}, {Sobrino
  Figaredo}, {Catalan}  \& {Lewis}}{{Fian} et~al.}{2022}]{Fian2022}
{Fian} C.,  {Chelouche} D.,  {Kaspi} S.,  {Sobrino Figaredo} C.,  {Catalan} S.,
    {Lewis} T.,  2022, \mn@doi [\aap] {10.1051/0004-6361/202141509}, \href
  {https://ui.adsabs.harvard.edu/abs/2022A&A...659A..13F} {659, A13}

\bibitem[\protect\citeauthoryear{{Filippenko}, {Li}, {Treffers}  \&
  {Modjaz}}{{Filippenko} et~al.}{2001}]{Filippenko01}
{Filippenko} A.~V.,  {Li} W.~D.,  {Treffers} R.~R.,   {Modjaz} M.,  2001, in
  {Paczynski} B.,  {Chen} W.-P.,   {Lemme} C.,  eds,  Astronomical Society of
  the Pacific Conference Series Vol. 246, IAU Colloq. 183: Small Telescope
  Astronomy on Global Scales. p.~121

\bibitem[\protect\citeauthoryear{{Fitzpatrick}}{{Fitzpatrick}}{1999}]{Fitzpatrick:1999}
{Fitzpatrick} E.~L.,  1999, \mn@doi [\pasp] {10.1086/316293}, \href
  {https://ui.adsabs.harvard.edu/abs/1999PASP..111...63F} {111, 63}

\bibitem[\protect\citeauthoryear{{Fukugita}, {Ichikawa}, {Gunn}, {Doi},
  {Shimasaku}  \& {Schneider}}{{Fukugita} et~al.}{1996}]{Fukugita1996}
{Fukugita} M.,  {Ichikawa} T.,  {Gunn} J.~E.,  {Doi} M.,  {Shimasaku} K.,
  {Schneider} D.~P.,  1996, \mn@doi [\aj] {10.1086/117915}, \href
  {https://ui.adsabs.harvard.edu/abs/1996AJ....111.1748F} {111, 1748}

\bibitem[\protect\citeauthoryear{{Gaskell} \& {Peterson}}{{Gaskell} \&
  {Peterson}}{1987}]{Gaskell87}
{Gaskell} C.~M.,  {Peterson} B.~M.,  1987, \mn@doi [\apjs] {10.1086/191216},
  \href {https://ui.adsabs.harvard.edu/abs/1987ApJS...65....1G} {65, 1}

\bibitem[\protect\citeauthoryear{{Gonz{\'a}lez-Buitrago}, {Hern{\'a}ndez
  Santisteban}, {Barth}, {Jimenez-Bail{\'o}n}, {Li}, {Garc{\'\i}a-D{\'\i}az},
  {Lopez Vargas}  \& {Herrera-Endoqui}}{{Gonz{\'a}lez-Buitrago}
  et~al.}{2022}]{Gonzalez2022}
{Gonz{\'a}lez-Buitrago} D.~H.,  {Hern{\'a}ndez Santisteban} J.~V.,  {Barth}
  A.~J.,  {Jimenez-Bail{\'o}n} E.,  {Li} Y.-R.,  {Garc{\'\i}a-D{\'\i}az} M.~T.,
   {Lopez Vargas} A.,   {Herrera-Endoqui} M.,  2022, \mn@doi [\mnras]
  {10.1093/mnras/stac1945}, \href
  {https://ui.adsabs.harvard.edu/abs/2022MNRAS.515.2890G} {515, 2890}

\bibitem[\protect\citeauthoryear{{Gonz{\'a}lez-Buitrago},
  {Garc{\'\i}a-D{\'\i}az}, {Pozo Nu{\~n}ez}  \& {Guo}}{{Gonz{\'a}lez-Buitrago}
  et~al.}{2023}]{Gonzalez-Buitrago2023}
{Gonz{\'a}lez-Buitrago} D.~H.,  {Garc{\'\i}a-D{\'\i}az} M.~T.,  {Pozo
  Nu{\~n}ez} F.,   {Guo} H.,  2023, \mn@doi [\mnras] {10.1093/mnras/stad2483},
  \href {https://ui.adsabs.harvard.edu/abs/2023MNRAS.tmp.2401G} {}

\bibitem[\protect\citeauthoryear{{Gonzalez}, {Gallo}, {Miller}, {Kammoun},
  {Ghosh}  \& {Pottie}}{{Gonzalez} et~al.}{2024}]{Gonzalez2023}
{Gonzalez} A.~G.,  {Gallo} L.~C.,  {Miller} J.~M.,  {Kammoun} E.~S.,  {Ghosh}
  A.,   {Pottie} B.~A.,  2024, \mn@doi [\mnras] {10.1093/mnras/stae1353}, \href
  {https://ui.adsabs.harvard.edu/abs/2024MNRAS.531.3729G} {531, 3729}

\bibitem[\protect\citeauthoryear{{Gravity Collaboration} et~al.,}{{Gravity
  Collaboration} et~al.}{2018}]{GravityColl:2018}
{Gravity Collaboration} et~al., 2018, \mn@doi [\nat]
  {10.1038/s41586-018-0731-9}, \href
  {https://ui.adsabs.harvard.edu/abs/2018Natur.563..657G} {563, 657}

\bibitem[\protect\citeauthoryear{{Gravity Collaboration} et~al.,}{{Gravity
  Collaboration} et~al.}{2020}]{GravityColl:2020}
{Gravity Collaboration} et~al., 2020, \mn@doi [\aap]
  {10.1051/0004-6361/202039067}, \href
  {https://ui.adsabs.harvard.edu/abs/2020A&A...643A.154G} {643, A154}

\bibitem[\protect\citeauthoryear{{Grier} et~al.,}{{Grier}
  et~al.}{2013}]{Grier13}
{Grier} C.~J.,  et~al., 2013, \mn@doi [\apj] {10.1088/0004-637X/773/2/90},
  \href {https://ui.adsabs.harvard.edu/abs/2013ApJ...773...90G} {773, 90}

\bibitem[\protect\citeauthoryear{{Guo}, {Wang}, {Cai}  \& {Sun}}{{Guo}
  et~al.}{2017}]{Guo17}
{Guo} H.,  {Wang} J.,  {Cai} Z.,   {Sun} M.,  2017, \mn@doi [\apj]
  {10.3847/1538-4357/aa8d71}, \href
  {https://ui.adsabs.harvard.edu/abs/2017ApJ...847..132G} {847, 132}

\bibitem[\protect\citeauthoryear{{Guo}, {Shen}  \& {Wang}}{{Guo}
  et~al.}{2018}]{guo:pyqsofit}
{Guo} H.,  {Shen} Y.,   {Wang} S.,  2018, {PyQSOFit: Python code to fit the
  spectrum of quasars} (\mn@eprint {ascl} {1809.008})

\bibitem[\protect\citeauthoryear{{Guo} et~al.,}{{Guo}
  et~al.}{2022a}]{Guo:paschen2022}
{Guo} H.,  et~al., 2022a, \mn@doi [\apj] {10.3847/1538-4357/ac4bc6}, \href
  {https://ui.adsabs.harvard.edu/abs/2022ApJ...927...60G} {927, 60}

\bibitem[\protect\citeauthoryear{{Guo}, {Li}, {Zhang}, {Ho}  \& {Wang}}{{Guo}
  et~al.}{2022b}]{WGuo2022}
{Guo} W.-J.,  {Li} Y.-R.,  {Zhang} Z.-X.,  {Ho} L.~C.,   {Wang} J.-M.,  2022b,
  \mn@doi [\apj] {10.3847/1538-4357/ac4e84}, \href
  {https://ui.adsabs.harvard.edu/abs/2022ApJ...929...19G} {929, 19}

\bibitem[\protect\citeauthoryear{{Guo}, {Barth}  \& {Wang}}{{Guo}
  et~al.}{2022c}]{HGuo2022}
{Guo} H.,  {Barth} A.~J.,   {Wang} S.,  2022c, \mn@doi [\apj]
  {10.3847/1538-4357/ac96ec}, \href
  {https://ui.adsabs.harvard.edu/abs/2022ApJ...940...20G} {940, 20}

\bibitem[\protect\citeauthoryear{{Henden}}{{Henden}}{2019}]{henden2019}
{Henden} A.~A.,  2019, Journal of the American Association of Variable Star
  Observers (JAAVSO), \href
  {https://ui.adsabs.harvard.edu/abs/2019JAVSO..47..130H} {47, 130}

\bibitem[\protect\citeauthoryear{{Hern{\'a}ndez Santisteban}
  et~al.,}{{Hern{\'a}ndez Santisteban} et~al.}{2020}]{Hernandez20}
{Hern{\'a}ndez Santisteban} J.~V.,  et~al., 2020, \mn@doi [\mnras]
  {10.1093/mnras/staa2365}, \href
  {https://ui.adsabs.harvard.edu/abs/2020MNRAS.498.5399H} {498, 5399}

\bibitem[\protect\citeauthoryear{{Hlabathe} et~al.,}{{Hlabathe}
  et~al.}{2020}]{hlabathe2020}
{Hlabathe} M.~S.,  et~al., 2020, \mn@doi [\mnras] {10.1093/mnras/staa2171},
  \href {https://ui.adsabs.harvard.edu/abs/2020MNRAS.497.2910H} {497, 2910}

\bibitem[\protect\citeauthoryear{{Hodgkin}, {Irwin}, {Hewett}  \&
  {Warren}}{{Hodgkin} et~al.}{2009}]{Hodgkin09}
{Hodgkin} S.~T.,  {Irwin} M.~J.,  {Hewett} P.~C.,   {Warren} S.~J.,  2009,
  \mn@doi [\mnras] {10.1111/j.1365-2966.2008.14387.x}, \href
  {https://ui.adsabs.harvard.edu/abs/2009MNRAS.394..675H} {394, 675}

\bibitem[\protect\citeauthoryear{{Homayouni} et~al.,}{{Homayouni}
  et~al.}{2019}]{Homayouni19}
{Homayouni} Y.,  et~al., 2019, \mn@doi [\apj] {10.3847/1538-4357/ab2638}, \href
  {https://ui.adsabs.harvard.edu/abs/2019ApJ...880..126H} {880, 126}

\bibitem[\protect\citeauthoryear{{H{\"o}nig}}{{H{\"o}nig}}{2014}]{Honig14}
{H{\"o}nig} S.~F.,  2014, \mn@doi [\apjl] {10.1088/2041-8205/784/1/L4}, \href
  {https://ui.adsabs.harvard.edu/abs/2014ApJ...784L...4H} {784, L4}

\bibitem[\protect\citeauthoryear{Hunter}{Hunter}{2007}]{Hunter:2007}
Hunter J.~D.,  2007, \mn@doi [Computing In Science \& Engineering]
  {10.1109/MCSE.2007.55}, 9, 90

\bibitem[\protect\citeauthoryear{{Jha}, {Joshi}, {Chand}, {Wu}, {Ho}, {Rastogi}
   \& {Ma}}{{Jha} et~al.}{2022}]{Jha2022}
{Jha} V.~K.,  {Joshi} R.,  {Chand} H.,  {Wu} X.-B.,  {Ho} L.~C.,  {Rastogi} S.,
    {Ma} Q.,  2022, \mn@doi [\mnras] {10.1093/mnras/stac109}, \href
  {https://ui.adsabs.harvard.edu/abs/2022MNRAS.511.3005J} {511, 3005}

\bibitem[\protect\citeauthoryear{{Jiang} et~al.,}{{Jiang}
  et~al.}{2017}]{Jiang17}
{Jiang} Y.-F.,  et~al., 2017, \mn@doi [\apj] {10.3847/1538-4357/aa5b91}, \href
  {https://ui.adsabs.harvard.edu/abs/2017ApJ...836..186J} {836, 186}

\bibitem[\protect\citeauthoryear{{Jiang}, {Wu}, {Ma}, {Gu}  \& {Wen}}{{Jiang}
  et~al.}{2024}]{Jiang2024}
{Jiang} Y.,  {Wu} X.-B.,  {Ma} Q.,  {Gu} H.,   {Wen} Y.,  2024, \mn@doi [\apj]
  {10.3847/1538-4357/ad36c0}, \href
  {https://ui.adsabs.harvard.edu/abs/2024ApJ...966..149J} {966, 149}

\bibitem[\protect\citeauthoryear{Jones, Oliphant, Peterson  \& {others}}{Jones
  et~al.}{2001}]{Scipy}
Jones E.,  Oliphant T.,  Peterson P.,   {others} 2001, SciPy: Open Source
  Scientific Tools for Python

\bibitem[\protect\citeauthoryear{{Kammoun}, {Papadakis}  \&
  {Dov{\v{c}}iak}}{{Kammoun} et~al.}{2021}]{Kammoun2021}
{Kammoun} E.~S.,  {Papadakis} I.~E.,   {Dov{\v{c}}iak} M.,  2021, \mn@doi
  [\mnras] {10.1093/mnras/stab725}, \href
  {https://ui.adsabs.harvard.edu/abs/2021MNRAS.503.4163K} {503, 4163}

\bibitem[\protect\citeauthoryear{{Kara} et~al.,}{{Kara} et~al.}{2021}]{Kara21}
{Kara} E.,  et~al., 2021, \mn@doi [\apj] {10.3847/1538-4357/ac2159}, \href
  {https://ui.adsabs.harvard.edu/abs/2021ApJ...922..151K} {922, 151}

\bibitem[\protect\citeauthoryear{{Kara} et~al.,}{{Kara}
  et~al.}{2023}]{Kara2023}
{Kara} E.,  et~al., 2023, \mn@doi [\apj] {10.3847/1538-4357/acbcd3}, \href
  {https://ui.adsabs.harvard.edu/abs/2023ApJ...947...62K} {947, 62}

\bibitem[\protect\citeauthoryear{{Kasliwal}, {Vogeley}  \&
  {Richards}}{{Kasliwal} et~al.}{2015}]{Kasliwal15}
{Kasliwal} V.~P.,  {Vogeley} M.~S.,   {Richards} G.~T.,  2015, \mn@doi [\mnras]
  {10.1093/mnras/stv1230}, \href
  {https://ui.adsabs.harvard.edu/abs/2015MNRAS.451.4328K} {451, 4328}

\bibitem[\protect\citeauthoryear{{Kelly}, {Bechtold}  \&
  {Siemiginowska}}{{Kelly} et~al.}{2009}]{Kelly:2009}
{Kelly} B.~C.,  {Bechtold} J.,   {Siemiginowska} A.,  2009, \mn@doi [\apj]
  {10.1088/0004-637X/698/1/895}, \href
  {https://ui.adsabs.harvard.edu/abs/2009ApJ...698..895K} {698, 895}

\bibitem[\protect\citeauthoryear{{Kochanek}}{{Kochanek}}{2020}]{Kochanek:2020}
{Kochanek} C.~S.,  2020, \mn@doi [\mnras] {10.1093/mnras/staa344}, \href
  {https://ui.adsabs.harvard.edu/abs/2020MNRAS.493.1725K} {493, 1725}

\bibitem[\protect\citeauthoryear{{Korista} \& {Goad}}{{Korista} \&
  {Goad}}{2000}]{Korista00}
{Korista} K.~T.,  {Goad} M.~R.,  2000, \mn@doi [\apj] {10.1086/308930}, \href
  {https://ui.adsabs.harvard.edu/abs/2000ApJ...536..284K} {536, 284}

\bibitem[\protect\citeauthoryear{{Korista} \& {Goad}}{{Korista} \&
  {Goad}}{2001}]{Korista01}
{Korista} K.~T.,  {Goad} M.~R.,  2001, \mn@doi [\apj] {10.1086/320964}, \href
  {https://ui.adsabs.harvard.edu/abs/2001ApJ...553..695K} {553, 695}

\bibitem[\protect\citeauthoryear{{Korista} \& {Goad}}{{Korista} \&
  {Goad}}{2019}]{Korista19}
{Korista} K.~T.,  {Goad} M.~R.,  2019, \mn@doi [\mnras]
  {10.1093/mnras/stz2330}, \href
  {https://ui.adsabs.harvard.edu/abs/2019MNRAS.489.5284K} {489, 5284}

\bibitem[\protect\citeauthoryear{{Koz{\l}owski}}{{Koz{\l}owski}}{2016}]{Kozlowski:2016}
{Koz{\l}owski} S.,  2016, \mn@doi [\mnras] {10.1093/mnras/stw819}, \href
  {https://ui.adsabs.harvard.edu/abs/2016MNRAS.459.2787K} {459, 2787}

\bibitem[\protect\citeauthoryear{{Koz{\l}owski}}{{Koz{\l}owski}}{2017}]{Kozlowski:2017}
{Koz{\l}owski} S.,  2017, \mn@doi [\aap] {10.1051/0004-6361/201629890}, \href
  {https://ui.adsabs.harvard.edu/abs/2017A&A...597A.128K} {597, A128}

\bibitem[\protect\citeauthoryear{{Kubota} \& {Done}}{{Kubota} \&
  {Done}}{2018}]{KubotaDone18}
{Kubota} A.,  {Done} C.,  2018, \mn@doi [\mnras] {10.1093/mnras/sty1890}, \href
  {https://ui.adsabs.harvard.edu/abs/2018MNRAS.480.1247K} {480, 1247}

\bibitem[\protect\citeauthoryear{{Kumari}, {Pal}, {Naik}, {Jana}, {Jaisawal}
  \& {kushwaha}}{{Kumari} et~al.}{2021}]{Kumari2021}
{Kumari} N.,  {Pal} M.,  {Naik} S.,  {Jana} A.,  {Jaisawal} G.~K.,   {kushwaha}
  P.,  2021, arXiv e-prints, \href
  {https://ui.adsabs.harvard.edu/abs/2021arXiv210711994K} {p. arXiv:2107.11994}

\bibitem[\protect\citeauthoryear{{Landsman}}{{Landsman}}{1993}]{Landsman1993}
{Landsman} W.~B.,  1993, in {Hanisch} R.~J.,  {Brissenden} R.~J.~V.,   {Barnes}
  J.,  eds,  Astronomical Society of the Pacific Conference Series Vol. 52,
  Astronomical Data Analysis Software and Systems II. p.~246

\bibitem[\protect\citeauthoryear{{Lang}, {Hogg}, {Mierle}, {Blanton}  \&
  {Roweis}}{{Lang} et~al.}{2010}]{Lang10}
{Lang} D.,  {Hogg} D.~W.,  {Mierle} K.,  {Blanton} M.,   {Roweis} S.,  2010,
  \mn@doi [\aj] {10.1088/0004-6256/139/5/1782}, \href
  {https://ui.adsabs.harvard.edu/abs/2010AJ....139.1782L} {139, 1782}

\bibitem[\protect\citeauthoryear{{Lawrence}}{{Lawrence}}{2018}]{Lawrence:2018}
{Lawrence} A.,  2018, \mn@doi [Nature Astronomy] {10.1038/s41550-017-0372-1},
  \href {https://ui.adsabs.harvard.edu/abs/2018NatAs...2..102L} {2, 102}

\bibitem[\protect\citeauthoryear{{Lewin}, {Kara}, {Cackett}, {Wilkins},
  {Panagiotou}, {Garcia}  \& {Gelbord}}{{Lewin} et~al.}{2023}]{Lewin2023}
{Lewin} C.,  {Kara} E.,  {Cackett} E.~M.,  {Wilkins} D.,  {Panagiotou} C.,
  {Garcia} J.~A.,   {Gelbord} J.,  2023, \mn@doi [arXiv e-prints]
  {10.48550/arXiv.2307.11145}, \href
  {https://ui.adsabs.harvard.edu/abs/2023arXiv230711145L} {p. arXiv:2307.11145}

\bibitem[\protect\citeauthoryear{{Lewin} et~al.,}{{Lewin}
  et~al.}{2024}]{Lewin2024}
{Lewin} C.,  et~al., 2024, \mn@doi [\apj] {10.3847/1538-4357/ad6b08}, \href
  {https://ui.adsabs.harvard.edu/abs/2024ApJ...974..271L} {974, 271}

\bibitem[\protect\citeauthoryear{{Li}, {Wang}, {Hu}, {Du}  \& {Bai}}{{Li}
  et~al.}{2014}]{Li2014}
{Li} Y.-R.,  {Wang} J.-M.,  {Hu} C.,  {Du} P.,   {Bai} J.-M.,  2014, \mn@doi
  [\apjl] {10.1088/2041-8205/786/1/L6}, \href
  {https://ui.adsabs.harvard.edu/abs/2014ApJ...786L...6L} {786, L6}

\bibitem[\protect\citeauthoryear{{Li} et~al.,}{{Li} et~al.}{2021}]{Li2021}
{Li} T.,  et~al., 2021, \mn@doi [\apjl] {10.3847/2041-8213/abf9aa}, \href
  {https://ui.adsabs.harvard.edu/abs/2021ApJ...912L..29L} {912, L29}

\bibitem[\protect\citeauthoryear{{Mahmoud} \& {Done}}{{Mahmoud} \&
  {Done}}{2020}]{Mahmoud2020}
{Mahmoud} R.~D.,  {Done} C.,  2020, \mn@doi [\mnras] {10.1093/mnras/stz3196},
  \href {https://ui.adsabs.harvard.edu/abs/2020MNRAS.491.5126M} {491, 5126}

\bibitem[\protect\citeauthoryear{{Mandal}, {Woo}  \& {Wang}}{{Mandal}
  et~al.}{2025}]{Mandal:2025}
{Mandal} A.~K.,  {Woo} J.-H.,   {Wang} S.,  2025, \mn@doi [\apj]
  {10.3847/1538-4357/adc56e}, \href
  {https://ui.adsabs.harvard.edu/abs/2025ApJ...985...30M} {985, 30}

\bibitem[\protect\citeauthoryear{{McHardy} et~al.,}{{McHardy}
  et~al.}{2016}]{McHardy2016}
{McHardy} I.~M.,  et~al., 2016, \mn@doi [Astronomische Nachrichten]
  {10.1002/asna.201612337}, \href
  {https://ui.adsabs.harvard.edu/abs/2016AN....337..500M} {337, 500}

\bibitem[\protect\citeauthoryear{{McHardy} et~al.,}{{McHardy}
  et~al.}{2018}]{McHardy18}
{McHardy} I.~M.,  et~al., 2018, \mn@doi [\mnras] {10.1093/mnras/sty1983}, \href
  {https://ui.adsabs.harvard.edu/abs/2018MNRAS.480.2881M} {480, 2881}

\bibitem[\protect\citeauthoryear{{McHardy} et~al.,}{{McHardy}
  et~al.}{2023}]{McHardy2023}
{McHardy} I.~M.,  et~al., 2023, \mn@doi [\mnras] {10.1093/mnras/stac3651},
  \href {https://ui.adsabs.harvard.edu/abs/2023MNRAS.519.3366M} {519, 3366}

\bibitem[\protect\citeauthoryear{{Montano}, {Guo}, {Barth}, {U}, {Remigio},
  {Gonz{\'a}lez-Buitrago}  \& {Hern{\'a}ndez Santisteban}}{{Montano}
  et~al.}{2022}]{Montano2022}
{Montano} J.~W.,  {Guo} H.,  {Barth} A.~J.,  {U} V.,  {Remigio} R.,
  {Gonz{\'a}lez-Buitrago} D.~H.,   {Hern{\'a}ndez Santisteban} J.~V.,  2022,
  \mn@doi [\apjl] {10.3847/2041-8213/ac7e54}, \href
  {https://ui.adsabs.harvard.edu/abs/2022ApJ...934L..37M} {934, L37}

\bibitem[\protect\citeauthoryear{{Morgan}, {Kochanek}, {Morgan}  \&
  {Falco}}{{Morgan} et~al.}{2010}]{Morgan10}
{Morgan} C.~W.,  {Kochanek} C.~S.,  {Morgan} N.~D.,   {Falco} E.~E.,  2010,
  \mn@doi [\apj] {10.1088/0004-637X/712/2/1129}, \href
  {https://ui.adsabs.harvard.edu/abs/2010ApJ...712.1129M} {712, 1129}

\bibitem[\protect\citeauthoryear{{Mudd} et~al.,}{{Mudd} et~al.}{2018}]{Mudd18}
{Mudd} D.,  et~al., 2018, \mn@doi [\apj] {10.3847/1538-4357/aac9bb}, \href
  {https://ui.adsabs.harvard.edu/abs/2018ApJ...862..123M} {862, 123}

\bibitem[\protect\citeauthoryear{{Mushotzky}, {Edelson}, {Baumgartner}  \&
  {Gandhi}}{{Mushotzky} et~al.}{2011}]{Mushotzky11}
{Mushotzky} R.~F.,  {Edelson} R.,  {Baumgartner} W.,   {Gandhi} P.,  2011,
  \mn@doi [\apjl] {10.1088/2041-8205/743/1/L12}, \href
  {https://ui.adsabs.harvard.edu/abs/2011ApJ...743L..12M} {743, L12}

\bibitem[\protect\citeauthoryear{{Netzer}}{{Netzer}}{2022}]{Netzer2022}
{Netzer} H.,  2022, \mn@doi [\mnras] {10.1093/mnras/stab3133}, \href
  {https://ui.adsabs.harvard.edu/abs/2022MNRAS.509.2637N} {509, 2637}

\bibitem[\protect\citeauthoryear{{Netzer} et~al.,}{{Netzer}
  et~al.}{2024}]{Netzer2024}
{Netzer} H.,  et~al., 2024, \mn@doi [\apj] {10.3847/1538-4357/ad8160}, \href
  {https://ui.adsabs.harvard.edu/abs/2024ApJ...976...59N} {976, 59}

\bibitem[\protect\citeauthoryear{{Neustadt} \& {Kochanek}}{{Neustadt} \&
  {Kochanek}}{2022}]{Neustadt2022}
{Neustadt} J.~M.~M.,  {Kochanek} C.~S.,  2022, \mn@doi [\mnras]
  {10.1093/mnras/stac888}, \href
  {https://ui.adsabs.harvard.edu/abs/2022MNRAS.513.1046N} {513, 1046}

\bibitem[\protect\citeauthoryear{{Onken}, {Ferrarese}, {Merritt}, {Peterson},
  {Pogge}, {Vestergaard}  \& {Wandel}}{{Onken} et~al.}{2004}]{Onken04}
{Onken} C.~A.,  {Ferrarese} L.,  {Merritt} D.,  {Peterson} B.~M.,  {Pogge}
  R.~W.,  {Vestergaard} M.,   {Wandel} A.,  2004, \mn@doi [\apj]
  {10.1086/424655}, \href
  {https://ui.adsabs.harvard.edu/abs/2004ApJ...615..645O} {615, 645}

\bibitem[\protect\citeauthoryear{{Pal} \& {Naik}}{{Pal} \&
  {Naik}}{2018}]{Pal2018}
{Pal} M.,  {Naik} S.,  2018, \mn@doi [\mnras] {10.1093/mnras/stx3103}, \href
  {https://ui.adsabs.harvard.edu/abs/2018MNRAS.474.5351P} {474, 5351}

\bibitem[\protect\citeauthoryear{{Pei} et~al.,}{{Pei} et~al.}{2014}]{Pei14}
{Pei} L.,  et~al., 2014, \mn@doi [\apj] {10.1088/0004-637X/795/1/38}, \href
  {https://ui.adsabs.harvard.edu/abs/2014ApJ...795...38P} {795, 38}

\bibitem[\protect\citeauthoryear{{Peterson}, {Wanders}, {Horne}, {Collier},
  {Alexander}, {Kaspi}  \& {Maoz}}{{Peterson} et~al.}{1998}]{Peterson98}
{Peterson} B.~M.,  {Wanders} I.,  {Horne} K.,  {Collier} S.,  {Alexander} T.,
  {Kaspi} S.,   {Maoz} D.,  1998, \mn@doi [\pasp] {10.1086/316177}, \href
  {https://ui.adsabs.harvard.edu/abs/1998PASP..110..660P} {110, 660}

\bibitem[\protect\citeauthoryear{{Peterson} et~al.,}{{Peterson}
  et~al.}{2004}]{Peterson04}
{Peterson} B.~M.,  et~al., 2004, \mn@doi [\apj] {10.1086/423269}, \href
  {https://ui.adsabs.harvard.edu/abs/2004ApJ...613..682P} {613, 682}

\bibitem[\protect\citeauthoryear{{Pozo Nu{\~n}ez} et~al.,}{{Pozo Nu{\~n}ez}
  et~al.}{2019}]{PozoNunez2019}
{Pozo Nu{\~n}ez} F.,  et~al., 2019, \mn@doi [\mnras] {10.1093/mnras/stz2830},
  \href {https://ui.adsabs.harvard.edu/abs/2019MNRAS.490.3936P} {490, 3936}

\bibitem[\protect\citeauthoryear{{Roming} et~al.,}{{Roming}
  et~al.}{2005}]{Roming05}
{Roming} P. W.~A.,  et~al., 2005, \mn@doi [\ssr] {10.1007/s11214-005-5095-4},
  \href {https://ui.adsabs.harvard.edu/abs/2005SSRv..120...95R} {120, 95}

\bibitem[\protect\citeauthoryear{{Runnoe}, {Brotherton}  \& {Shang}}{{Runnoe}
  et~al.}{2012}]{Runnoe2012}
{Runnoe} J.~C.,  {Brotherton} M.~S.,   {Shang} Z.,  2012, \mn@doi [\mnras]
  {10.1111/j.1365-2966.2012.20620.x}, \href
  {https://ui.adsabs.harvard.edu/abs/2012MNRAS.422..478R} {422, 478}

\bibitem[\protect\citeauthoryear{{Sakata} et~al.,}{{Sakata}
  et~al.}{2010}]{Sakata2010}
{Sakata} Y.,  et~al., 2010, \mn@doi [\apj] {10.1088/0004-637X/711/1/461}, \href
  {https://ui.adsabs.harvard.edu/abs/2010ApJ...711..461S} {711, 461}

\bibitem[\protect\citeauthoryear{{Schlafly} \& {Finkbeiner}}{{Schlafly} \&
  {Finkbeiner}}{2011}]{Schlafly11}
{Schlafly} E.~F.,  {Finkbeiner} D.~P.,  2011, \mn@doi [\apj]
  {10.1088/0004-637X/737/2/103}, \href
  {https://ui.adsabs.harvard.edu/abs/2011ApJ...737..103S} {737, 103}

\bibitem[\protect\citeauthoryear{{Sergeev}, {Doroshenko}, {Golubinskiy},
  {Merkulova}  \& {Sergeeva}}{{Sergeev} et~al.}{2005}]{Sergeev05}
{Sergeev} S.~G.,  {Doroshenko} V.~T.,  {Golubinskiy} Y.~V.,  {Merkulova} N.~I.,
    {Sergeeva} E.~A.,  2005, \mn@doi [\apj] {10.1086/427820}, \href
  {https://ui.adsabs.harvard.edu/abs/2005ApJ...622..129S} {622, 129}

\bibitem[\protect\citeauthoryear{{Shakura} \& {Sunyaev}}{{Shakura} \&
  {Sunyaev}}{1973}]{ShakuraSunyaev73}
{Shakura} N.~I.,  {Sunyaev} R.~A.,  1973, \aap, \href
  {https://ui.adsabs.harvard.edu/abs/1973A&A....24..337S} {500, 33}

\bibitem[\protect\citeauthoryear{{Sharp} et~al.,}{{Sharp}
  et~al.}{2024}]{Sharp:2024}
{Sharp} H.~W.,  et~al., 2024, \mn@doi [\apj] {10.3847/1538-4357/ad0cea}, \href
  {https://ui.adsabs.harvard.edu/abs/2024ApJ...961...93S} {961, 93}

\bibitem[\protect\citeauthoryear{{Starkey}, {Horne}  \& {Villforth}}{{Starkey}
  et~al.}{2016a}]{Starkey16}
{Starkey} D.~A.,  {Horne} K.,   {Villforth} C.,  2016a, \mn@doi [\mnras]
  {10.1093/mnras/stv2744}, \href
  {https://ui.adsabs.harvard.edu/abs/2016MNRAS.456.1960S} {456, 1960}

\bibitem[\protect\citeauthoryear{{Starkey}, {Horne}  \& {Villforth}}{{Starkey}
  et~al.}{2016b}]{Starkey:2016}
{Starkey} D.~A.,  {Horne} K.,   {Villforth} C.,  2016b, \mn@doi [\mnras]
  {10.1093/mnras/stv2744}, \href
  {https://ui.adsabs.harvard.edu/abs/2016MNRAS.456.1960S} {456, 1960}

\bibitem[\protect\citeauthoryear{{Su} et~al.,}{{Su} et~al.}{2024}]{Su2024}
{Su} Z.-B.,  et~al., 2024, \mn@doi [\apj] {10.3847/1538-4357/ad86bc}, \href
  {https://ui.adsabs.harvard.edu/abs/2024ApJ...976..155S} {976, 155}

\bibitem[\protect\citeauthoryear{{Sun}, {Grier}  \& {Peterson}}{{Sun}
  et~al.}{2018}]{pyccf}
{Sun} M.,  {Grier} C.~J.,   {Peterson} B.~M.,  2018, {PyCCF: Python Cross
  Correlation Function for reverberation mapping studies} (\mn@eprint {ascl}
  {1805.032})

\bibitem[\protect\citeauthoryear{{The Astropy Collaboration} et~al.,}{{The
  Astropy Collaboration} et~al.}{2013}]{Astropy:2013}
{The Astropy Collaboration} et~al., 2013, \mn@doi [A\&A]
  {10.1051/0004-6361/201322068}, 558, A33

\bibitem[\protect\citeauthoryear{{Tie} \& {Kochanek}}{{Tie} \&
  {Kochanek}}{2018}]{Tie2018}
{Tie} S.~S.,  {Kochanek} C.~S.,  2018, \mn@doi [\mnras]
  {10.1093/mnras/stx2348}, \href
  {https://ui.adsabs.harvard.edu/abs/2018MNRAS.473...80T} {473, 80}

\bibitem[\protect\citeauthoryear{{Tsuzuki}, {Kawara}, {Yoshii}, {Oyabu},
  {Tanab{\'e}}  \& {Matsuoka}}{{Tsuzuki} et~al.}{2006}]{Tsuzuki06}
{Tsuzuki} Y.,  {Kawara} K.,  {Yoshii} Y.,  {Oyabu} S.,  {Tanab{\'e}} T.,
  {Matsuoka} Y.,  2006, \mn@doi [\apj] {10.1086/506376}, \href
  {https://ui.adsabs.harvard.edu/abs/2006ApJ...650...57T} {650, 57}

\bibitem[\protect\citeauthoryear{{Urry} \& {Padovani}}{{Urry} \&
  {Padovani}}{1995}]{Urry1995}
{Urry} C.~M.,  {Padovani} P.,  1995, \mn@doi [\pasp] {10.1086/133630}, \href
  {https://ui.adsabs.harvard.edu/abs/1995PASP..107..803U} {107, 803}

\bibitem[\protect\citeauthoryear{{Valenti} et~al.,}{{Valenti}
  et~al.}{2015}]{Valenti15}
{Valenti} S.,  et~al., 2015, \mn@doi [\apjl] {10.1088/2041-8205/813/2/L36},
  \href {https://ui.adsabs.harvard.edu/abs/2015ApJ...813L..36V} {813, L36}

\bibitem[\protect\citeauthoryear{{Vasudevan} \& {Fabian}}{{Vasudevan} \&
  {Fabian}}{2009}]{Vasudevan:2009}
{Vasudevan} R.~V.,  {Fabian} A.~C.,  2009, \mn@doi [\mnras]
  {10.1111/j.1365-2966.2008.14108.x}, \href
  {https://ui.adsabs.harvard.edu/abs/2009MNRAS.392.1124V} {392, 1124}

\bibitem[\protect\citeauthoryear{{Vestergaard} \& {Wilkes}}{{Vestergaard} \&
  {Wilkes}}{2001}]{Vestergaard01}
{Vestergaard} M.,  {Wilkes} B.~J.,  2001, \mn@doi [\apjs] {10.1086/320357},
  \href {https://ui.adsabs.harvard.edu/abs/2001ApJS..134....1V} {134, 1}

\bibitem[\protect\citeauthoryear{{Vincentelli} et~al.,}{{Vincentelli}
  et~al.}{2021}]{Vincentelli:2021}
{Vincentelli} F.~M.,  et~al., 2021, \mn@doi [\mnras] {10.1093/mnras/stab1033},
  \href {https://ui.adsabs.harvard.edu/abs/2021MNRAS.504.4337V} {504, 4337}

\bibitem[\protect\citeauthoryear{{Watson} et~al.,}{{Watson}
  et~al.}{2012}]{Watson12}
{Watson} A.~M.,  et~al., 2012, in {Stepp} L.~M.,  {Gilmozzi} R.,   {Hall}
  H.~J.,  eds,  Society of Photo-Optical Instrumentation Engineers (SPIE)
  Conference Series Vol. 8444, Ground-based and Airborne Telescopes IV. p.
  84445L, \mn@doi{10.1117/12.926927}

\bibitem[\protect\citeauthoryear{{Weaver} \& {Horne}}{{Weaver} \&
  {Horne}}{2022}]{Weaver22}
{Weaver} J.~R.,  {Horne} K.,  2022, \mn@doi [\mnras] {10.1093/mnras/stac248},
  \href {https://ui.adsabs.harvard.edu/abs/2022MNRAS.512..899W} {512, 899}

\bibitem[\protect\citeauthoryear{{White} \& {Peterson}}{{White} \&
  {Peterson}}{1994}]{White:1994}
{White} R.~J.,  {Peterson} B.~M.,  1994, \mn@doi [\pasp] {10.1086/133456},
  \href {https://ui.adsabs.harvard.edu/abs/1994PASP..106..879W} {106, 879}

\bibitem[\protect\citeauthoryear{{Yao}, {Secunda}, {Jiang}, {Greene}  \&
  {Villar}}{{Yao} et~al.}{2023}]{Yao2023}
{Yao} P.~Z.,  {Secunda} A.,  {Jiang} Y.-F.,  {Greene} J.~E.,   {Villar} A.,
  2023, \mn@doi [\apj] {10.3847/1538-4357/acde7e}, \href
  {https://ui.adsabs.harvard.edu/abs/2023ApJ...953...43Y} {953, 43}

\bibitem[\protect\citeauthoryear{{Yu} et~al.,}{{Yu} et~al.}{2020a}]{Yu2020:DES}
{Yu} Z.,  et~al., 2020a, \mn@doi [\apjs] {10.3847/1538-4365/ab5e7a}, \href
  {https://ui.adsabs.harvard.edu/abs/2020ApJS..246...16Y} {246, 16}

\bibitem[\protect\citeauthoryear{{Yu}, {Kochanek}, {Peterson}, {Zu}, {Brandt},
  {Cackett}, {Fausnaugh}  \& {McHardy}}{{Yu} et~al.}{2020b}]{Yu2020:sim}
{Yu} Z.,  {Kochanek} C.~S.,  {Peterson} B.~M.,  {Zu} Y.,  {Brandt} W.~N.,
  {Cackett} E.~M.,  {Fausnaugh} M.~M.,   {McHardy} I.~M.,  2020b, \mn@doi
  [\mnras] {10.1093/mnras/stz3464}, \href
  {https://ui.adsabs.harvard.edu/abs/2020MNRAS.491.6045Y} {491, 6045}

\bibitem[\protect\citeauthoryear{{Yuan} et~al.,}{{Yuan}
  et~al.}{2020}]{Yuan2020}
{Yuan} W.,  et~al., 2020, \mn@doi [\apj] {10.3847/1538-4357/abb377}, \href
  {https://ui.adsabs.harvard.edu/abs/2020ApJ...902...26Y} {902, 26}

\bibitem[\protect\citeauthoryear{{Zhou} et~al.,}{{Zhou}
  et~al.}{2024}]{Zhou2024}
{Zhou} S.,  et~al., 2024, \mn@doi [arXiv e-prints] {10.48550/arXiv.2408.11292},
  \href {https://ui.adsabs.harvard.edu/abs/2024arXiv240811292Z} {p.
  arXiv:2408.11292}

\bibitem[\protect\citeauthoryear{{Zu}, {Kochanek}  \& {Peterson}}{{Zu}
  et~al.}{2011}]{Zu11}
{Zu} Y.,  {Kochanek} C.~S.,   {Peterson} B.~M.,  2011, \mn@doi [\apj]
  {10.1088/0004-637X/735/2/80}, \href
  {https://ui.adsabs.harvard.edu/abs/2011ApJ...735...80Z} {735, 80}

\bibitem[\protect\citeauthoryear{{Zu}, {Kochanek}, {Koz{\l}owski}  \&
  {Peterson}}{{Zu} et~al.}{2016}]{Zu16}
{Zu} Y.,  {Kochanek} C.~S.,  {Koz{\l}owski} S.,   {Peterson} B.~M.,  2016,
  \mn@doi [\apj] {10.3847/0004-637X/819/2/122}, \href
  {https://ui.adsabs.harvard.edu/abs/2016ApJ...819..122Z} {819, 122}

\bibitem[\protect\citeauthoryear{{van Dokkum}}{{van
  Dokkum}}{2001}]{vanDokkum01}
{van Dokkum} P.~G.,  2001, \mn@doi [\pasp] {10.1086/323894}, \href
  {https://ui.adsabs.harvard.edu/abs/2001PASP..113.1420V} {113, 1420}

\makeatother
\end{thebibliography}



\appendix

\section{facilities and software}

\textbf{Facilities:} LCOGT, FTN, FTS, Liverpool:2m, OANSPM:HJT, BYU:0.9m, Asiago:Schmidt, KAIT, MLO:1m, Nickel, \sw\,, HST (STIS)

\noindent\textbf{Software:} Astropy \citep{Astropy:2013, Astropy2018}, Matplotlib \citep{Hunter:2007}, Scipy \citep{Scipy}, \javelin \citep{Zu16}, CREAM \citep{Starkey:2016}, \pyccf \citep{pyccf},  PyCALI \citep{Li2014}, The IDL Astronomy User's Library \citep{Landsman1993}.

\section*{Affiliations}
\noindent
$^{1}$Universidad Nacional Aut\'onoma de M\'exico, Instituto de Astronom\'ia, AP 106,  Ensenada 22860, BC, M\'exico\\
$^{2}$Department of Physics and Astronomy, 4129 Frederick Reines 
Hall, University of California, Irvine, CA, 92697-4575, USA\\
$^{3}$Eureka Scientific, 2452 Delmer Street Suite 100, Oakland, CA 94602-3017, USA\\
$^{4}$SUPA Physics and Astronomy, University of St. Andrews, 
Fife, KY16 9SS, Scotland, UK\\
$^5$Key Laboratory for Particle Astrophysics, Institute of High Energy Physics, Chinese Academy of Sciences, 19B Yuquan Road,\\ Beijing 100049, People's Republic of China\\
$^6$Shanghai Astronomical Observatory, Chinese Academy of Sciences, 80 Nandan Road, Shanghai 200030, People's Republic of China\\
$^7$Department of Physics and Astronomy, N284 ESC, Brigham Young University, Provo, UT 84602, USA\\
$^{8}$Department of Physics and Astronomy, Wayne State University, 666 W.\ Hancock St, Detroit, MI, 48201, USA\\
$^{9}$Spectral Sciences Inc., 30 Fourth Ave. Suite 2, Burlington MA 01803, USA\\
$^{10}$Department of Physics and Astronomy, Georgia State University, Atlanta, GA 30303, USA\\
$^{11}$Department of Astronomy and Astrophysics, 525 Davey Lab, The Pennsylvania State University, University Park, PA 16802, USA\\
$^{12}$Institute for Gravitation and the Cosmos, The Pennsylvania State University, University Park, PA 16802, USA\\
$^{13}$Department of Physics, 104 Davey Laboratory, The Pennsylvania State University, University Park, PA 16802, USA\\
$^{14}$University of Leicester, Department of Physics and Astronomy, Leicester, LE1 7RH, UK\\
$^{15}$DARK, The Niels Bohr Institute, University of Copenhagen, Jagtvej 155, DK-2200 Copenhagen N, Denmark\\
$^{16}$Steward Observatory and Dept. of Astronomy, University of Arizona, 933 N Cherry Avenue, Tucson, AZ 85721, USA\\
$^{17}$Department of Physics; Western Michigan University; Kalamazoo, MI 49008-5252, USA\\
$^{18}$University of Bath, Department of Physics, Claverton Down, Bath, BA27AY, UK\\
$^{19}$University College London, Mullard Space Science Laboratory, Holmbury St. Mary, Dorking, Surrey, RH5 6NT, UK\\
$^{20}$Department of Astronomy, University of California, Berkeley, CA 94720-3411, USA\\
$^{21}$Dipartimento di Fisica e Astronomia ``G. Galilei”, Universit\`a di Padova, Vicolo dell’Osservatorio 3, I-35122 Padova, Italy\\
$^{22}$INAF -- Osservatorio Astronomico di Padova, Vicolo dell'Osservatorio 5, I-35122 Padova, Italy\\
$^{23}$Jeremiah Horrocks Institute, University of Central Lancashire, Preston, PR1 2HE, UK\\
$^{24}$Department of Physics and Astronomy, University of Kentucky, Lexington, KY 40506, USA\\
$^{25}$Institute for Astrophysical Research, Boston University, 725 Commonwealth Avenue, Boston, MA 02215, USA\\
$^{26}$Department of Astronomy, San Diego State University, San Diego, CA 92812, USA\\
$^{27}$Department of Astronomy and Atmospheric Sciences, Kyungpook National University, Daegu 41566, Republic of Korea\\
$^{28}$Instituto de Astronomia, Universidad Nacional Autonoma de Mexico, 04510 Mexico City, Mexico\\
$^{29}$Space Telescope Science Institute, 3700 San Martin Drive, Baltimore, MD
21218, USA\\  
$^{30}$Department of Astronomy, University of Virginia, Charlottesville, VA 22904, USA\\
$^{31}$Space Telescope Science Institute, 3700 San Martin Drive, Baltimore, MD
21218, USA\\
$^{32}$Instituto de Astronom\'ia y Ciencias Planetarias, Universidad de Atacama, Copayapu 485, Copiap\'o, Chile\\
$^{33}$Astronomy department, Ohio State University, Columbus, OH 43210, USA.\\
$^{34}$Center for Cosmology and AstroParticle Physics, Ohio State University, Columbus, OH 43210, USA\\
$^{35}$School of Physics and Astronomy and the Wise Observatory, The Raymond and Beverly Sackler Faculty of Exact Sciences, Tel-Aviv University, Tel-Aviv 69978, Israel.\\
$^{36}$Korea Astronomy and Space Science Institute, 776 Daedeokdae-ro, Yuseong-gu, Daejeon 34055, Republic of Korea\\
$^{37}$Department of Physics and Astronomy, University of Leicester, Leicester LE1 7RH, UK.\\
$^{38}$Department of Physics, University of Johannesburg, P.O. Box 524, 2006 Auckland Park, South Africa \\
$^{39}$Bengier-Winslow-Eustace Specialist in Astronomy\\

\bsp	
\label{lastpage}
\end{document}